\documentclass[article]{elsarticle}

\usepackage{lineno,hyperref}
\modulolinenumbers[1]

\usepackage{bm}
\usepackage{amsmath}
\usepackage{subfigure}
\usepackage{booktabs}
\usepackage{threeparttable}
\usepackage{multirow}
\usepackage{makecell}
\usepackage{color}
\usepackage{appendix}
\usepackage[left=2.5cm,right=2.5cm,top=3.5cm,bottom=3.5cm]{geometry}
\journal{Communications in Nonlinear Science and Numerical Simulation}

\begin{document}
	
	\begin{frontmatter}
		
		\title{A novel Cercignani-Lampis boundary model for discrete velocity methods in predicting rarefied and multi-scale flows}
		
		\author[a]{Jianfeng Chen}
		\ead[Jianfeng Chen]{chenjf@mail.nwpu.edu.cn}
		
		\author[a,b,c]{Sha Liu\corref{mycorrespondingauthor}}
		\cortext[mycorrespondingauthor]{Corresponding author}
		\ead[Sha Liu]{shaliu@nwpu.edu.cn}
		
		\author[a]{Rui Zhang}
		
		\author[a]{Hao Jin}
		
		\author[a,b,c]{Congshan Zhuo}
		
		\author[d,e]{Ming Fang}
		
		\author[d,e]{Yanguang Yang}
		
		\author[a,b,c]{Chengwen Zhong}
		\ead[Chengwen Zhong]{zhongcw@nwpu.edu.cn}
		
		\address[a]{School of Aeronautics, Northwestern Polytechnical University, Xi’an, Shaanxi 710072, China}
		\address[b]{Institute of Extreme Mechanics, Northwestern Polytechnical University, Xi’an, Shaanxi 710072, China}
		\address[c]{National Key Laboratory of Aircraft Configuration Design, Northwestern Polytechnical University, Xi’an, Shaanxi 710072, China}
		\address[d]{National Key Laboratory of Aerospace Physics in Fluids, Mianyang, Sichuan 621000, China}
		\address[e]{China Aerodynamics Research and Development Center, Mianyang, Sichuan 621000, China}

		\begin{abstract}
		To extend the discrete velocity method (DVM) and unified methods to more realistic boundary conditions, a Cercignani-Lampis (CL) boundary with different momentum and thermal energy accommodations is proposed and integrated into the DVM framework.
		By giving the macroscopic flux from the numerical quadrature of the incident molecular distribution, the reflected macroscopic flux can be obtained for the given accommodation coefficients. Then, an anisotropic Gaussian distribution can be found for the reflected molecules, whose parameters are determined by the calculated reflected macroscopic flux. These macroscopic flux and microscopic Gaussian distribution form a complete physical process for the reflected molecules.
		Furthermore, the CL boundary is integrated into the unified gas-kinetic scheme (UGKS), making it suitable for the simulation of both monatomic and diatomic gas flows, and it accommodates both the conventional Cartesian velocity space and the recently developed efficient unstructured velocity space.
		Moreover, this new GSI boundary is suitable for both explicit and implicit schemes, offering better performance for flow prediction.
		Finally, the performance of the new boundary is validated through a series of numerical tests covering a wide range of Knudsen and Mach numbers.		
		\end{abstract}
		
		\begin{keyword}
			\texttt Multi-scale flows\sep
			Gas-surface interaction\sep
			Discrete velocity method\sep 
			Accommodation coefficient
		\end{keyword}
		
	\end{frontmatter}
	
	
	\section{Introduction}\label{sec:introduction}
	\par
	Multi-scale flows from earth surface to outer space (or from macroscale to microscale) are common in scenarios such as near-space vehicles and micro-electro-mechanical systems (MEMS), where multiple flow regimes (including continuum, slip, transitional, and free molecular ones) often coexist within a single flow field, leading to complex dynamic process, and challenging the physical modeling and numerical predictions.	
	It is necessary to employ multiple flow models to describe gas-gas interaction (GGI). Additionally, the gas-surface interaction (GSI), which serves as a wall boundary condition \cite{sharipov1998data}, plays an important role in the prediction of aerodynamic forces and heat transfer \cite{cercignani1971kinetic, hedahl1995comparisons}. As the rarefaction level increases, the impact of the GSI boundary becomes increasingly significant \cite{santos2004dsmc}.
	Consequently, the GSI boundary has important applications in aerospace engineering, vacuum technology, microelectronics manufacturing, and surface science. To accurately predict these complex multi-scale flows, it is essential to effectively tackle both GGI and GSI challenges.
	\par
	There are two basic approaches to develop a numerical method for multi-scale flows: the domain decomposition strategy and the unified strategy \cite{liu2022progress}. The unified strategy can be further classified into the deterministic methods, also referred to as the discrete velocity method (DVM) \cite{mieussens2000discrete, li2009gas}, based on discrete velocity space, and the stochastic methods based on model particles \cite{sun2005evaluation,fei2017particle,liu2020unified,liu2020simplified}.
	Deterministic methods provide substantial benefits in simulating low-speed multi-scale flows without statistical fluctuations. However, they require extensive computational resources for high-speed multi-scale flow simulations \cite{yuan2020conservative, chen2019conserved, zhang2024conservative}.
	On the other hand, stochastic methods provide notable advantages in simulating high-speed multi-scale flows. Nevertheless, they face challenges related to statistical fluctuations and require lengthy time-averaging processes for low-speed multi-scale flow simulations \cite{ho2019comparative}.
	In recent years, a class of unified methods based on discrete velocity space has been proposed, such as the unified gas-kinetic scheme (UGKS) \cite{xu2010unified, xu2014direct}, discrete unified gas-kinetic scheme (DUGKS) \cite{guo2013discrete, guo2015discrete}, the general synthetic iteration scheme (GSIS) \cite{su2020can, su2020fast}, the gas-kinetic unified algorithm (GKUA) \cite{li2009gas,peng2016implicit} the improved discrete velocity method (IDVM) \cite{yang2018improved,yang2019improved}. These unified methods make it possible to solve multi-scale flow problems using a unified numerical method.	
	At the present stage, unified methods have been successfully extended to other multi-scale physics, such as radiation of photons \cite{luo2018multiscale,song2020discrete}, phonon heat transfer \cite{guo2016discrete, luo2017discrete}, and plasma gas transfer \cite{pan2018unified}.
	After a decade of development, numerous numerical techniques have been devised and incorporated into these unified methods to improve computational efficiency and reduce memory costs \cite{chen2012unified, liu2012modified, liu2014unified, zhu2016discrete, wang2019arbitrary, chen2017unified, zhu2016implicit, zhang2022unified_parallelization, zhong2020simplified, chen2024global}. 
	\par
	In comparison to the significant attention given to the GGI models in simulating multi-scale flows, the effort devoted to the GSI boundary is insufficient \cite{wu2017assessment}.
	Although several GSI boundaries have been developed, including the Maxwell boundary \cite{kennard1938kinetic, maxwell1879vii}, the Cercignani–Lampis–Lord (CLL) boundary \cite{lord1991some, lord1995some}, and other derivative boundaries \cite{struchtrup2013maxwell, brull2016nanoscale, wu2017assessment, yamamoto2007scattering, hossein2014gas}, almost all the deterministic methods have only applied the diffuse reflection boundary condition with full thermal accommodation, which can be viewed as a rough Maxwell boundary with a fixed accommodation coefficient (AC) \cite{hedahl1995comparisons, peddakotla2019molecular} at unity, deviating from the real value.
	The Maxwell boundary, which combines the diffuse and specular reflection models by a fraction, is the first and simplest GSI boundary. It is widely recognized that the Maxwell boundary has a single AC and cannot simultaneously describe the different momentum and thermal energy accommodations of reflected gas molecules \cite{deng2022modified}.
	In fact, the AC is a physical property of a solid surface, which affects the friction force and heat transport on it, being crucial for aerodynamic and aerothermal predictions.
	Typically, ACs can be calibrated based on experimental data \cite{kinefuchi2017incident, liu2021dsmc} or the results of molecular dynamics (MD) simulations \cite{chirita1997non, reinhold2014molecular, lim2016simulation, andric2018molecular}.
	Common ACs include the normal and tangential momentum ACs (NMAC and TMAC), the energy AC (EAC), the normal and tangential energy ACs (NEAC and TEAC), and the rotational energy AC (REAC).
	Cercignani and Lampis proposed a phenomenological GSI boundary (CL boundary) that uses two independent scattering kernels and two independent ACs (NEAC and TMAC) to describe the normal and tangential velocity components of reflected gas molecules \cite{cercignani1971kinetic}.
	Later, Lord expanded Cercignani and Lampis's boundary (known as the CLL boundary) and implemented it in the direct simulation Monte Carlo (DSMC) method, one of the most famous stochastic methods, making it a popular tool for theoretical and computational studies of rarefied gas flows \cite{lord1991some, lord1995some}.	
	In the stochastic methods based on model particles, the wall boundary condition is applied through the reflected gas molecules, obtained via sampling methods based on the scattering kernel of a GSI boundary \cite{shen2013rarefied}. However, in the DVM methods, the wall boundary condition is applied through the reflected distribution function, which requires integrating the scattering kernel over the velocity space, making the application of GSI boundaries more challenging within the DVM framework.	
	It is worth noting that the original CL boundary and CLL boundary depends on the velocity of incident particles, presenting a challenge in deriving the corresponding reflected distribution function. In contrast, the scattering kernel of the diffuse reflection model is independent of the incident particle velocity, simplifying the derivation of the corresponding reflected distribution function.
	As a result, the stochastic methods can seamlessly apply the CLL boundary by taking advantage of the straightforward sampling of reflection velocity. However, applying the GSI boundary to the DVM methods, which require calculating the reflected distribution function, remains challenging.
	Consequently, in previous research, the DVM methods have only been able to adopt the full accommodation diffuse reflection model.	
	More recently, the author \cite{CHEN2024570} proposed a Maxwell boundary algorithm within the DVM framework, enabling both the DVM and unified methods to apply more precise boundary conditions for predicting the behavior of multi-scale flows and other physical fields.	
	In this paper, we propose a new CL type boundary within the DVM framework that features different momentum and thermal energy accommodations and is independent of the velocity of incident particles.
	The new boundary is designed to accommodate both the recently developed efficient unstructured velocity space \cite{yuan2020conservative, chen2019conserved, chen2020compressible} and the conventional Cartesian velocity space.
	Additionally, the proposed CL boundary enables simulations of both monatomic gases and diatomic gases with internal degrees of freedom.
	Finally, this boundary is integrated into the implicit UGKS with a simplified multi-scale numerical flux \cite{zhang2023unified}.
	\par
	The remainder of this paper is organized as follows: Section~\ref{sec:CL boundary} introduces the proposed CL boundary within the DVM framework, including a brief overview of the UGKS and a detailed description of the new boundary. Section~\ref{sec:Numerical experiment} presents and analyzes several classical numerical simulations to validate the performance of the proposed CL boundary. Finally, the concluding remarks are given in Section~\ref{sec:Conclusion}.
	\par

	\section{A novel CL boundary for the DVM framework}\label{sec:CL boundary}
	\par 
	This paper proposes a novel CL boundary for the DVM framework and implements it within the UGKS with simplified multi-scale flux \cite{chen2019conserved, zhang2023unified}. Then, the main body of this section focuses on the details of the new CL boundary. For completeness, this boundary suits the implicit scheme and diatomic gases. 
	\subsection{Unified gas-kinetic scheme}
	The UGKS adopts the gas-kinetic relaxation model equation in the following form
	\begin{equation}\label{B_EQ1}
		\frac{\partial f}{\partial t}+\bm{\xi }\cdot \nabla f=\Omega \equiv \frac{g-f}{\tau},
	\end{equation}
	where $f=f\left( \bm{x},\bm{\xi},\bm{\eta },e,t \right)$ is the distribution function for particles moving in D-dimensional physical space with a velocity of $\bm{\xi }=\left( \xi _{1}^{{}},...,\xi _{D}^{{}} \right)$ at position $\bm{x}=\left( x_{1}^{{}},...,x_{D}^{{}} \right)$ and time $t$. Here, $\bm{\eta }=\left( \xi _{D+1}^{{}},...,\xi _{3}^{{}} \right)$ is the dummy velocity consisting of the rest components of the particle velocity in three-dimensional space, $e$ represents molecular rotational energy. $\Omega$ is the collision operator. $\tau$ is the relaxation time relating to the dynamic viscosity $\mu$ and pressure $p$ with $\tau = \mu/p$. $g$ is the equilibrium distribution function, such as the Maxwellian equilibrium distribution $g^{eq}$, the Shakhov \cite{shakhov1968generalization} equilibrium distribution $g^{S}$, and the Rykov \cite{rykov1975model} equilibrium distribution $g^{R}$
	\begin{equation}\label{geq}
		{{g}^{eq}}=\frac{\rho }{{{\left( 2\pi RT \right)}^{\left( 3+K \right)/2}}}\exp \left( -\frac{{{c}^{2}}+{{\eta }^{2}}+{{e}^{2}}}{2RT} \right),
	\end{equation}
	\begin{equation}\label{gS}
		{{g}^{S}}={{g}^{eq}}\left[ 1+\left( 1-\Pr  \right)\frac{\bm{c}\cdot \bm{q}}{5pRT}\left( \frac{{{c}^{2}}+{{\eta }^{2}}}{RT}-5 \right) \right],
	\end{equation}
	\begin{equation}\label{gR}
		{{g}^{R}}=\left( 1-\frac{1}{{{Z}_{rot}}} \right){{g}^{tr}}+\frac{1}{{{Z}_{rot}}}{{g}^{rot}},
	\end{equation}
	where $\rho$ is the density, $R$ is the gas constant, $T$ is the temperature, $Pr$ is the Prandtl number, $\bm{c}=\bm{\xi}-\bm{U}$ is the peculiar velocity with $\bm{U}$ being the macroscopic flow velocity, $\bm{q}$ is the heat flux, and ${Z}_{rot}$ is the rotational collision number. The distribution functions $g^{tr}$ and $g^{rot}$ in Eq. (\ref{gR}) are given by 
	\begin{equation}\label{gtr}
		\begin{aligned}
			& {{g}^{tr}}=n{{\left( \frac{1}{2\pi R{{T}_{tr}}} \right)}^{\frac{3}{2}}}{{\exp }^{-\frac{{{c}^{2}}+{{\eta }^{2}}}{2R{{T}_{tr}}}}}\frac{1}{mR{{T}_{rot}}}{{\exp }^{-\frac{e}{mR{{T}_{rot}}}}} \\ 
			& \ \ \ \ \ \times \left[ 1+\frac{\bm{c}\cdot {{\bm{q}}_{tr}}}{15R{{T}_{tr}}{{p}_{tr}}}\left( \frac{{{c}^{2}}+{{\eta }^{2}}}{R{{T}_{tr}}}-5 \right)+\left( 1-\delta  \right)\frac{\bm{c}\cdot {{\bm{q}}_{rot}}}{R{{T}_{tr}}{{p}_{rot}}}\left( \frac{e}{mR{{T}_{rot}}}-1 \right) \right], \\ 
		\end{aligned}
	\end{equation} 
	\begin{equation}\label{grot}
		\begin{aligned}
			& {{g}^{rot}}=n{{\left( \frac{1}{2\pi RT} \right)}^{\frac{3}{2}}}{{\exp }^{-\frac{{{c}^{2}}+{{\eta }^{2}}}{2RT}}}\frac{1}{mRT}{{\exp }^{-\frac{e}{mRT}}} \\ 
			& \ \ \ \ \ \ \times \left[ 1+{{\omega }_{0}}\frac{\bm{c}\cdot {{\bm{q}}_{tr}}}{15RTp}\left( \frac{{{c}^{2}}+{{\eta }^{2}}}{RT}-5 \right)+{{\omega }_{1}}\left( 1-\delta  \right)\frac{\bm{c}\cdot {{\bm{q}}_{rot}}}{RTp}\left( \frac{e}{mRT}-1 \right) \right],\  \\ 
		\end{aligned}	
	\end{equation}
	where $n$ is the molecular number density, $m$ is the molecular mass, $T_{tr}$ and $T_{rot}$ are the translational and rotational temperature, respectively, $p_{tr}$ and $p_{rot}$ are the pressure corresponding to $T_{tr}$ and $T_{rot}$, respectively, $\bm{q}_{tr}$ and $\bm{q}_{rot}$ are the translational and rotational heat flux, respectively. The other coefficients are $\delta =1/1.55$, ${{\omega }_{0}}=0.2354$, and ${{\omega }_{1}}=0.3049$ for nitrogen \cite{zhang2023unified}.
	In this work, the Shakhov equilibrium distribution $g^{S}$ and the Rykov equilibrium distribution $g^{R}$ are adopted for the simulations of monatomic and diatomic gas flows, respectively.	
	\par
	Integrating Eq. (\ref{B_EQ1}) over control volume (cell) $j$ from time $t_{n}$ to $t_{n+1}$, the discrete governing equation can be written as
	\begin{equation}\label{B_EQ2}
		f _{j}^{n+1}-f _{j}^{n}+\frac{\Delta t}{\left| V_{j}^{{}} \right|}F_{j}^{n+1/2}=\frac{\Delta t}{2}\left( \Omega _{j}^{n+1}+\Omega _{j}^{n} \right)
	\end{equation}
	where $\left| V_{j} \right|$ is the volume of cell $j$, ${\Delta t=t_{n+1}-t_{n}}$ is the time step, and $F_{j}^{n+1/2}$ is the microscopic flux
	\begin{equation}\label{micro_flux}
		{F}_{j}^{n+1/2} = \sum\limits_{k}{\bm{\xi }\cdot \bm{A}_{j}^{k} f\left( \bm{x}_{j}^{k}, \bm{\xi }, t_{n+1/2} \right)},
	\end{equation}
	where $\bm{A}_{j}^{k}$ is the outward normal vector of the $k$th face of cell $j$
	with an area of $\left|{A}_{j}^{k}\right|$, and $\bm{x}_{j}^{k}$ is the center of this face.	
	In this study, a simplified multi-scale flux is employed. Integrating Eq. (\ref{B_EQ1}) along the characteristic line (in the direction of particle velocity) from $t_{n}$ to $t_{n+1/2}$, the interface distribution function $f \left( \bm{x}_{j}^{k}, \bm{\xi }, t_{n+1/2} \right)$ can be expressed as
	\begin{equation}\label{half_f}
		f \left( \bm{x}_{j}^{k}, \bm{\xi }, t_{n+1/2} \right)=\frac{{2{\tau }^{n+1/2}}}{{2{\tau }^{n+1/2}}+\Delta t}{f}\left( \bm{x}_{j}^{k}-\bm{\xi }\frac{\Delta t}{2},\bm{\xi },{{t}_{n}} \right)+\frac{\Delta t}{{2{\tau }^{n+1/2}}+\Delta t}{{g}}\left( \bm{x}_{j}^{k},\bm{\xi },{{t}_{n+1/2}} \right),
	\end{equation}		
	where the free transport distribution (representing the microscopic mechanism) and equilibrium distribution (representing the macroscopic mechanism) are coupled rationally, leading to the multi-scale property of the present UGKS.
	\par
	To run the microscopic evolution for getting $f$, the equilibrium state $g$ is needed, which is depended on the macroscopic variables $\bm{W}=\left(\rho, \rho \bm{U}, \rho E, \rho E_{rot} \right)^T$, where $\rho$, $\rho \bm{U}$, $\rho E$ and $\rho E_{rot}$ are the density, momentum, total energy and rotational energy, respectively. 
	Therefore, take the moments of Eq. (\ref{B_EQ2}), the macroscopic evolution equation can be found
	\begin{equation}\label{update_W}
		\bm{W}_{j}^{n+1}=\bm{W}_{j}^{n}-\frac{\Delta t}{\left| {{V}_{j}} \right|}\int{\bm{\psi} F_{j}^{n+1/2}d\bm{\xi }} + \frac{1}{2}\left( \bm{S}_{j}^{n+1}+\bm{S}_{j}^{n} \right),
	\end{equation}
	where $\int{ \bm{\psi} F_{j}^{n+1/2}d\bm{\xi }}$ is the macroscopic flux, ${\bm{\psi }}$ is the collision invariant, and $\bm{S}$ is the source term \cite{zhang2024conservative}.
	Once the macroscopic variables are updated, the implicit equation Eq. (\ref{B_EQ2}) can be explicitly written as
	\begin{equation}\label{update_f}
		f _{j}^{n+1}=\left( 1+\frac{\Delta t}{2\tau _{j}^{n+1}} \right)_{{}}^{-1}\left[ f _{j}^{n}-\frac{\Delta t}{\left| V_{j}^{{}} \right|}F_{j}^{n+1/2}+\frac{\Delta t}{2}\left( \frac{g_{j}^{n+1}}{\tau _{j}^{n+1}}+\frac{g_{j}^{n}-f _{j}^{n}}{\tau _{j}^{n}} \right) \right].
	\end{equation}
	Therefore, in UGKS the macroscopic variables and distribution functions are updated sequentially by Eqs. (\ref{update_W}) and (\ref{update_f}), respectively.
	\par
	
	\subsection{Details of modeling the new boundary}~{}
	In this section, the details of modeling the new CL boundary are discussed.
	The main idea is to first derive the reflected macroscopic flux from the definition of the given ACs ($Q_{AC}$). Next, a specialized Gaussian distribution with unknown parameters is designed to describe the behavior of the reflected gas molecules, and the Gaussian reflected macroscopic flux ($Q_G$), which is expressed in terms of these parameters, can be obtained. The unknown parameters are then determined by equating the two fluxes ($Q_G = Q_{AC}$). Finally, the Gaussian reflected distribution function can be obtained based on these parameters.	
	\par
	
	\subsubsection{The reflected macroscopic flux dependent on AC}
	According to Ref. \cite{shen2013rarefied, liu2021dsmc}, the definition of AC is
	\begin{equation}\label{AC}
		\alpha_{p} =\frac{{{Q}_{I,p}}-{{Q}_{AC,p}}}{{{Q}_{I,p}}-{{Q}_{D,p}}},
	\end{equation}
	where $\alpha_{p}$ is the AC of the flow variables $p$. For example, $\alpha_{mn}$ and $\alpha_{mt}$ stand for the AC of normal momentum and tangential momentum, respectively, which are referred to as NMAC and TMAC.
	${Q}_{I,p}$, ${Q}_{D,p}$, and ${Q}_{AC,p}$ are the incident macroscopic flux, diffuse reflected macroscopic flux, and reflected macroscopic flux dependent on AC of the flow variables $p$, respectively.
	According to Eq. (\ref{AC}), the reflected macroscopic flux dependent on $\alpha_{p}$ can be calculated as
	\begin{equation}\label{QRAC}
		{{Q}_{AC,p}}=\left( 1-\alpha_{p}  \right){{Q}_{I,p}}+\alpha_{p} {{Q}_{D,p}}.
	\end{equation}	
	The quantity ${Q}_{I,p}$ can be obtained by integrating the incident distribution function (Eq. (\ref{half_f})), while ${Q}_{D,p}$ is obtained using the diffuse reflection model \cite{CHEN2024570}.
	Therefore, once $\alpha_{p}$ is determined from either experimental data or MD simulation results, the reflected macroscopic flux can be obtained.	
	Specially, when the full accommodation diffuse reflection model is employed, the reflected macroscopic flux equals the diffuse reflected macroscopic flux (${Q}_{AC,p} = {Q}_{D,p}$). Consequently, all the ACs are equal to 1 for the diffuse reflection model.
	\par
	
	\subsubsection{The Gaussian reflected distribution function}
	It is well known that the Maxwell boundary is a combination of the diffuse and specular reflection models, and the diffuse reflection model follows a Maxwellian distribution, which is an isotropic Gaussian distribution.
	Inspired by this, a simplified anisotropic Gaussian distribution is used to replace the Maxwellian distribution in the diffuse reflection model, allowing for a description of different accommodation of the normal and tangential components of momentum and energy. Therefore, the new boundary can be viewed as a combination of anisotropic Gaussian reflection and specular reflection.	
	Furthermore, to avoid the complex treatment of the specular model in velocity space \cite{CHEN2024570}, a tangential reflected velocity is introduced and combined with the simplified anisotropic Gaussian distribution. Finally, the mathematical expression of the proposed CL boundary can be stated as follows
	\begin{equation}\label{CL_GSI}
		{{f}_{G}}={{\rho }_{w}}\frac{1}{{{\left( 2\pi  \right)}^{3/2}}{{\left( {{\sigma }_{1}}{{\sigma }_{2}}^{2} \right)}^{1/2}}}\exp \left[ -\frac{1}{2}\left( \frac{{{v}_{1}}^{2}}{{{\sigma }_{1}}}+\frac{{{\left( {{v}_{2}}-{{V}_{s,2}} \right)}^{2}}+{{\left( {{v}_{3}}-{{V}_{s,3}} \right)}^{2}}}{{{\sigma }_{2}}} \right) \right],{{v}_{1}}>0,
	\end{equation}
	where $f_{G}$ is the Gaussian reflected distribution function. ${v}_{i}$ represents the component of the reflected molecular velocity in the local surface coordinate system, where ${v}_{1}$ is the normal component pointing toward the flow from the surface, ${v}_{2}$ and ${v}_{3}$ are the tangential components. ${\rho}_{w}$ and ${V}_{s,i}$ are the surface density and the reflected velocity, respectively. ${\sigma }_{i}$ is related to the anisotropic temperature $T_{i}$ with ${\sigma }_{i} = RT_{i}$. $T_{1}$ and $T_{2}$ are the normal temperature and tangential temperature, respectively. $R$ is the gas constant.
	It should be noted that the parameters ${\rho}_{w}$, ${\sigma }_{i}$, and ${V}_{s,i}$ are unknown and need to be determined.
	Specifically, when ${V}_{s,i} = 0$ and $T_{i} = T_{w}$ (wall temperature), the present boundary reverts to the diffuse reflection model
	\begin{equation}\label{Diffuse_GSI}
		{{f}_{D}}={{\rho }_{w}}\frac{1}{{\left( 2\pi {\sigma }_{w} \right)}^{3/2}}\exp \left[ -\frac{1}{2}\left( \frac{{v}_{1} ^{2}+{v}_{2} ^{2}+{v}_{3} ^{2}}{{\sigma }_{w}} \right) \right],{{v}_{1}}>0.
	\end{equation}
	For diatomic gas flows \cite{liu2014unified}, the rotational reflected distribution function can be calculated as
	\begin{equation}\label{CL_rot}	
		{{f}_{R}}=\sigma_{r}{{f}_{G}},{{v}_{1}}>0,
	\end{equation}
	where $\sigma_{r}$ is related to the rotational temperature $T_{r}$ with $\sigma_{r}=RT_{r}$.
	\par
	According to Ref. \cite{shen2013rarefied}, the distribution function of the diffuse reflection model on the discrete velocity can be decomposed into
	\begin{equation}\label{Mf}
		f_{D}={{\rho }_{w}}f_{D}\left( {{v}_{1}} \right)f_{D}\left( {{v}_{2}} \right)f_{D}\left( {{v}_{3}} \right),
	\end{equation}
	where $f_{D}\left( {v}_{i}\right)$ follows a normal distribution
	\begin{equation}\label{Mf123}
		f_{D}\left( {{v}_{i}} \right)=\frac{1}{{{\left( 2\pi {{\sigma }_{w}} \right)}^{1/2}}}\exp \left( -\frac{{{v}_{i}}^{2}}{2{{\sigma }_{w}}} \right),i=1,2,3.
	\end{equation}
	Similarly, the distribution function of proposed CL boundary (Eq. (\ref{CL_GSI})) on the discrete velocity can be decomposed into
	\begin{equation}\label{1f123}
		f_{G}={{\rho }_{w}}f_{G}\left( {{v}_{1}} \right)f_{G}\left( {{v}_{2}} \right)f_{G}\left( {{v}_{3}} \right),
	\end{equation}
	where
	\begin{equation}\label{2f123}
		\begin{aligned}
			& f_{G}\left( {{v}_{1}} \right)=\frac{1}{{{\left( 2\pi {{\sigma }_{1}} \right)}^{1/2}}}\exp \left( -\frac{{{v}_{1}}^{2}}{2{{\sigma }_{1}}} \right), \\ 
			& f_{G}\left( {{v}_{i}} \right)=\frac{1}{{{\left( 2\pi {{\sigma }_{2}} \right)}^{1/2}}}\exp \left( -\frac{{{\left( {{v}_{i}}-{{V}_{s,i}} \right)}^{2}}}{2{{\sigma }_{2}}} \right),i=2,3. \\  
		\end{aligned}
	\end{equation}
	Obviously, the distribution function in Eq. (\ref{2f123}) also follows a normal distribution.
	Furthermore, it is important to emphasize that, unlike the diffuse reflection model where the three sub-distributions are identical, the present boundary features distinct sub-distributions. This distinction allows the present boundary to simultaneously capture different momentum and thermal energy accommodations for reflected gas molecules.
	\par
	According to Eqs. (\ref{CL_GSI}) and (\ref{CL_rot}), the Gaussian reflected macroscopic flux $Q_G$ can be calculated as
	\begin{equation}\label{maroflux}
		\begin{aligned}
			& {{Q}_{G,\rho}}=\left\lfloor f_{G} \right\rfloor ={\rho }_{w}{{\left( \frac{{{\sigma }_{1}}}{2\pi } \right)}^{1/2}},\text{        } \\ 
			& {{Q}_{G,mn}}=\left\lfloor {{v}_{1}}f_{G} \right\rfloor ={\rho }_{w}\frac{{{\sigma }_{1}}}{2}, \\ 
			& {{Q}_{G,mt_2}}=\left\lfloor {{v}_{2}}f_{G} \right\rfloor ={\rho }_{w}{{V}_{s,2}}{{\left( \frac{{{\sigma }_{1}}}{2\pi } \right)}^{1/2}}, \\ 
			& {{Q}_{G,mt_3}}=\left\lfloor {{v}_{3}}f_{G} \right\rfloor ={\rho }_{w}{{V}_{s,3}}{{\left( \frac{{{\sigma }_{1}}}{2\pi } \right)}^{1/2}}, \\ 
			& {{Q}_{G,E}}=\left\lfloor \frac{v_{1}^{2}+v_{2}^{2}+v_{3}^{2}}{2}f_{G}+f_{R} \right\rfloor ={\rho }_{w}\frac{{{\sigma }_{1}}^{1/2}\left( {{V}_{s,2}}^{2}+{{V}_{s,3}}^{2}+2\left( {{\sigma }_{1}}+{{\sigma }_{2}}+{{\sigma }_{r}} \right) \right)}{2{{\left( 2\pi  \right)}^{1/2}}}, \\ 
			& {{Q}_{G,En}}=\left\lfloor \frac{v_{1}^{2}}{2}f_{G} \right\rfloor ={\rho }_{w}\frac{{{\sigma }_{1}}^{3/2}}{{{\left( 2\pi  \right)}^{1/2}}}, \\ 
			& {{Q}_{G,Et}}=\left\lfloor \frac{v_{2}^{2}+v_{3}^{2}}{2}f_{G} \right\rfloor ={\rho }_{w}\frac{{{\sigma }_{1}}^{1/2}\left( {{V}_{s,2}}^{2}+{{V}_{s,3}}^{2}+2{{\sigma }_{2}} \right)}{2{{\left( 2\pi  \right)}^{1/2}}}, \\
			& {{Q}_{G,\epsilon}}=\left\lfloor f_{R} \right\rfloor ={\rho }_{w}{{\sigma }_{r}}{{\left( \frac{{{\sigma }_{1}}}{2\pi } \right)}^{1/2}}, \\ 
		\end{aligned}
	\end{equation}
	where $\left\lfloor \psi  \right\rfloor =\int\limits_{{{v}_{1}}>0}{{v}_{1}}\psi d\mathbf{v}$ represents the reflected macroscopic flux of $\psi$. ${Q}_{G,\rho}$, ${Q}_{G,mn}$, ${Q}_{G,mt_2}$, ${Q}_{G,mt_3}$, ${Q}_{G,E}$, ${Q}_{G,En}$, ${Q}_{G,Et}$ and ${Q}_{G,\epsilon}$ are the reflected mass flux, normal momentum flux, tangential momentum flux related to $v_{2}$, tangential momentum flux related to $v_{3}$, total energy flux, normal energy flux, tangential energy flux, and rotational energy flux, respectively.
	It is evident that the total energy flux is the sum of the normal, tangential, and rotational energy fluxes (${Q}_{G,E} = {Q}_{G,En} + {Q}_{G,Et} + {Q}_{G,\epsilon}$). Therefore, knowing any three of these four fluxes enables determination of the fourth.
	Once the unknown parameters in Eqs. (\ref{CL_GSI}), (\ref{CL_rot}), and (\ref{maroflux}) are determined, the reflected distribution function, the reflected microscopic flux, and the reflected macroscopic flux can be obtained.
	\par
	
	\subsubsection{Solving the unknown parameters}	
	The macroscopic fluxes that need to be calculated in numerical simulations are mass flux, normal momentum flux, tangential momentum flux, total energy flux, and rotational energy flux. The expressions for these macroscopic fluxes form a closed system of equations with six unknown parameters, thus we have
	\begin{equation}
		\left\{ \begin{matrix} {{\rho }_{w}} & {{\sigma }_{1}} & {{\sigma }_{2}} & {{V}_{s,2}} & {{V}_{s,3}} & {{\sigma }_{r}} \end{matrix} \right\}
		=Solve\left[ \begin{matrix} {{Q }_{G,\rho}} & {{Q }_{G,mn}} & {{Q }_{G,mt_2}} & {{Q }_{G,mt_3}} & {{Q }_{G,E}} & {{Q }_{G,\epsilon}} \end{matrix} \right],
	\end{equation} 
	where $Solve[G]$ represents solving the system of equations for the macroscopic flux G. More specifically, the expressions for these unknown parameters can be written as
	\begin{equation}
		\begin{aligned}
			& {{\rho }_{w}}=\frac{\pi {{Q}_{G,\rho }}^{2}}{{{Q}_{G,mn}}}, \\ 
			& {{\sigma }_{1}}=\frac{2{{Q}_{G,mn}}^{2}}{\pi {{Q}_{G,\rho }}}, \\ 
			& {{V}_{s,2}}= \frac{{{Q}_{G,m{{t}_{2}}}}}{{{Q}_{G,\rho }}}, \\ 
			& {{V}_{s,3}}= \frac{{{Q}_{G,m{{t}_{3}}}}}{{{Q}_{G,\rho }}}, \\ 
			& {{\sigma }_{2}}= \frac{\pi \left( 2{{Q}_{G,\rho }}\left( {{Q}_{G,E}}-{{Q}_{G,\varepsilon }} \right)-{{Q}_{G,m{{t}_{2}}}}^{2}-{{Q}_{G,m{{t}_{3}}}}^{2} \right)-4{{Q}_{G,mn}}^{2}}{2\pi {{Q}_{G,\rho }}^{2}}, \\ 
			& {{\sigma }_{r}}= \frac{{{Q}_{G,\varepsilon }}}{{{Q}_{G,\rho }}}. \\ 
		\end{aligned}
	\end{equation}
	\par
	
	\subsubsection{The linear small perturbation boundary}
	Since the present CL boundary is introduced for the first time, it is crucial to validate its performance.
	In this study, the results of the Maxwell boundary and the CLL boundary will serve as benchmarks to evaluate the proposed boundary.
	It is important to note that the ACs for the energy components—namely, the NEAC and TEAC—are also required for both the Maxwell and CLL boundaries. In the Maxwell boundary, all ACs are identical, whereas in the CLL boundary, they may differ.
	In this paper, a linear small perturbation CL boundary is proposed to meet these ACs
	\begin{equation}\label{CL_GSI_a}
		{{\widetilde{f}}_{G}}={{f}_{G}}\left( 1+a\left( {{v}_{1}}+{{\left( \frac{\pi {{\sigma }_{1}}}{2} \right)}^{1/2}} \right) \right),\text{ }{{v}_{1}}>0,
	\end{equation}
	where a is the new unknown parameter. It should be noted that the linear small perturbation CL boundary can only be employed when the following condition is satisfied
	\begin{equation}\label{Delta}
		\Delta =\pi {{Q}_{G,mn}}^{2}+4\left( \pi -4 \right)\left( 3\pi -8 \right){{Q}_{G,En}}{{Q}_{G,\rho}}>0.
	\end{equation}
	In this case, the unknown parameters can be determined by
	\begin{equation}\label{free_parameter}
		\left\{ \begin{matrix} a & {{\rho }_{w}} & {{\sigma }_{1}} & {{\sigma }_{2}} & {{V}_{s,2}} & {{V}_{s,3}} & {{\sigma }_{r}} \end{matrix} \right\}
		=Solve\left[ \begin{matrix} {{Q }_{G,\rho}} & {{Q }_{G,mn}} & {{Q }_{G,mt_2}} & {{Q }_{G,mt_3}} & {{Q }_{G,E}} & {{Q }_{G,En}} & {{Q }_{G,\epsilon}} \end{matrix} \right].
	\end{equation} 
	If the condition (\ref{Delta}) is not satisfied, the parameter a in Eq. (\ref{CL_GSI_a}) is set to 0, indicating that the linear small perturbation Eq. (\ref{CL_GSI_a}) reverts to Eq. (\ref{CL_GSI}). And the remaining unknown parameters for the Maxwell boundary can be calculated as follows
	\begin{equation}\label{M_parameter}
		\begin{aligned}
			& \left\{ \begin{matrix} {{\rho }_{w}} & {{\sigma }_{1}} & {{\sigma }_{2}} & {{V}_{s,2}} & {{V}_{s,3}} & {{\sigma }_{r}} \end{matrix} \right\}
			=\chi Solve\left[ \begin{matrix} {{Q }_{G,\rho}} & {{Q }_{G,mn}} & {{Q }_{G,mt_2}} & {{Q }_{G,mt_3}} & {{Q }_{G,E}} & {{Q }_{G,\epsilon}} \end{matrix} \right] \\
			& \text{                          } + \left( 1-\chi \right)Solve\left[ \begin{matrix} {{Q }_{G,\rho}} & {{Q }_{G,mt_2}} & {{Q }_{G,mt_3}} & {{Q }_{G,E}} & {{Q }_{G,En}} & {{Q }_{G,\epsilon}} \end{matrix} \right],
		\end{aligned}
	\end{equation}	
	here, $\chi$ serves as the weight coefficient and is set as the single AC ($\alpha_{M}$ in Eq. (\ref{MAC})) of the Maxwell boundary in this work.
	For the CLL boundary, the unknown parameters can be calculated as
	\begin{equation}\label{C_parameter}
		\left\{ \begin{matrix} {{\rho }_{w}} & {{\sigma }_{1}} & {{\sigma }_{2}} & {{V}_{s,2}} & {{V}_{s,3}} & {{\sigma }_{r}} \end{matrix} \right\}
		=Solve\left[ \begin{matrix} {{Q }_{G,\rho}} & {{Q }_{G,mt_2}} & {{Q }_{G,mt_3}} & {{Q }_{G,E}} & {{Q }_{G,En}} & {{Q }_{G,\epsilon}} \end{matrix} \right].
	\end{equation}
	The expressions for these parameters in Eqs. (\ref{free_parameter}), (\ref{M_parameter}), and (\ref{C_parameter}) are provided in \ref{sec:appendix_A}
	\par
	
	\subsubsection{Discussion on the NMAC}
\begin{figure}[!htp]
		\centering
		\includegraphics[width=0.6\textwidth]{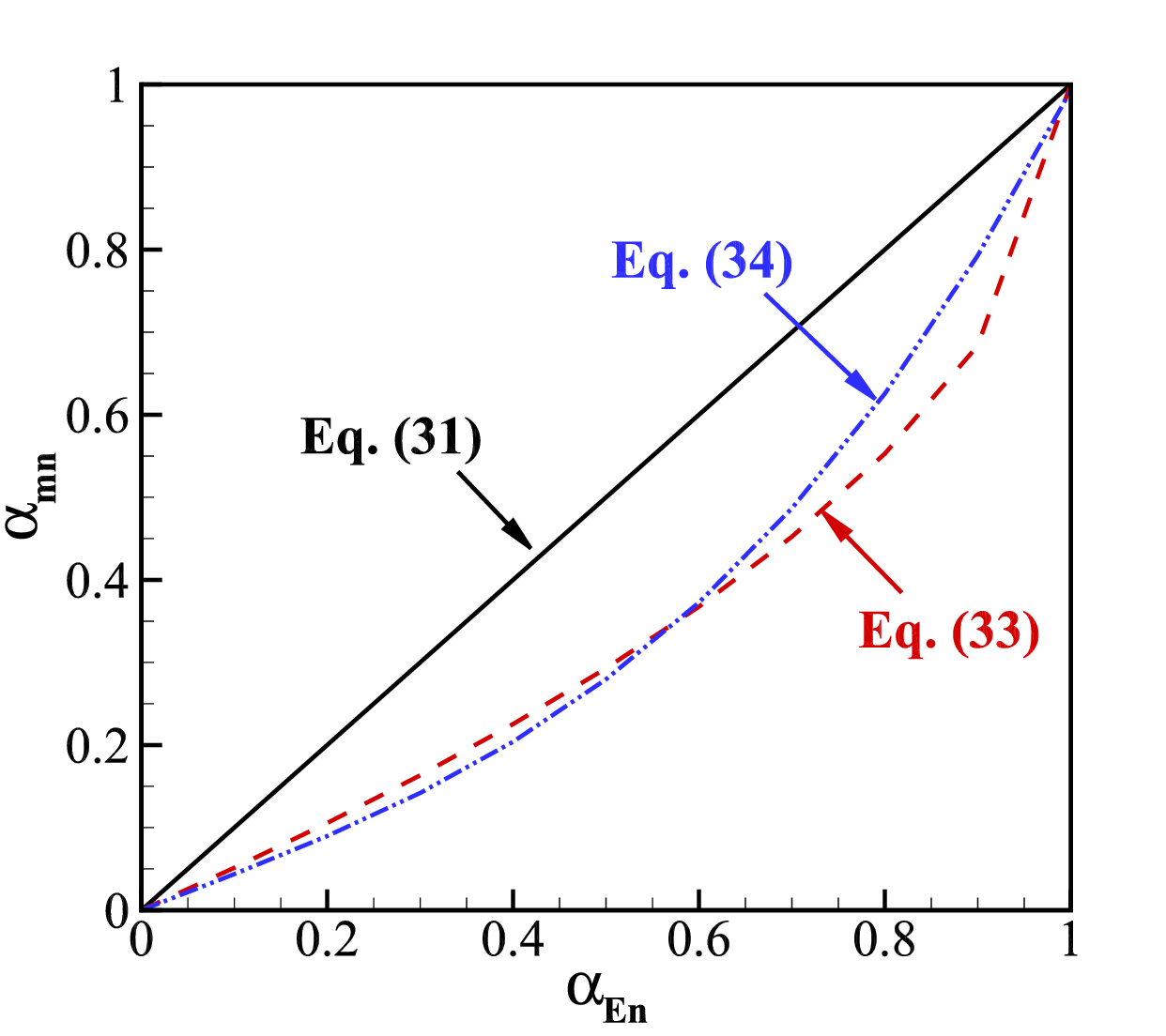}
		\caption{\label{NMAC}{The curves of NMAC and NEAC.}}		
	\end{figure}

	In this paper, the ACs are represented by $\alpha_{mn}$, $\alpha_{mt}$, $\alpha_{E}$, $\alpha_{En}$, $\alpha_{Et}$ and $\alpha_{\epsilon}$, corresponding to NMAC, TMAC, EAC, NEAC, TEAC, and REAC, respectively.		
	For the single-parameter Maxwell boundary, all the ACs are equal to the same value, $\alpha_{M}$
	\begin{equation}\label{MAC}
		\alpha_{mn} = \alpha_{mt}= \alpha_{E} = \alpha_{En} = \alpha_{Et} = \alpha_{\epsilon} = \alpha_{M}.
	\end{equation}
	For the CLL boundary, the accommodation of momentum and thermal energy differ in the tangential and normal components. There are two independent parameters for monatomic gas flows ($\alpha_{mt}$ and $\alpha_{En}$) and three independent parameters for diatomic gas flows ($\alpha_{mt}$, $\alpha_{En}$, and $\alpha_{\epsilon}$). And the TEAC is calculated from the TMAC \cite{shen2013rarefied, liu2021dsmc}
	\begin{equation}\label{TEAC}
		\alpha_{Et} = \alpha_{mt}(2 - \alpha_{mt}).
	\end{equation}
	However, there is no unique method to evaluate the NMAC. In Ref. \cite{yousefi2019determination} and \cite{liu2023modeling}, the NMAC is considered equal to the NEAC
	\begin{equation}\label{NMAC1}
		\alpha_{mn} = \alpha_{En}.
	\end{equation}
	While in Ref. \cite{walker2014drag}, the NMAC and NEAC have a relationship similar to that of the TMAC and TEAC	
	\begin{equation}\label{NMAC2}
		\alpha_{En} = \alpha_{mn}(2 - \alpha_{mn}),
	\end{equation}
	or
	\begin{equation}\label{NMAC3}
		\alpha_{mn} = 1 - \sqrt{1 - \alpha_{En}}.
	\end{equation}
	In this paper, the following empirical formula is used to evaluate the NMAC for the proposed CL boundary when compared to the CLL boundary
	\begin{equation}\label{NMAC4}
		{{\alpha }_{mn}}=0.6233{{\alpha }_{En}}^{3}-0.0682{{\alpha }_{En}}^{2}+0.4375{{\alpha }_{En}}.
	\end{equation}
	The curves of NMAC and NEAC, as given by Eqs. (\ref{NMAC1}), (\ref{NMAC3}), and (\ref{NMAC4}), are plotted in Fig. \ref{NMAC}.	
		
	In conclusion, we summarize the entire computation procedure for the reflected distribution function of the proposed CL boundary as follows
	\par
	\textbf{Step 1.} Calculate the incident macroscopic flux ${Q}_{I}$ and the diffuse reflected macroscopic flux ${Q}_{D}$.
	\par
	\textbf{Step 2.} Calculate the reflected macroscopic flux ${Q}_{AC}$ dependent on AC according to Eq. (\ref{QRAC}).
	\par
	\textbf{Step 3.} Calculate the unknown parameters using Eq. (\ref{free_parameter}), (\ref{M_parameter}) or (\ref{C_parameter}) as needed.
	\par
	\textbf{Step 4.} Calculate the Gaussian reflected distribution function using Eq. (\ref{CL_GSI}).
	\par
	In other words, once the incident distribution function is obtained through the numerical method within the DVM framework (with the UGKS method used in this paper), its macroscopic flux (${Q}_{I}$) can be calculated by integrating it in velocity space. Next, the classical diffuse reflection model is applied to calculate the diffuse reflection distribution function and the corresponding macroscopic flux (${Q}_{D}$). Finally, the reflected distribution function of the proposed CL model is computed by following these four steps.
	\par
	
	\section{Numerical experiment}\label{sec:Numerical experiment}
	\par
	In this section, three test cases are conducted to validate the proposed CL boundary within the DVM framework.
	The simulation results from DS2V, a widely recognized software developed by G. A. Bird \cite{bird1994molecular} using the DSMC method for two-dimensional steady and unsteady flows, along with literature data, are used as reference benchmarks.
	  
	\par
	Generally, four independent characteristic variables are introduced in the non-dimensional reference system, namely, reference length $L_{ref}=L_{c}$, reference temperature $T_{ref}=T_{\infty}$, reference density $\rho_{ref}=\rho_{\infty}$ and reference speed $U_{ref}=\sqrt{2RT_{ref}}$, where $L_{c}$ is the characteristic length scale of the flow, $T_{\infty}$ and $\rho_{\infty}$ are temperature and density of the freestream, respectively.
	Thus, the following basic non-dimensional quantities can be obtained
	\begin{equation}\label{non-dimensional}
		\hat{L}=\frac{L}{L_{ref}},   \hat{T}=\frac{T}{T_{ref}}, \hat{\rho}=\frac{\rho}{\rho_{ref}}, \hat{U}=\frac{U}{U_{ref}}.
	\end{equation}
	One can obtain a complete non-dimensional system by employing these basic quantities. Unless declared otherwise, all variables in the following that lack a ``hat'' are non-dimensional quantities for simplicity's sake.
	\par
	
	\subsection{Supersonic flow over a sharp flat plate} 
	The supersonic flow over a sharp flat plate is simulated to validate the performance of the proposed method in monatomic gas flows.
	The working gas is argon and the variable hard-sphere (VHS) model with a heat index of $\omega=0.81$ is employed. The dimensions of the flat plate are the same as the run34 case in Ref. \cite{tsuboi2005experimental}.
	Figure \ref{sharpplate_pmesh} illustrates the physical space mesh and geometric shape of the sharp flat plate. The height of the first layer (HFL) of the mesh on the surface of the plate is 0.2 $mm$. The flat plate has a thickness of 15 $mm$ and an upper surface length of 100 $mm$, forming a sharp angle of 30 degrees.
	The surface temperature of the flat plate is maintained at 290 K, and the Mach number ($Ma$) and temperature of freestream are 4.89 and 116 K, respectively.
	The Knudsen number ($Kn$) of the freestream, with the flat plate's length as the characteristic length, is 0.0078.
	Figure \ref{sharpplate_dvmesh} illustrates the unstructured velocity space mesh, comprising 896 cells.
	\par
	First, we test the performance of the proposed CL boundary in two extreme cases: full thermal accommodation (where all ACs are equal to 1, corresponding to the diffuse reflection model) and no thermal accommodation (where all ACs are equal to 0, corresponding to the specular reflection model).	
	Figures \ref{sharpplate_contour_10} and \ref{sharpplate_contour_0} show the contours of density, temperature, and horizontal velocity for the flow fields in these two cases. It is evident that the flow field properties in these two cases are quite different.
	Figures \ref{sharpplate_MAX_10} and \ref{sharpplate_MAX_0} present a comparison of the pressure coefficient, skin friction coefficient, and heat transfer coefficient on the surface of the flat plate for these two cases. The present results are in good agreement with those obtained from DS2V.	
	For ease of comparison, the surface coefficients are plotted in the S-coordinate system of the plate surface, progressing counterclockwise from the end of the upper surface to the lower surface.
	To further investigate the performance of the proposed boundary in the intermediate state between the two extreme cases, it is necessary to adjust the ACs for simulations and analyze the results. 	
	Figure \ref{sharpplate_MAX} shows the comparison of the pressure coefficient, skin friction coefficient, and heat transfer coefficient on the surface of a flat plate with the Maxwell boundary as the benchmark. The single AC parameter of the Maxwell boundary, $\alpha_{M}$, varies from 0.8 to 0.2, and the results are close to those of DS2V.
	Figure \ref{sharpplate_CLL} presents the comparison of the pressure coefficient, skin friction coefficient, and heat transfer coefficient on the surface of a flat plate with the CLL boundary as the benchmark. In this case, the two independent ACs are set to the same value, i.e., $\alpha_{mt}= \alpha_{En} = \alpha_{C}$, with $\alpha_{C}$ varying from 0.8 to 0.2. Once again, the present results are in good agreement with those of DS2V.
	In addition to analyzing the distribution of physical quantities on the surface, it is important to consider the influence of the GSI boundary on the flow field.
	Figures \ref{sharpplate_M_U} and \ref{sharpplate_M_T} present the horizontal velocity and temperature profiles over the flat plate at vertical positions X = 5 mm and 20 mm, with the Maxwell boundary as the benchmark.
	Figures \ref{sharpplate_C_U} and \ref{sharpplate_C_T} show the corresponding profiles with the CLL boundary as the benchmark.
	The results from the present simulation align well with those obtained from DS2V. Additionally, the slip velocity and jump temperature on the surface vary differently as the ACs change.
	This indicates that the GSI boundary has a significant influence on the flow field and aerodynamic properties, demonstrating that the proposed CL boundary can be effectively used for flow prediction, functioning as either a Maxwell or CLL boundary.
	\par
	
	\subsection{Supersonic flow in a microchannel}
	The simulation of supersonic microchannel flow is conducted to validate the performance of the present boundary in diatomic gas flows. The freestream flow conditions align with those reported in Ref. \cite{sebastiao2013gas}, and the results from the same source are used as benchmarks.
	The simulation configuration and computational domain are depicted in Fig. \ref{microchannel_geo}.
	The working gas is nitrogen and the VHS model with $\omega=0.74$ is employed. The rotational collision number is 3.5. The Mach number and freestream temperature are 4.15 and 300 K, respectively. The temperatures of the upper and lower surfaces are both 323 K.
	The aspect ratio of the microchannel is set to 5, corresponding to a height $H$ of 1.2 $\mu m$ and a length $L$ of 6.0 $\mu m$. The freestream upstream length $L_{u}$ is 0.6 $\mu m$.
	In this study, the characteristic length is defined as the microchannel height $H$, resulting in a Knudsen number of 0.062.
	Figures \ref{microchannel_pmesh} and \ref{microchannel_dvmesh} show the physical space mesh and velocity space mesh, which consist of 12348 and 1570 cells, respectively. The HFL of the physical space mesh on the surface is 0.005 $\mu m$.
	\par
	To simulate the GSI effects, the diffuse reflection model and the CLL boundary are employed for the lower and upper walls, respectively. In Ref. \cite{sebastiao2013gas}, the DSMC method was used to conduct this simulation.
	Since the flow is parallel to the wall, the NEAC is fixed at 1.0, and the REAC is also fixed at 1.0 in this case, while the TMAC varies from 1.0 to 0.2.
	Figure \ref{microchannel_contour} shows the flow contours of density, pressure, horizontal velocity, and translational temperature for $\alpha_{mt}=1.0$ (diffuse reflection model) and $\alpha_{mt}=0.2$, respectively. It can be seen that the flow structures with $\alpha_{mt}=0.2$ are noticeably different from those with $\alpha_{mt}=1.0$.
	In the case of $\alpha_{mt}=1.0$, two shock waves are observed at the inlet of the microchannel, resulting in strong coupling between the shock waves and boundary layers inside the microchannel.
	Moreover, as the TMAC of the upper surface decreases, the upper shock wave weakens.
	Figure \ref{microchannel_X_U} illustrates the horizontal velocity profiles at the vertical positions X = 2.4 $\mu m$ and X = 3.6 $\mu m$ for various TMAC. As the TMAC decreases, the velocity slip effect on the wall increases, leading to a gradual rise in horizontal velocity near the upper surface.	 
	Figure \ref{microchannel_X_rho} depicts the density profiles at the vertical positions X = 2.4 $\mu m$ and X = 3.6 $\mu m$ with varying TMAC. Notably, the density at the lower surface exhibits more significant variation compared to the upper surface, highlighting the impact of shock wave and boundary layer interactions.	 
	Figure \ref{microchannel_X_Tt} presents the translational temperature profiles at the vertical positions X = 2.4 $\mu m$ and X = 3.6 $\mu m$, showing that as the TMAC decreases, the translational temperature gradually decreases. This suggests that the surface's ability to heat the flow weakens, or, alternatively, the flow becomes less effective at cooling the surface.
	Overall, the present results align well with the references, confirming the effectiveness of the proposed CL boundary in diatomic gas flow simulations.
	\par
	
	\subsection{Hypersonic flow over a cylinder}
	The rarefied hypersonic flow over a cylinder, a classical example of multi-scale flow, is used to evaluate the performance of the proposed CL boundary across a wide range of Mach and Knudsen numbers. 
	Specifically, the study considers three Mach numbers (5.0, 10.0, and 15.0) and three Knudsen numbers (0.1, 1.0, and 10.0).
	The results from DS2V using the Maxwell and CLL boundaries, with ACs ranging from 0.8 to 0.2, are employed as benchmarks.	
	The working gas is argon, and the VHS model with $\omega=0.81$ is used. The temperatures of both the freestream and surface are 273 K, with the cylinder's radius of 1 $mm$ serving as the reference length.
	Figure \ref{cylinder_pmesh_10_10} illustrates the physical space mesh with an HFL of 0.01 $mm$ for 
	$Kn$ = 10.0. For $Kn$ = 1.0 and 0.1, the HFLs are 0.01 $mm$ and 0.005 $mm$, respectively. 
	Figure \ref{cylinder_dvmesh} shows the unstructured velocity space mesh with 2,391 cells for $Ma$ = 5.0, 2,606 cells for $Ma$ = 10.0, and 3,252 cells for $Ma$ = 15.0.
	\par
	The performance of the proposed CL boundary in representing both the Maxwell and CLL boundaries is examined in turn.
	First, the results obtained from DS2V using the Maxwell boundary are employed as reference.
	Figures \ref{cylinder_MAX_5_10}, \ref{cylinder_MAX_5_1}, and \ref{cylinder_MAX_5_01} show the pressure coefficient, friction coefficient, and heat transfer coefficient on the cylindrical surface for $Ma$ = 5.0 and $Kn$ = 10.0, 1.0, and 0.1, respectively.
	The simulation results obtained with the proposed method align closely with those from DS2V under different ACs.
	Comparisons of the pressure coefficient, friction coefficient, and heat transfer coefficient at $Ma$ = 10.0 are presented in Figures \ref{cylinder_MAX_10_10}, \ref{cylinder_MAX_10_1}, and \ref{cylinder_MAX_10_01}, corresponding to $Kn$ = 10.0, 1.0, and 0.1, respectively. 
	Figure \ref{cylinder_MAX_15_10} further illustrates the results of $Ma$ = 15.0.
	Again, the current results agree well with the DS2V results, demonstrating the effectiveness of the proposed CL boundary as a useful Maxwell boundary.
	Furthermore, the results demonstrate that the GSI boundary has a significant impact on both aerodynamic forces and aerodynamic heating.
	For the comparison of the proposed CL boundary with the CLL boundary, the TMAC and NEAC are set to the same value ($\alpha_{C}$) for simplicity, despite their inherent independence and the possibility of different values.
	Figures \ref{cylinder_CLL_5_10}, \ref{cylinder_CLL_5_1}, and \ref{cylinder_CLL_5_01} show the pressure coefficient, friction coefficient, and heat transfer coefficient at $Ma$ = 5, corresponding to $Kn$ = 10.0, 1.0, and 0.1, respectively. Similarly, Figs. \ref{cylinder_CLL_10_10}, \ref{cylinder_CLL_10_1}, and \ref{cylinder_CLL_10_01} present the comparisons at $Ma$ = 10, with $Kn$ = 10.0, 1.0, and 0.1, respectively.
	Figure \ref{cylinder_CLL_15_10} further illustrates the results of $Ma$ = 15.0.
	The results indicate that the proposed CL boundary aligns well with those obtained from DS2V using the CLL boundary, demonstrating that the present boundary can be considered a valid CLL boundary.
	\par
	Next, in order to ensure the completeness of the tests, we investigate the performance of model Eq. (\ref{CL_GSI}), although it has already been tested as part of the linear small perturbation model Eq. (\ref{CL_GSI_a}) in the previous simulations of this paper. Figures \ref{cylinder_MAX_15_10_fix} and \ref{cylinder_CLL_15_10_fix} show the results of model Eq. (\ref{CL_GSI}) for both the Maxwell and CLL models. It can be observed that the results of model Eq. (\ref{CL_GSI}) generally align well with those of the Maxwell model, but there is a noticeable deviation when compared to the pressure coefficient of the CLL model. It implies that model Eq. (\ref{CL_GSI_a}) is better to cover the CLL boundary (Figure \ref{cylinder_CLL_15_10}).
	Finally, we investigate the computational efficiency of the proposed CL boundary. The test is conducted using MPI parallelization with 2 cores on a computer with CPU: Intel Xeon  E5-2678 V3 @2.5 GHz.
	Table \ref{Time} presents the computational time for the cylinder simulation ($Ma = 15.0$, $Kn = 10.0$) with the diffuse reflection model and the proposed model, where all ACs are set to 1.0. The results show that the computational time for the proposed model is close to that of the diffuse reflection model, indicating good computational efficiency.
	This is because, compared to the diffuse reflection model, the additional computational load of the proposed model primarily consists of macroscopic flux integrations at the surface. Moreover, the number of surface cells is significantly smaller than that of the flow field cells; therefore, the additional computational load is negligible within the entire implicit UGKS algorithm.
    
	\par
	In summary, the simulations of hypersonic flow over a cylinder with Mach numbers of 5.0, 10.0, and 15.0, and Knudsen numbers of 10.0, 1.0, and 0.1, demonstrate that the proposed CL boundary can function as either a Maxwell boundary or a CLL boundary, depending on the chosen ACs.
	The proposed CL boundary allows both DVM and unified methods to apply more realistic boundary conditions, accounting for different momentum and energy accommodation in both normal and tangential directions, enhancing the prediction of multi-scale flows.
	\par
	
	\begin{table}
		\centering
		\caption{\label{Time}Computational time of the cylinder simulation ($Ma = 15.0$, $Kn = 10.0$).}
		\begin{threeparttable}
			\begin{tabular}{p{130pt}<{\centering}|p{80pt}<{\centering}|p{80pt}<{\centering}}
				\hline
				\hline
				-               & Steps        & Time (s)        \\
				\hline
				Diffuse reflection model   & 100          & 615.79         \\
				\hline  
				Present CL model   & 100          & 617.33          \\
				\hline
				\hline
			\end{tabular}
		\end{threeparttable}		
	\end{table}
	
	\par
	
	\section{Conclusion}\label{sec:Conclusion}	
	A novel CL boundary for the DVM framework is proposed in this work.
	This new boundary follows an anisotropic Gaussian distribution, enabling it to simultaneously capture different momentum and thermal energy accommodations for reflected gas molecules. Additionally, it can recover the classical full accommodation diffuse reflection model.
	Depending on the given ACs, this boundary can also recover either the Maxwell or CLL boundary in numerical simulations.
	It is compatible with both the conventional Cartesian velocity space and the recently developed efficient unstructured velocity space, and is integrated into both explicit and implicit UGKS, offering better performance for flow prediction.
	The performance of the new boundary has been validated through simulations of supersonic monatomic gas flow over a sharp flat plate and supersonic diatomic gas flow in a microchannel within the slip regime. The results are in good agreement with those of DSMC using Maxwell boundary and CLL boundary.
	Furthermore, the applicability of the CL boundary has been further explored by simulating hypersonic flow over a cylinder across a wide range of Mach and Knudsen numbers.
	This confirms that the proposed CL boundary effectively captures the gas-surface interactions and provides more accurate predictions. Consequently, it enables the DVM and unified methods to utilize more accurate boundary conditions for predicting the behavior of multi-scale flows and other physical phenomena.

	\section*{Acknowledgments}
	The authors thank Prof. Kun Xu in the Hong Kong University of Science and Technology and Prof. Zhaoli Guo in Huazhong University of Science and Technology for discussions of the UGKS, the DUGKS and multi-scale flow simulations.
	This work was supported by the National Natural Science Foundation of China (Grant Nos. 12172301 and 12072283) and the 111 Project of China (Grant No. B17037). This work is supported by the high performance computing power and technical support provided by Xi'an Future Artificial Intelligence Computing Center.

	\clearpage
	\begin{appendix}
		\section{Expressions for unknown parameters of the proposed boundary}\label{sec:appendix_A}
		\renewcommand\theequation{A\arabic{equation}} 
		The solution of Eq. (\ref{free_parameter}) are
		\begin{equation}
			\begin{aligned}
				& a=\frac{tc+ti}{td}, \\ 
				& {{\sigma }_{11}}=\frac{tf-th}{tg}, \\ 
				& {{U}_{2}}=\frac{{{Q}_{G,m{{t}_{2}}}}}{{{Q}_{G,\rho }}}, \\ 
				& {{U}_{3}}=\frac{{{Q}_{G,m{{t}_{3}}}}}{{{Q}_{G,\rho }}}, \\ 
				& {{\sigma }_{r}}=\frac{{{Q}_{G,\varepsilon }}}{{{Q}_{G,\rho }}}, \\ 
				& \rho ={{Q}_{G,\rho }}^{3/2}{{\left( \frac{2\pi }{{{Q}_{G,En}}} \right)}^{1/2}}, \\ 
				& {{\sigma }_{22}}={{\left( \frac{2\pi }{{{\sigma }_{11}}} \right)}^{1/2}}\frac{{{Q}_{G,E}}}{\rho }-\frac{1}{2}\left( {{U}_{2}}^{2}+{{U}_{3}}^{2} \right)-{{\sigma }_{11}}-{{\sigma }_{r}}+\frac{1}{4}a{{\left( 2\pi  \right)}^{1/2}}{{\sigma }_{11}}^{3/2}. \\				 
			\end{aligned}			
		\end{equation}
		where
		\begin{equation}
			\begin{aligned}
				& ta=\pi {{Q}_{G,mn}}^{2}+4\left( \pi -4 \right)\left( 3\pi -8 \right){{Q}_{G,En}}{{Q}_{G,\rho }}, \\ 
				& tb={{Q}_{G,mn}}^{2}+\left( \pi -4 \right){{Q}_{G,En}}{{Q}_{G,\rho }}, \\ 
				& tc=\pi {{Q}_{G,mn}}^{3}+\left( \pi -4 \right)\left( 7\pi -16 \right){{Q}_{G,En}}{{Q}_{G,mn}}{{Q}_{G,\rho }}, \\ 
				& td=2{{\left( \pi -4 \right)}^{3}}{{Q}_{G,En}}^{2}Qrr, \\ 
				& tf=\pi {{Q}_{G,mn}}^{2}+2\left( \pi -4 \right)\left( 3\pi -8 \right){{Q}_{G,En}}{{Q}_{G,\rho }}, \\ 
				& tg={{\left( 3\pi -8 \right)}^{2}}{{Q}_{G,\rho }}^{2}, \\ 
				& th=sgn \left( tb \right){{\left( \pi ta \right)}^{1/2}}{{Q}_{G,mn}}, \\ 
				& ti=\frac{tc+{{\left( \pi ta \right)}^{1/2}}\left| tb \right|}{td}. \\ 				
			\end{aligned}						
		\end{equation}
		
		The solution of Eq. (\ref{M_parameter}) are
		\begin{equation}
			\begin{aligned}
				& \rho =\chi \frac{\pi {{Q}_{G,\rho }}^{2}}{{{Q}_{G,mn}}}+\left( 1-\chi  \right){{Q}_{G,\rho }}^{3/2}{{\left( \frac{2\pi }{{{Q}_{G,En}}} \right)}^{1/2}}, \\ 
				& {{\sigma }_{11}}=\chi \frac{2{{Q}_{G,mn}}^{2}}{\pi {{Q}_{G,\rho }}^{2}}+\left( 1-\chi  \right)\frac{{{Q}_{G,En}}}{{{Q}_{G,\rho }}}, \\ 
				& {{U}_{2}}=\frac{{{Q}_{G,m{{t}_{2}}}}}{{{Q}_{G,\rho }}}, \\ 
				& {{U}_{3}}=\frac{{{Q}_{G,m{{t}_{3}}}}}{{{Q}_{G,\rho }}}, \\ 
				& {{\sigma }_{r}}=\frac{{{Q}_{G,\varepsilon }}}{{{Q}_{G,\rho }}}, \\ 
				& {{\sigma }_{22}}={{\left( \frac{2\pi }{{{\sigma }_{11}}} \right)}^{1/2}}\frac{{{Q}_{G,E}}}{\rho }-\frac{1}{2}\left( {{U}_{2}}^{2}+{{U}_{3}}^{2} \right)-{{\sigma }_{11}}-{{\sigma }_{r}}. \\ 
			\end{aligned}	    	
		\end{equation}
		
		The solution of Eq. (\ref{C_parameter}) are
		\begin{equation}
			\begin{aligned}
				& \rho ={{Q}_{G,\rho }}^{3/2}{{\left( \frac{2\pi }{{{Q}_{G,En}}} \right)}^{1/2}}, \\ 
				& {{\sigma }_{11}}=\frac{{{Q}_{G,En}}}{{{Q}_{G,\rho }}}, \\ 
				& {{U}_{2}}=\frac{{{Q}_{G,m{{t}_{2}}}}}{{{Q}_{G,\rho }}}, \\ 
				& {{U}_{3}}=\frac{{{Q}_{G,m{{t}_{3}}}}}{{{Q}_{G,\rho }}}, \\ 
				& {{\sigma }_{r}}=\frac{{{Q}_{G,\varepsilon }}}{{{Q}_{G,\rho }}}, \\ 
				& {{\sigma }_{22}}={{\left( \frac{2\pi }{{{\sigma }_{11}}} \right)}^{1/2}}\frac{{{Q}_{G,E}}}{\rho }-\frac{1}{2}\left( {{U}_{2}}^{2}+{{U}_{3}}^{2} \right)-{{\sigma }_{11}}-{{\sigma }_{r}}. \\ 
			\end{aligned}		
		\end{equation}
	\end{appendix}

	\clearpage
	\bibliography{CL_BC}

	\newpage
	\setcounter{table}{0}
	\renewcommand\thetable{\arabic{table}}
	\setcounter{figure}{0}
	\renewcommand\thefigure{\arabic{figure}}

	\clearpage	
	\begin{figure}[!htp]
		\centering
		\subfigure[]{\label{sharpplate_pmesh}\includegraphics[width=0.45\textwidth]{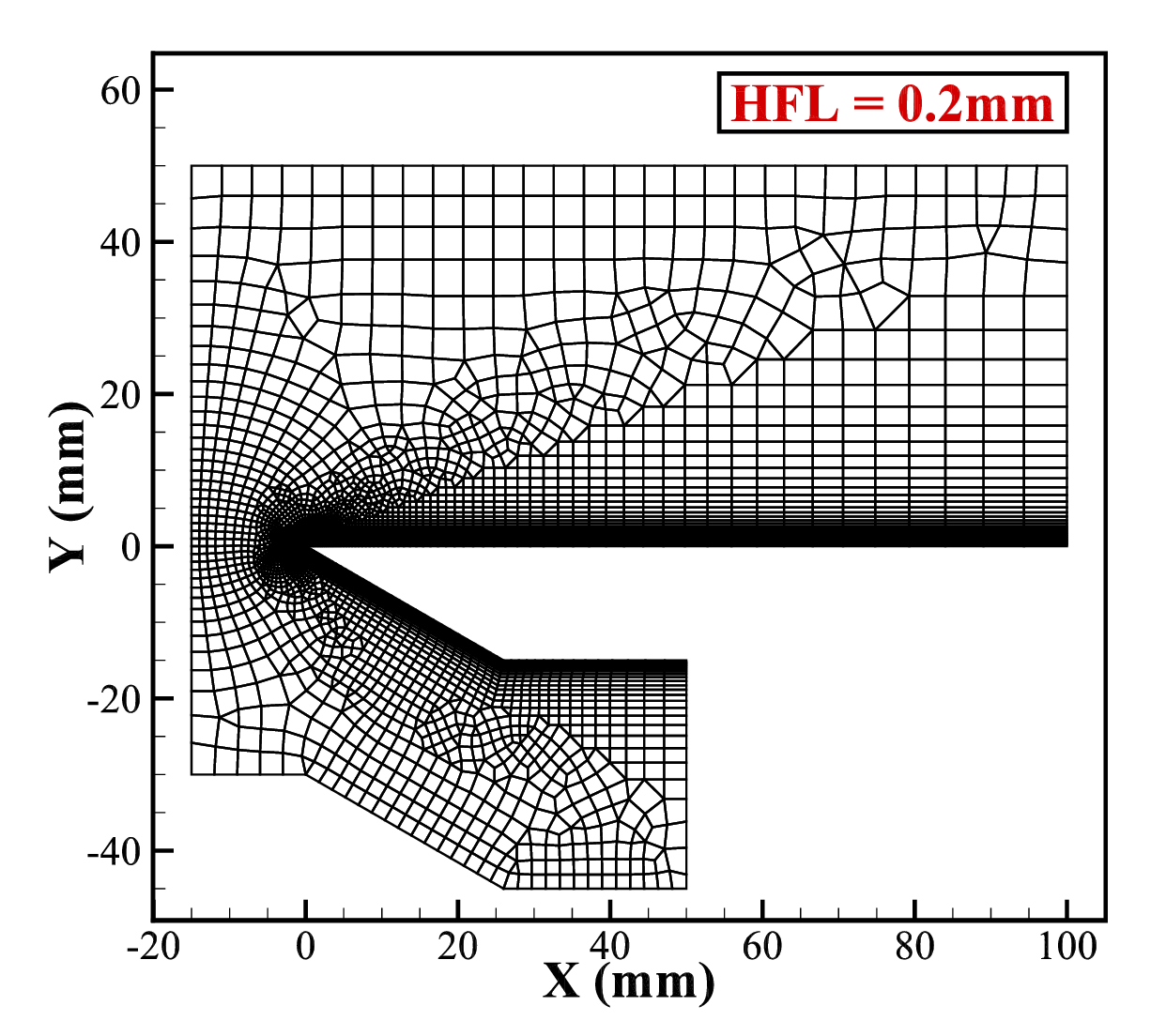}}
		\subfigure[]{\label{sharpplate_dvmesh}\includegraphics[width=0.45\textwidth]{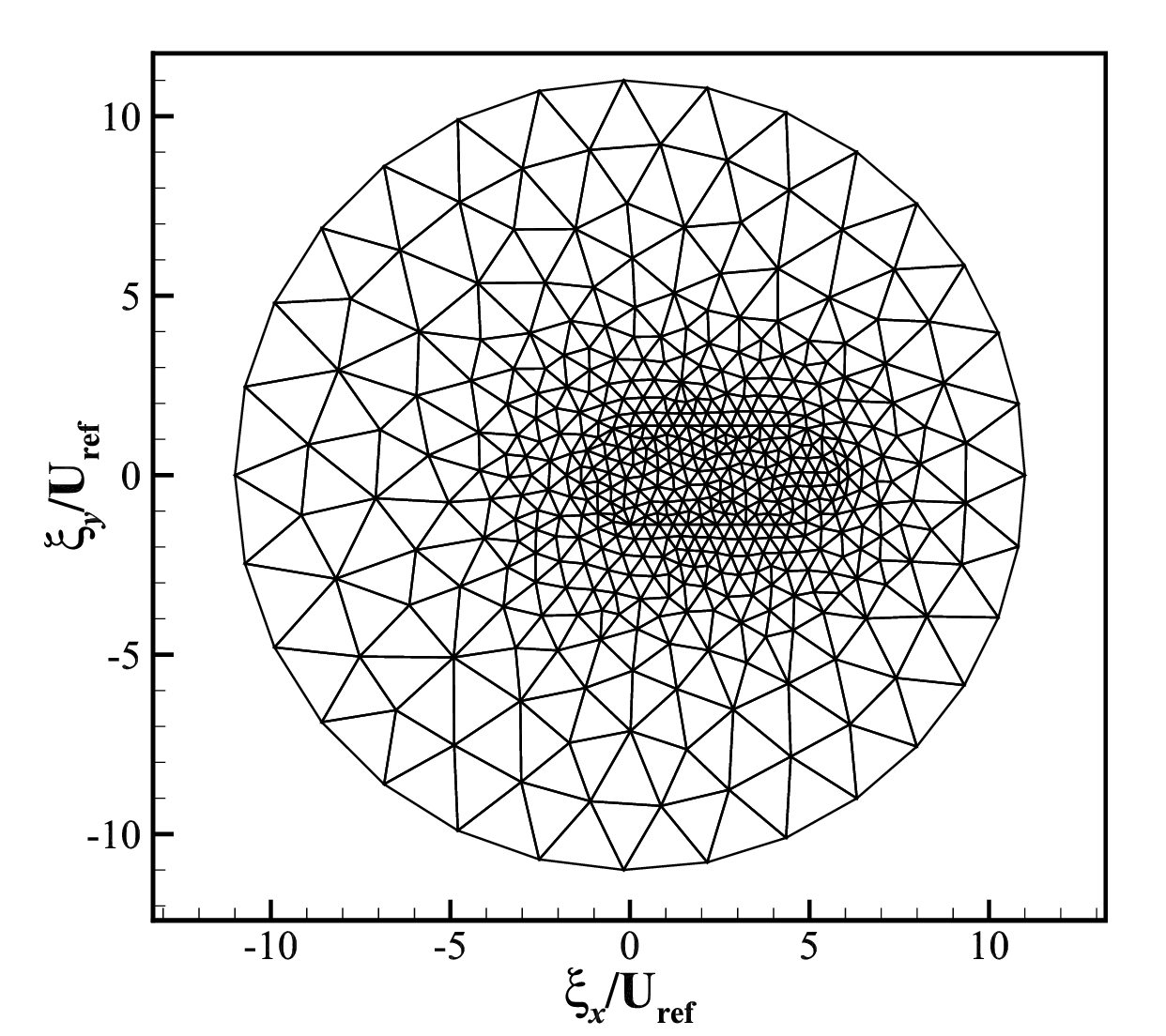}}
		\caption{\label{sharpplate_mesh}{The (a) unstructured physical mesh and (b) unstructured velocity mesh for the supersonic flow over a flat plate	($Ma = 4.89$, $Kn = 0.0078$, $T_{\infty} = 116 K$, $T_{w} = 290 K$).}}		
	\end{figure}
	
	\begin{figure}[!htp]
		\centering
		\subfigure[]{\label{sharpplate_contour_D_10}\includegraphics[width=0.32\textwidth]{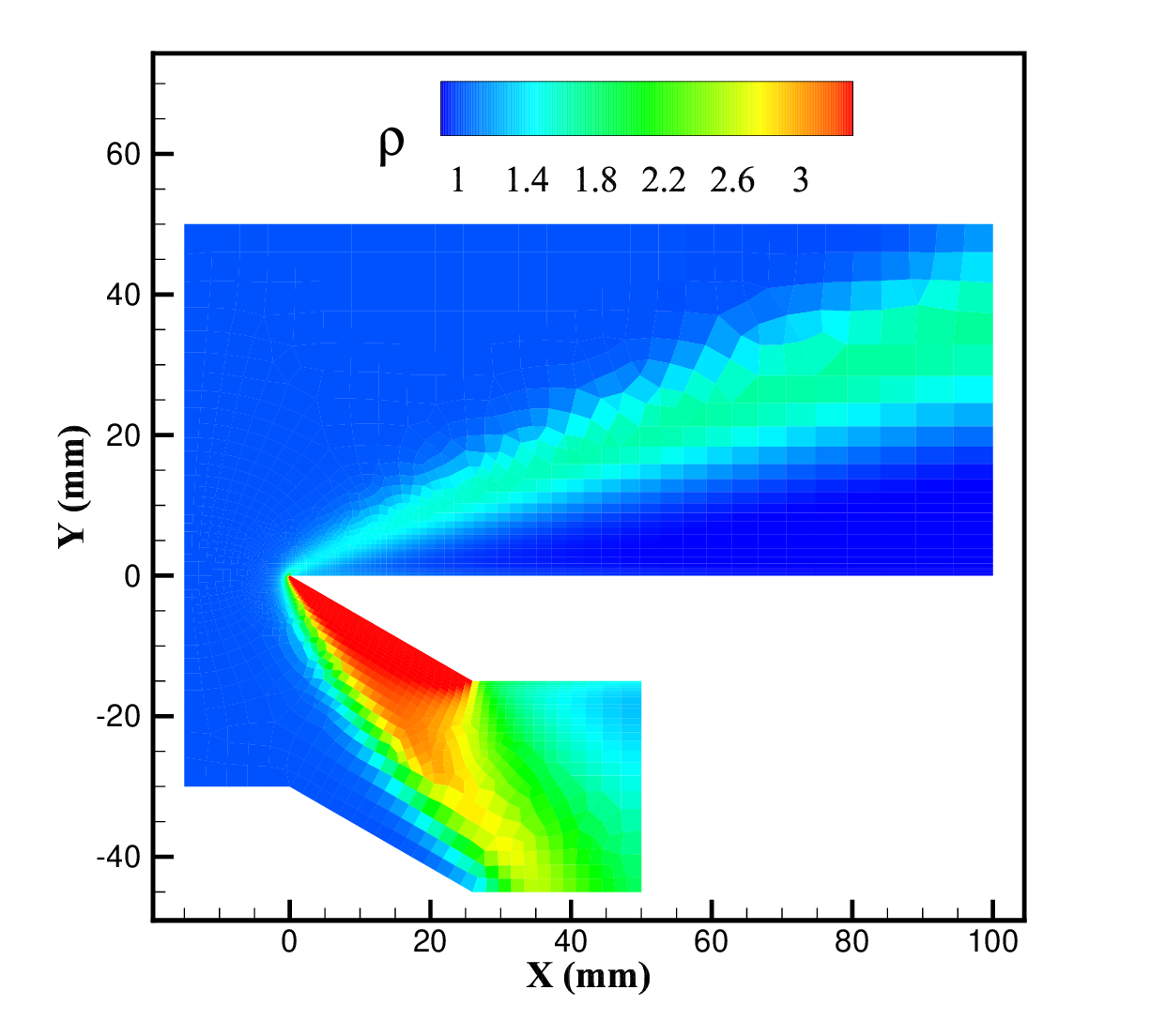}}
		\subfigure[]{\label{sharpplate_contour_U_10}\includegraphics[width=0.32\textwidth]{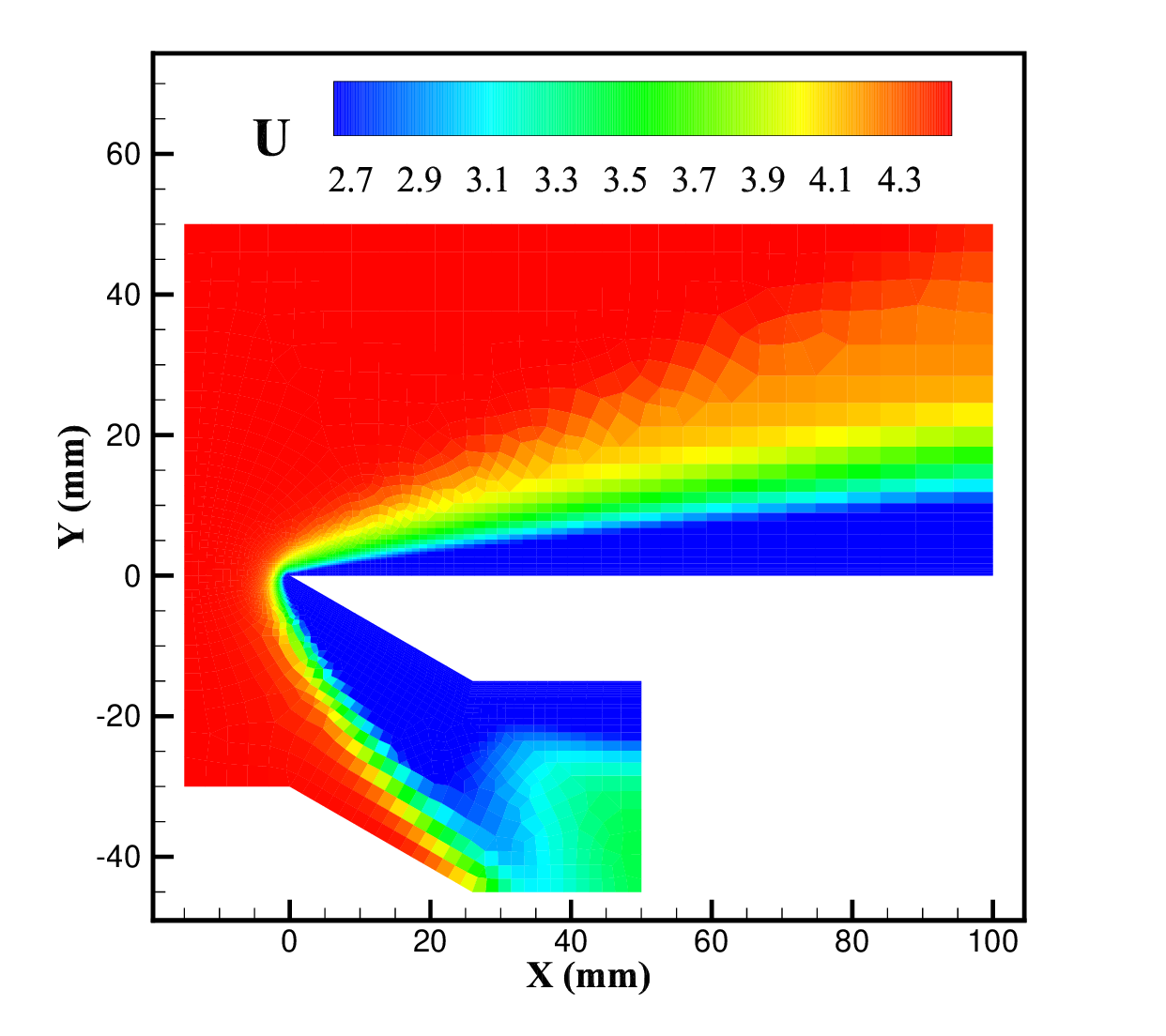}}
		\subfigure[]{\label{sharpplate_contour_T_10}\includegraphics[width=0.32\textwidth]{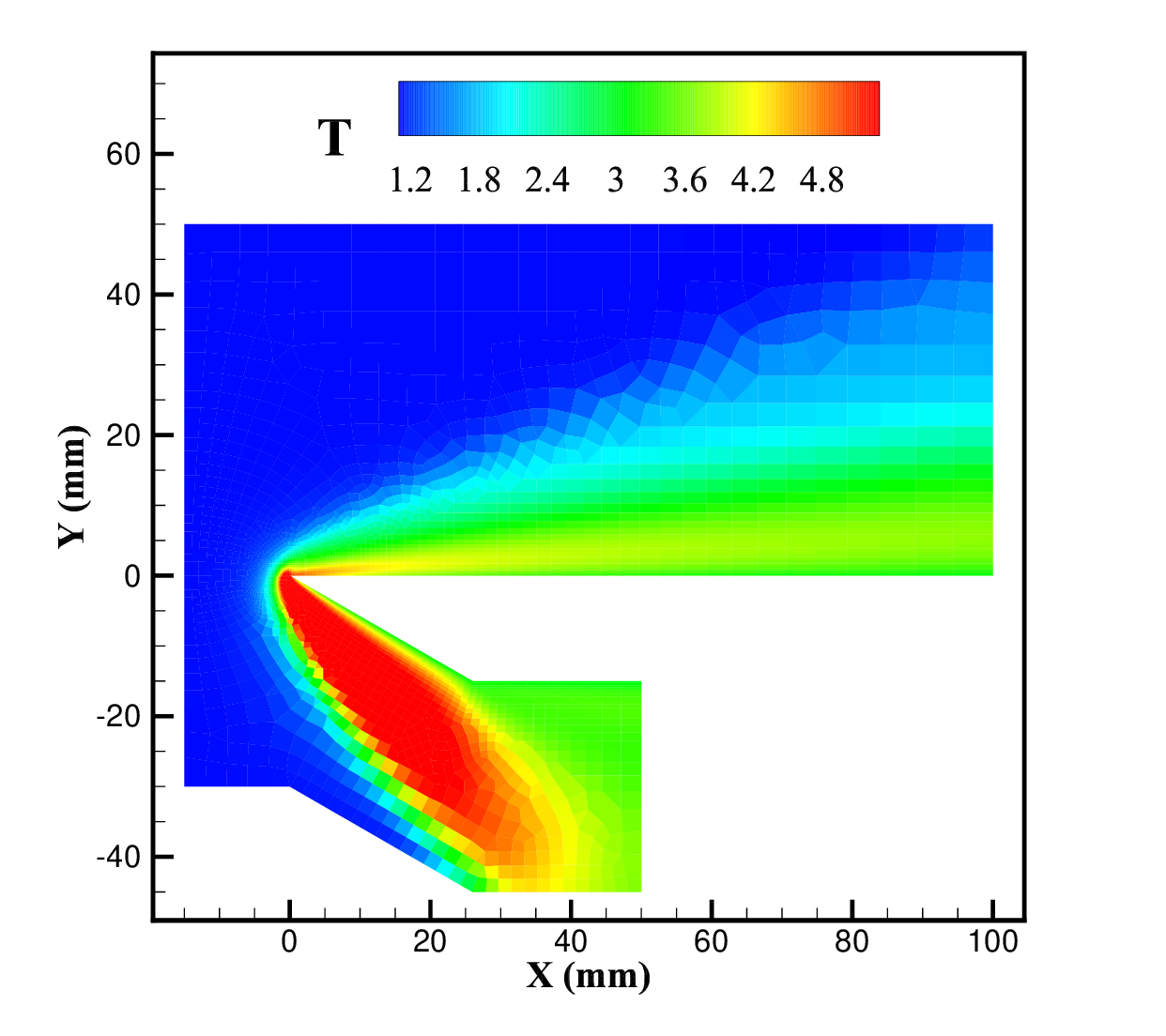}}
		\caption{\label{sharpplate_contour_10}{The contours of (a) density, (b) temperature, and (c) horizontal velocity of the supersonic flow over a flat plate, with all ACs set to 1 ($Ma = 4.89$, $Kn = 0.0078$, $T_{\infty} = 116 K$, $T_{w} = 290 K$).}}
	\end{figure}
	
	\begin{figure}[!htp]
		\centering
		\subfigure[]{\label{sharpplate_contour_D_0}\includegraphics[width=0.32\textwidth]{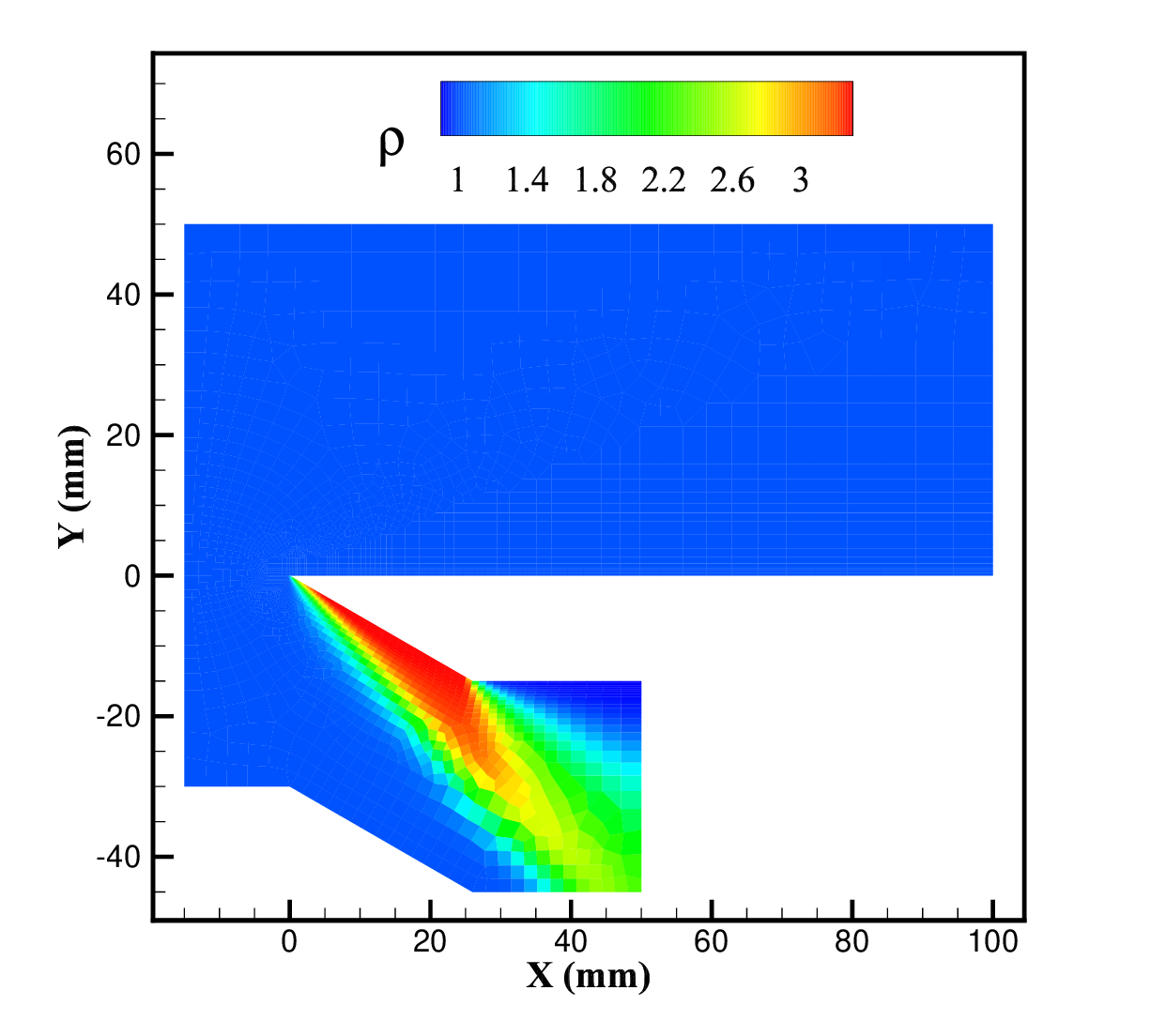}}
		\subfigure[]{\label{sharpplate_contour_U_0}\includegraphics[width=0.32\textwidth]{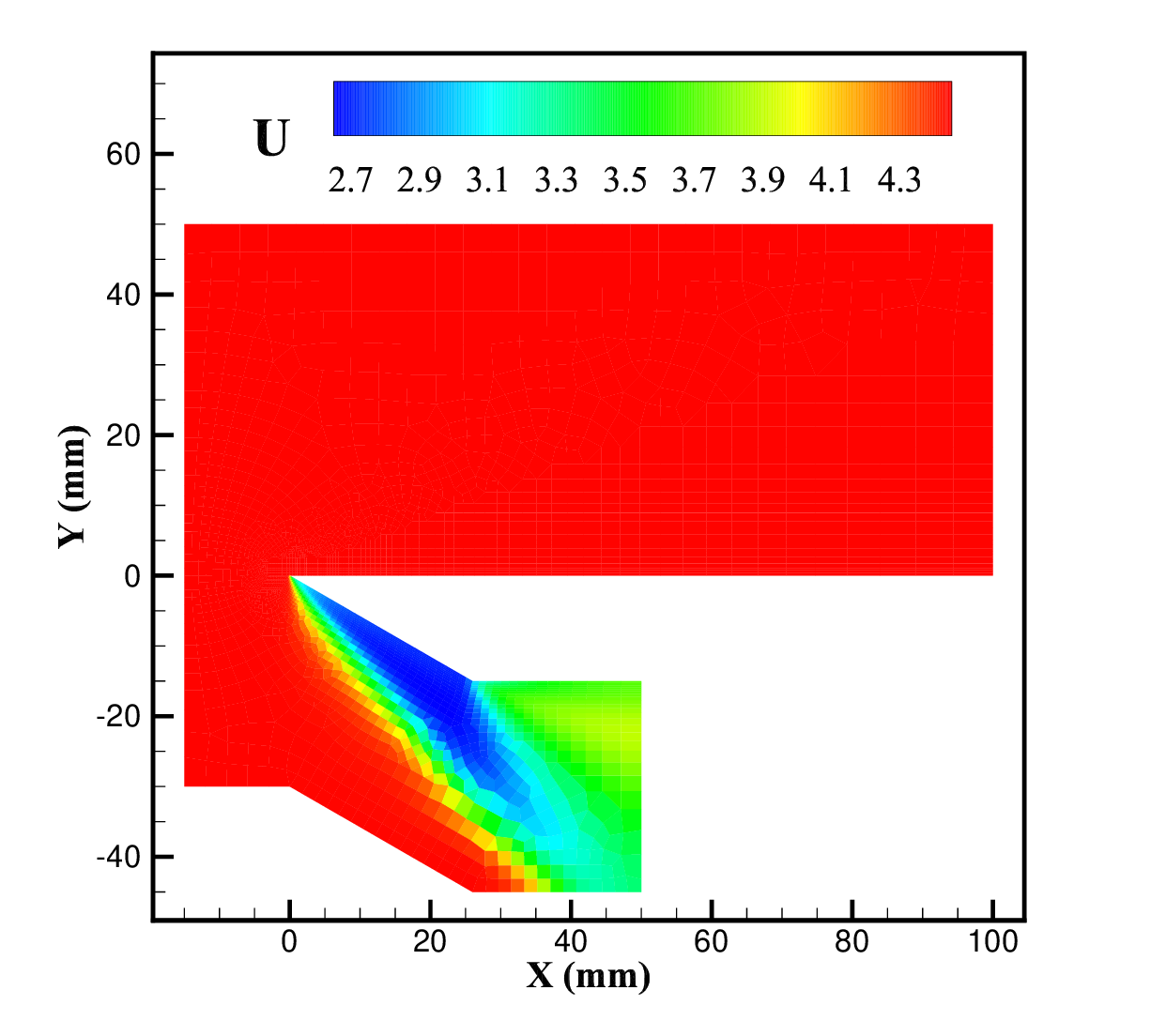}}
		\subfigure[]{\label{sharpplate_contour_T_0}\includegraphics[width=0.32\textwidth]{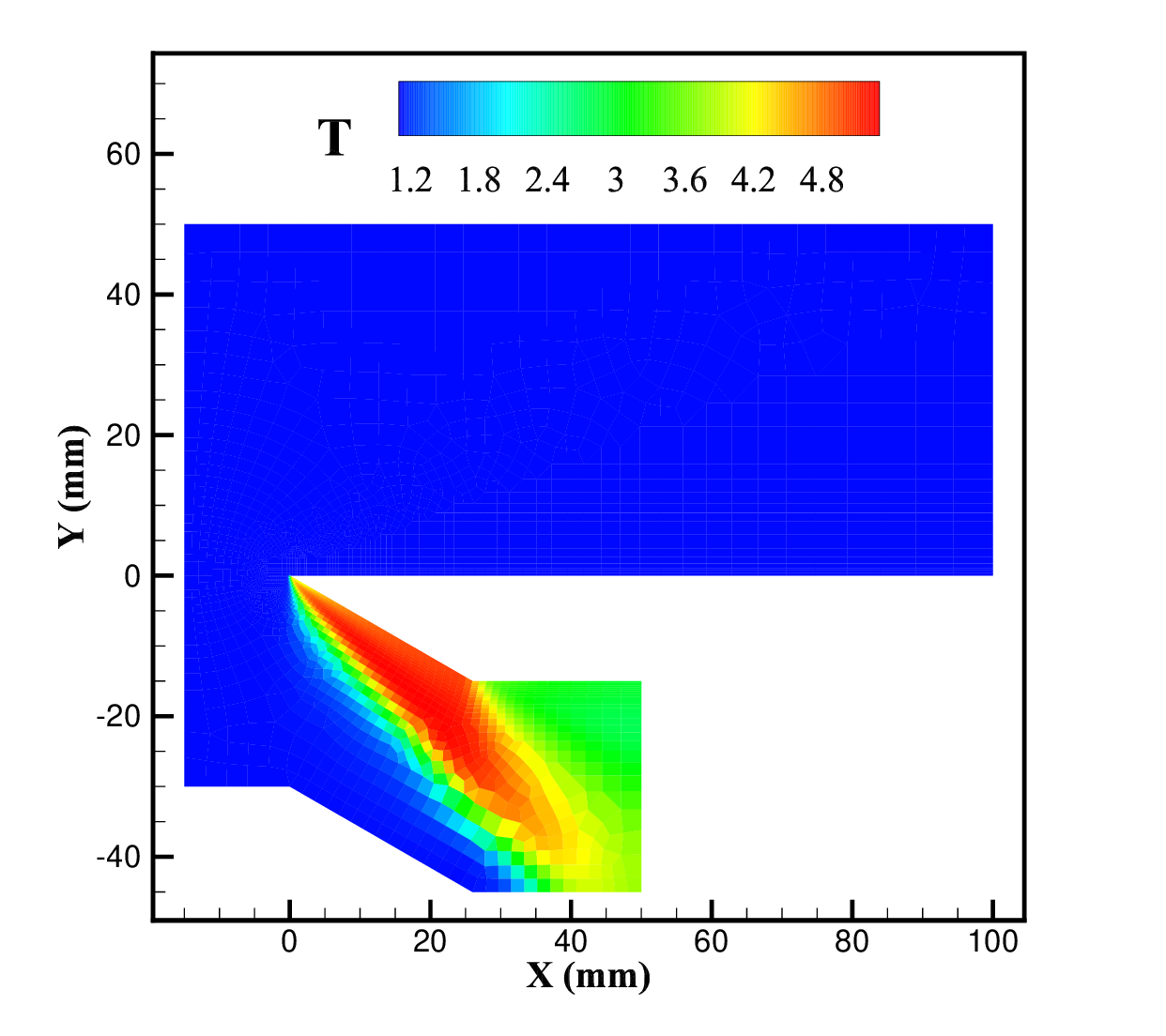}}
		\caption{\label{sharpplate_contour_0}{The contours of (a) density, (b) temperature, and (c) horizontal velocity of the supersonic flow over a flat plate, with all ACs set to 0 ($Ma = 4.89$, $Kn = 0.0078$, $T_{\infty} = 116 K$, $T_{w} = 290 K$).}}
	\end{figure}
	
	\begin{figure}[!htp]
		\centering
		\subfigure[]{\label{sharpplate_M_P_10}\includegraphics[width=0.32\textwidth]{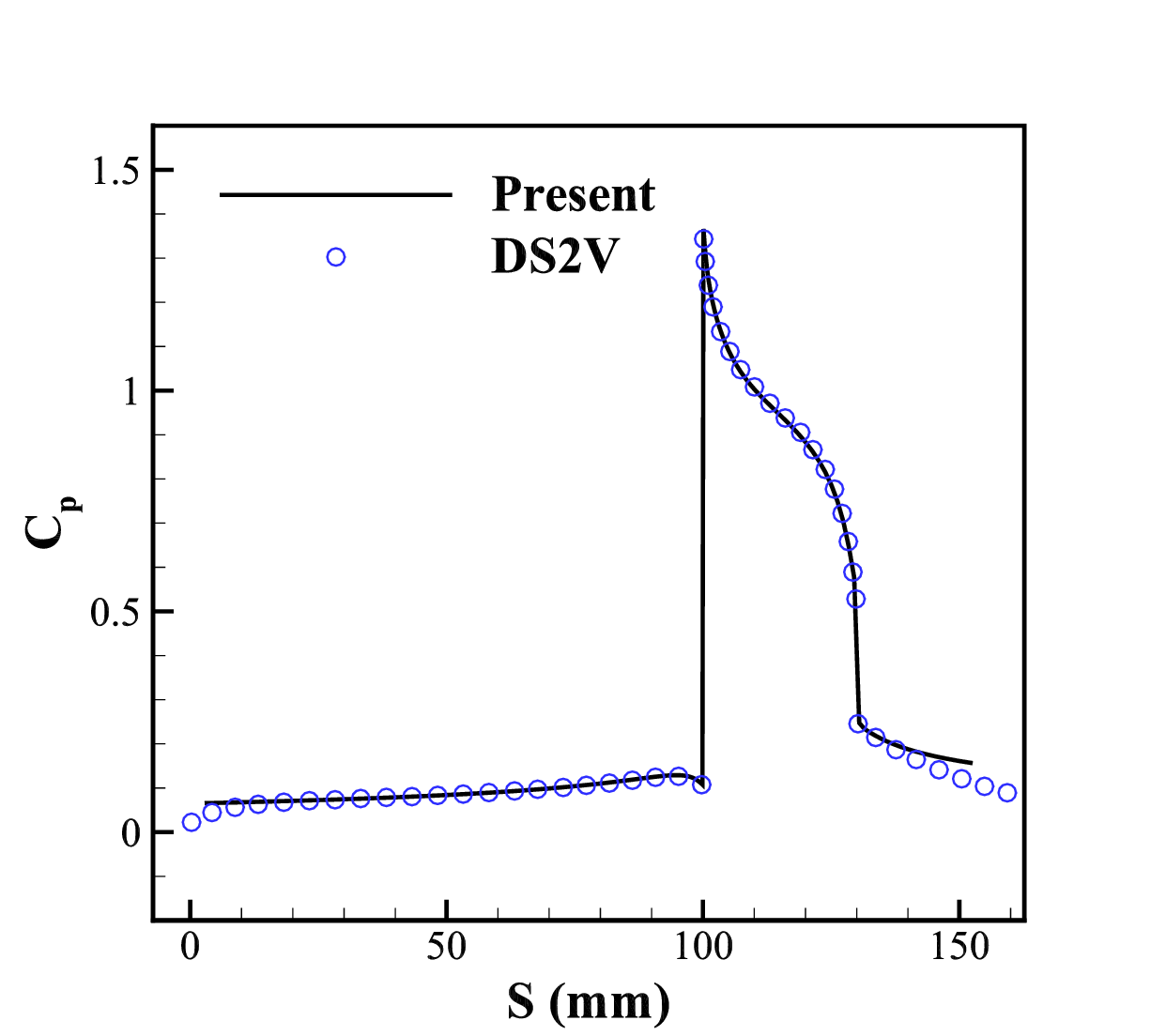}}
		\subfigure[]{\label{sharpplate_M_S_10}\includegraphics[width=0.32\textwidth]{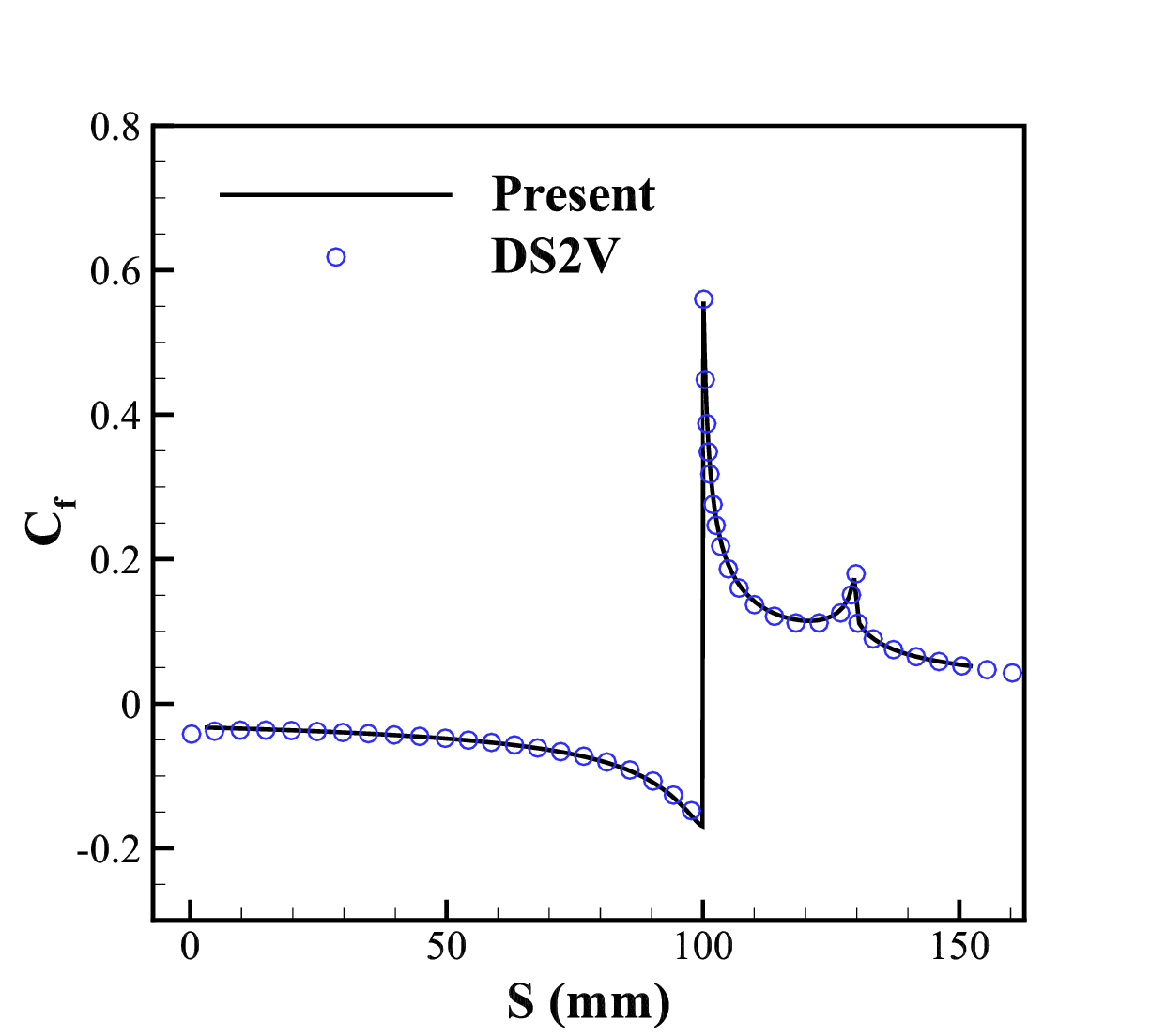}}
		\subfigure[]{\label{sharpplate_M_H_10}\includegraphics[width=0.32\textwidth]{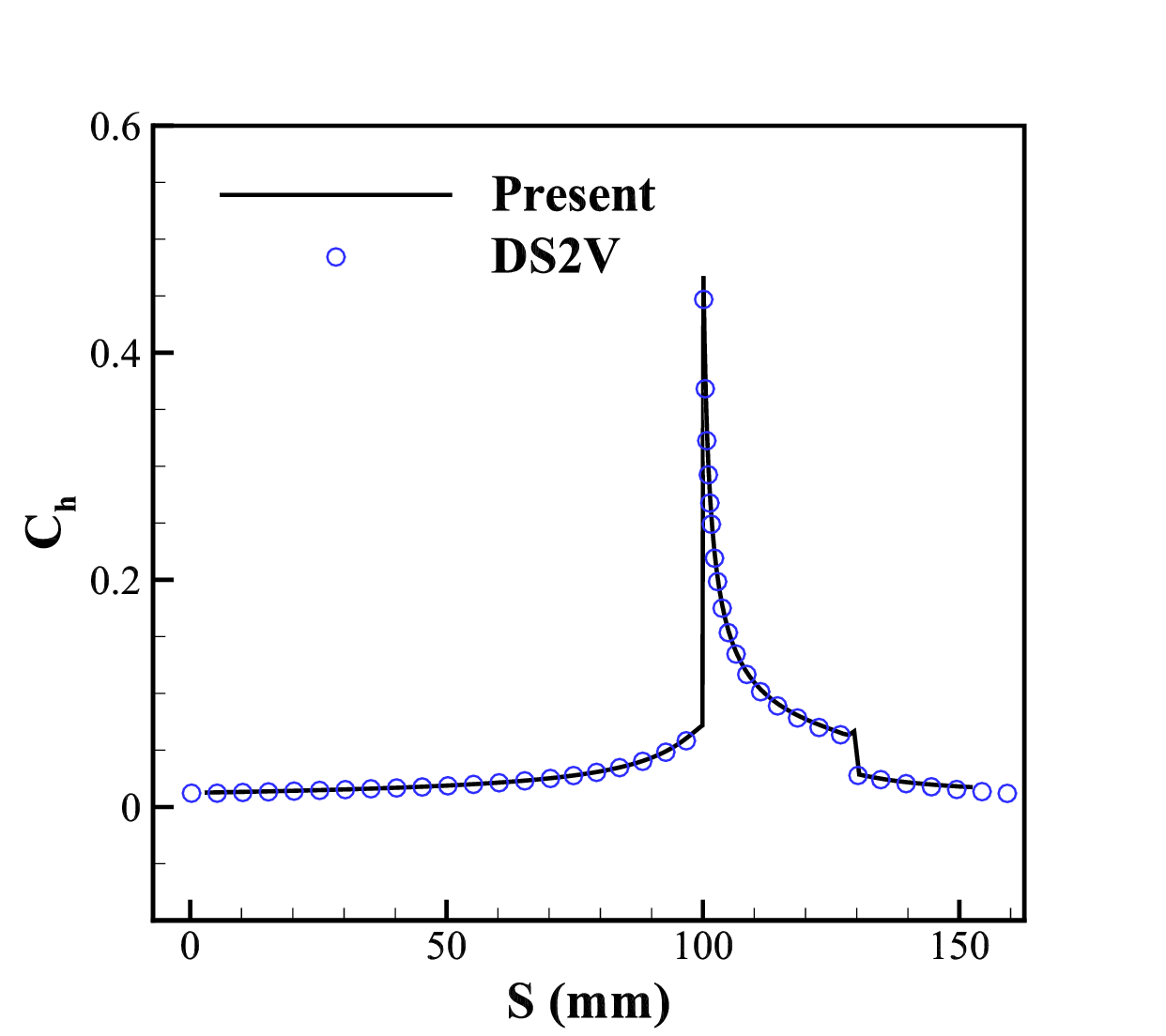}}
		\caption{\label{sharpplate_MAX_10}{Comparison of the (a) pressure coefficient, (b) skin friction coefficient, and (c) heat transfer coefficient on the surface of flat plate, with all ACs set to 1 ($Ma = 4.89$, $Kn = 0.0078$, $T_{\infty} = 116 K$, $T_{w} = 290 K$).}}
	\end{figure}
	
	\begin{figure}[!htp]
		\centering
		\subfigure[]{\label{sharpplate_M_P_0}\includegraphics[width=0.32\textwidth]{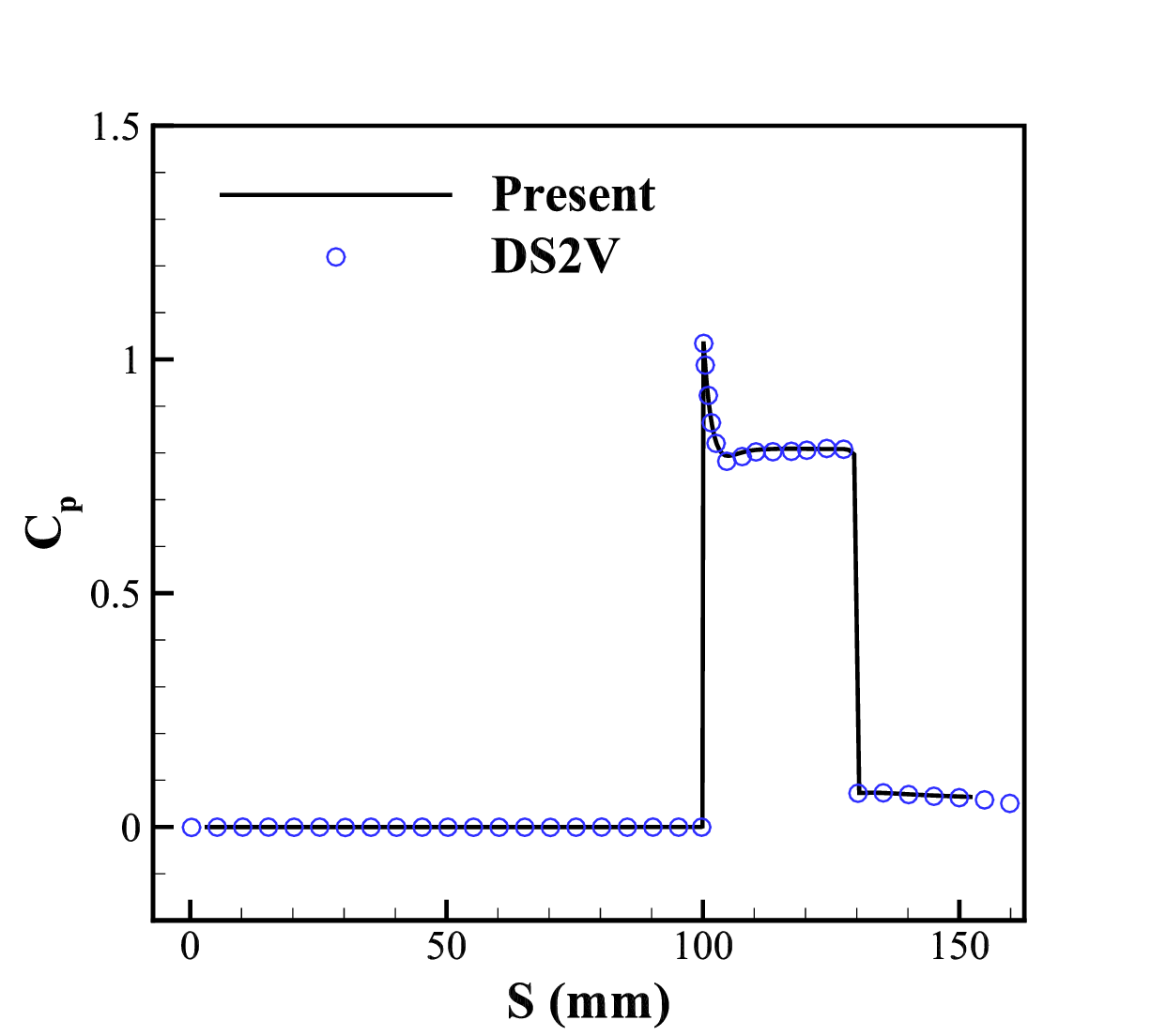}}
		\subfigure[]{\label{sharpplate_M_S_0}\includegraphics[width=0.32\textwidth]{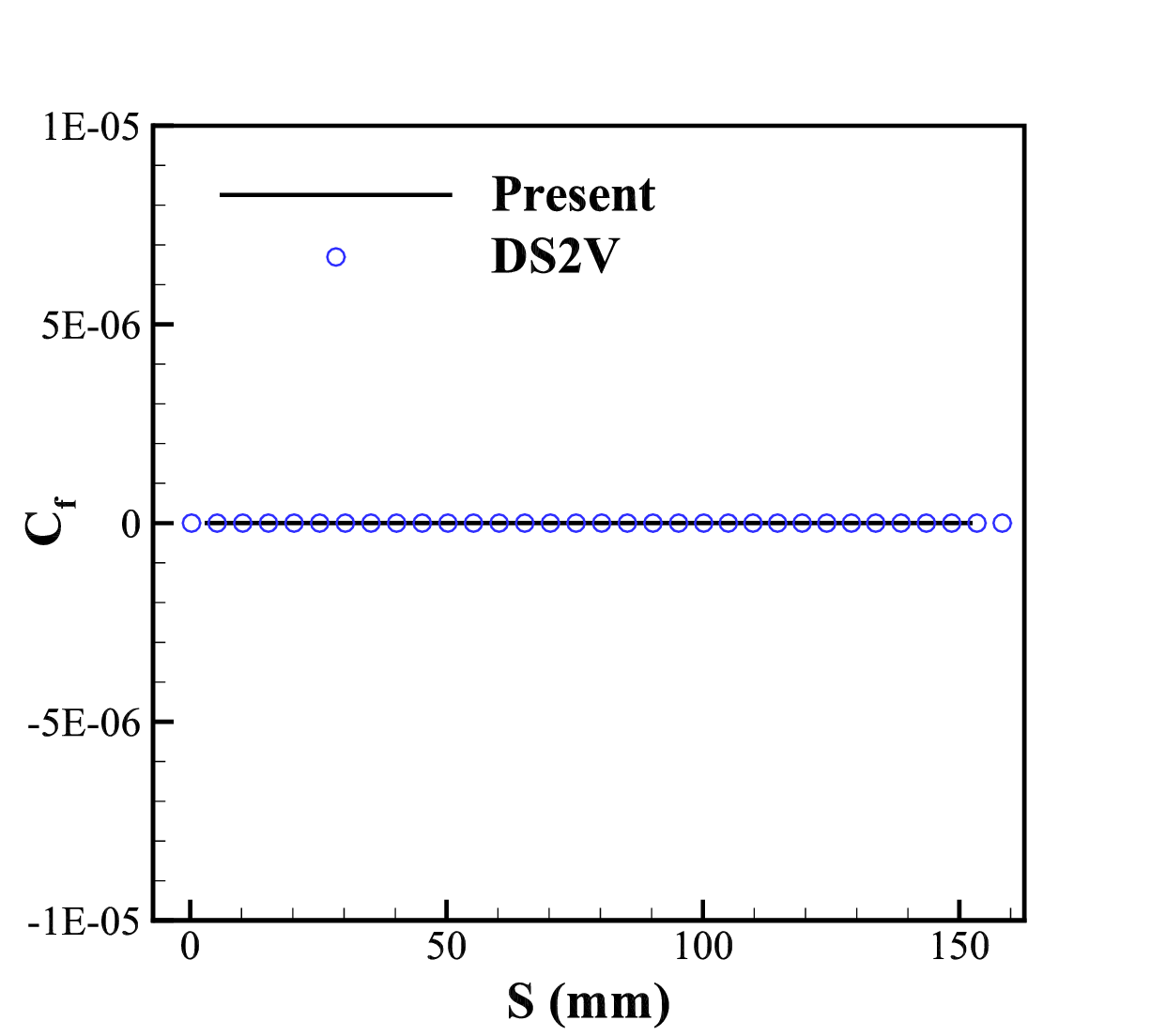}}
		\subfigure[]{\label{sharpplate_M_H_0}\includegraphics[width=0.32\textwidth]{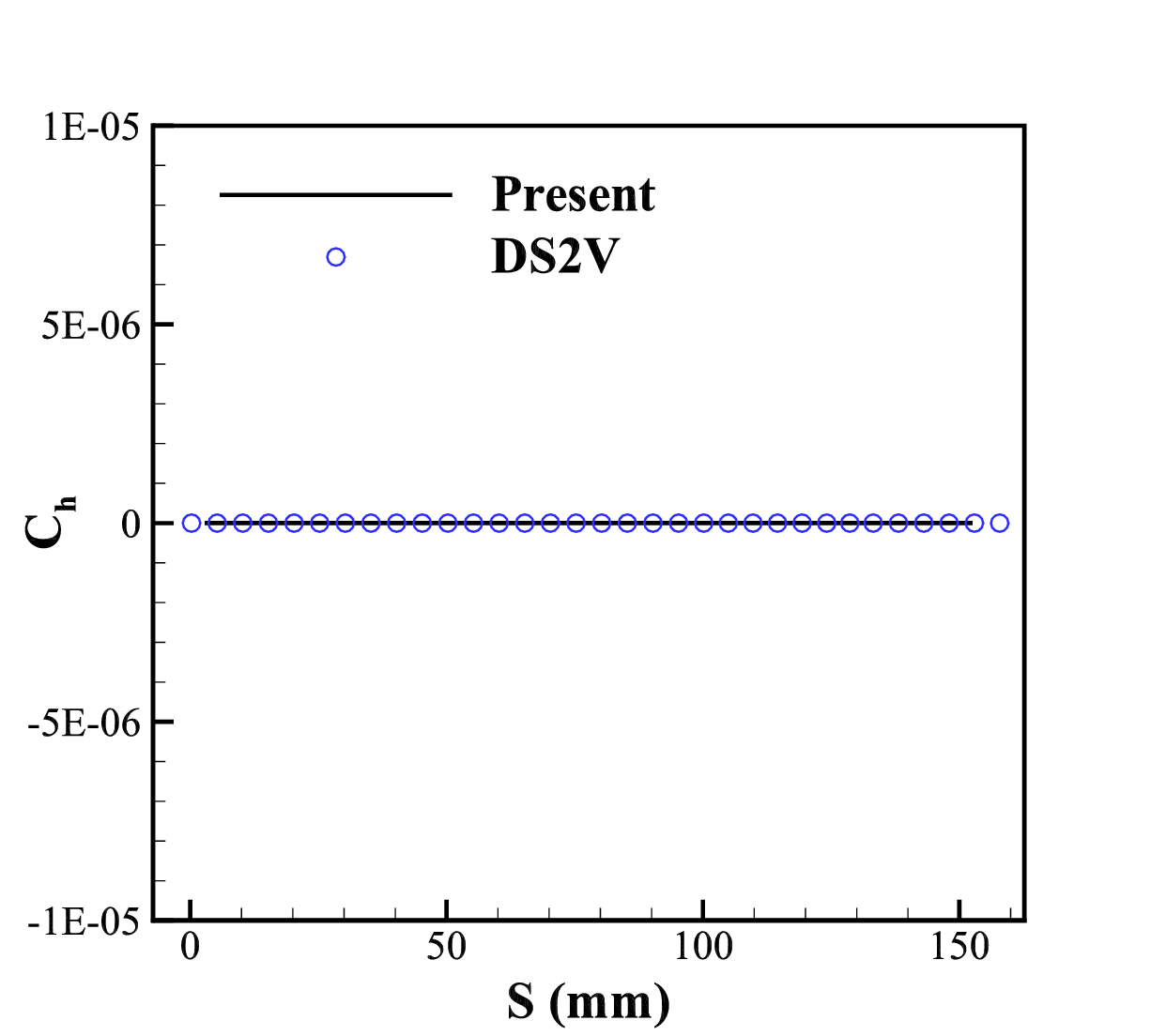}}
		\caption{\label{sharpplate_MAX_0}{Comparison of the (a) pressure coefficient, (b) skin friction coefficient, and (c) heat transfer coefficient on the surface of flat plate, with all ACs set to 0 ($Ma = 4.89$, $Kn = 0.0078$, $T_{\infty} = 116 K$, $T_{w} = 290 K$).}}
	\end{figure}
	
	\begin{figure}[!htp]
		\centering
		\subfigure[]{\label{sharpplate_M_P_a}\includegraphics[width=0.32\textwidth]{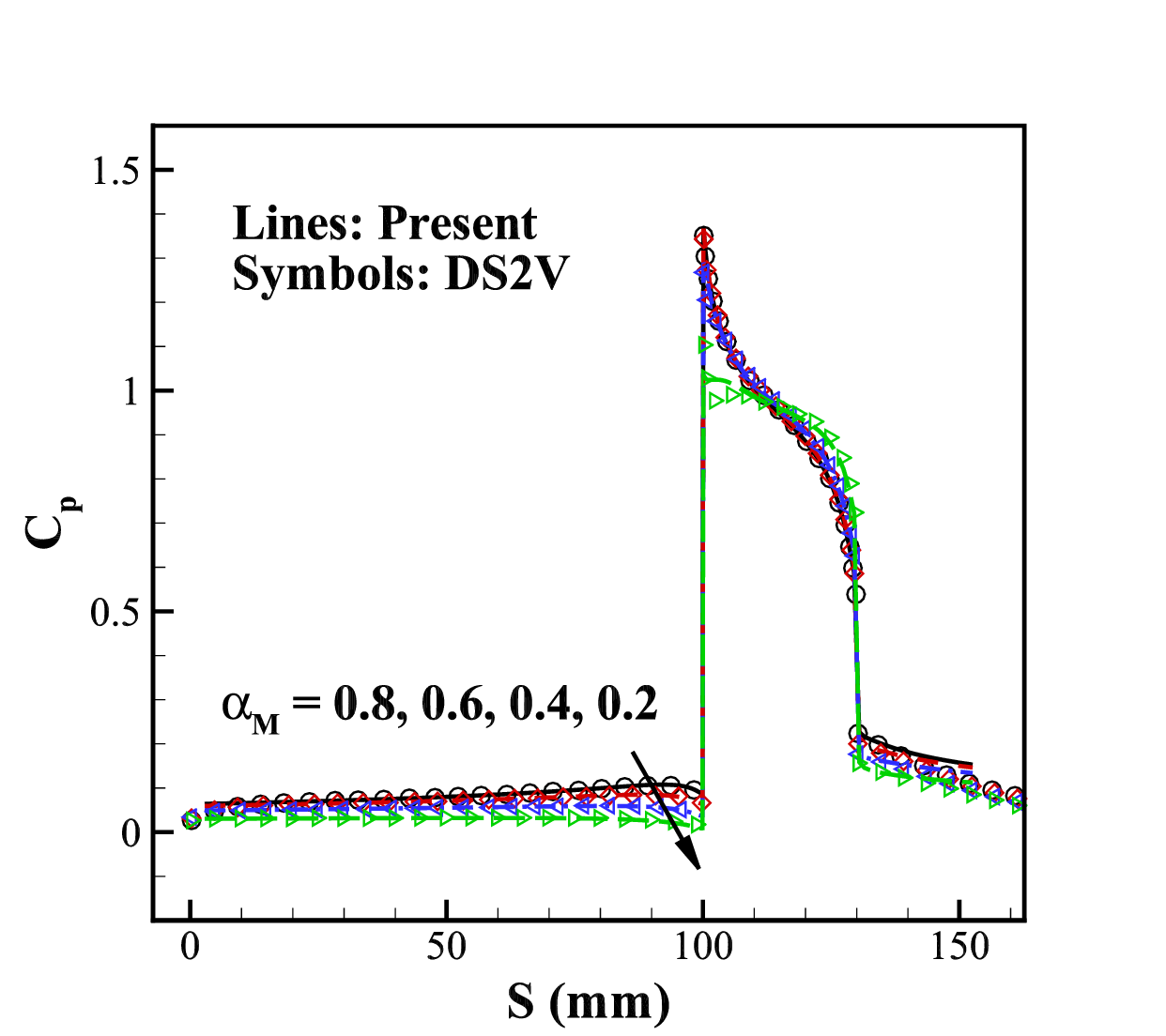}}
		\subfigure[]{\label{sharpplate_M_S_a}\includegraphics[width=0.32\textwidth]{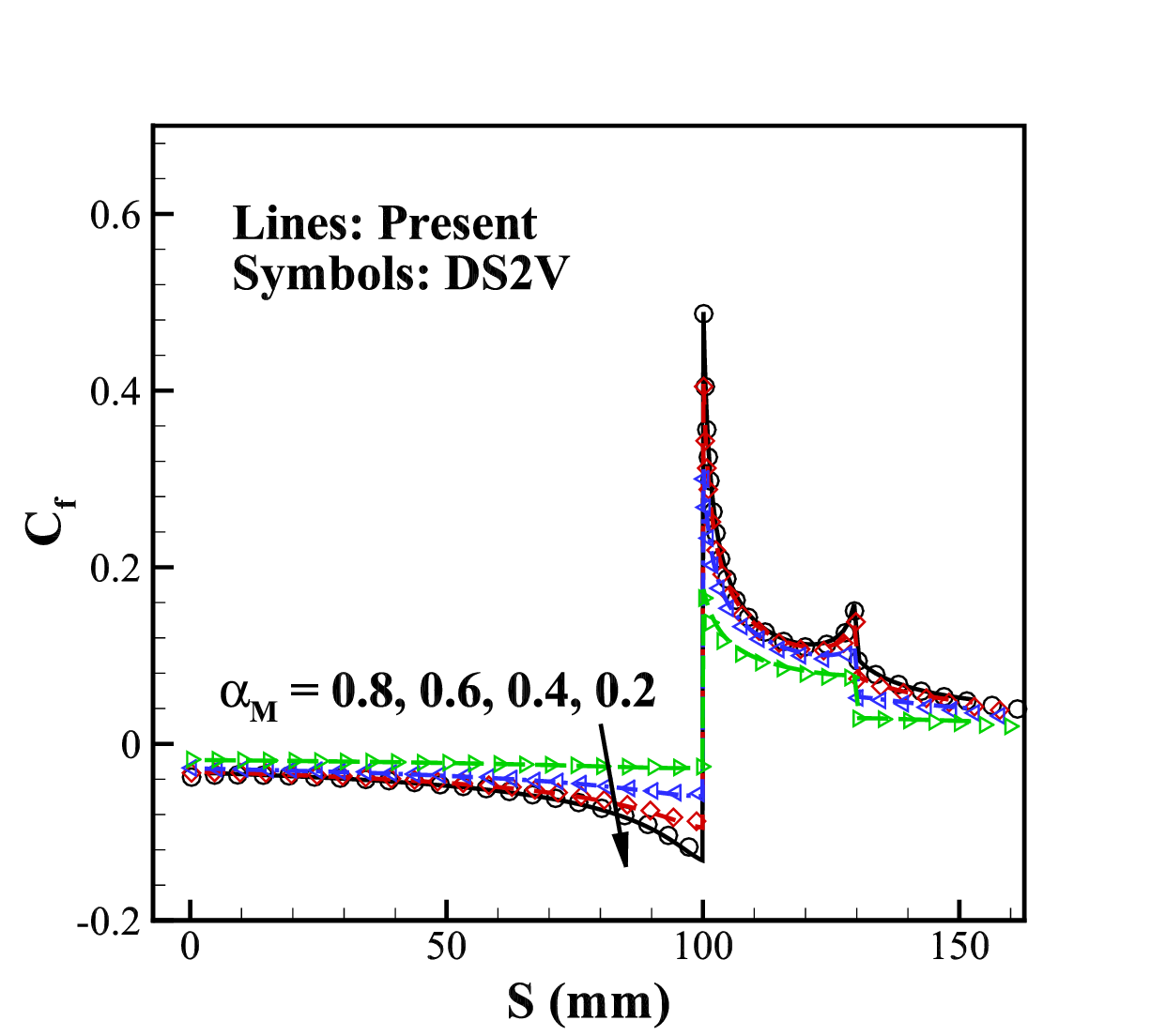}}
		\subfigure[]{\label{sharpplate_M_H_a}\includegraphics[width=0.32\textwidth]{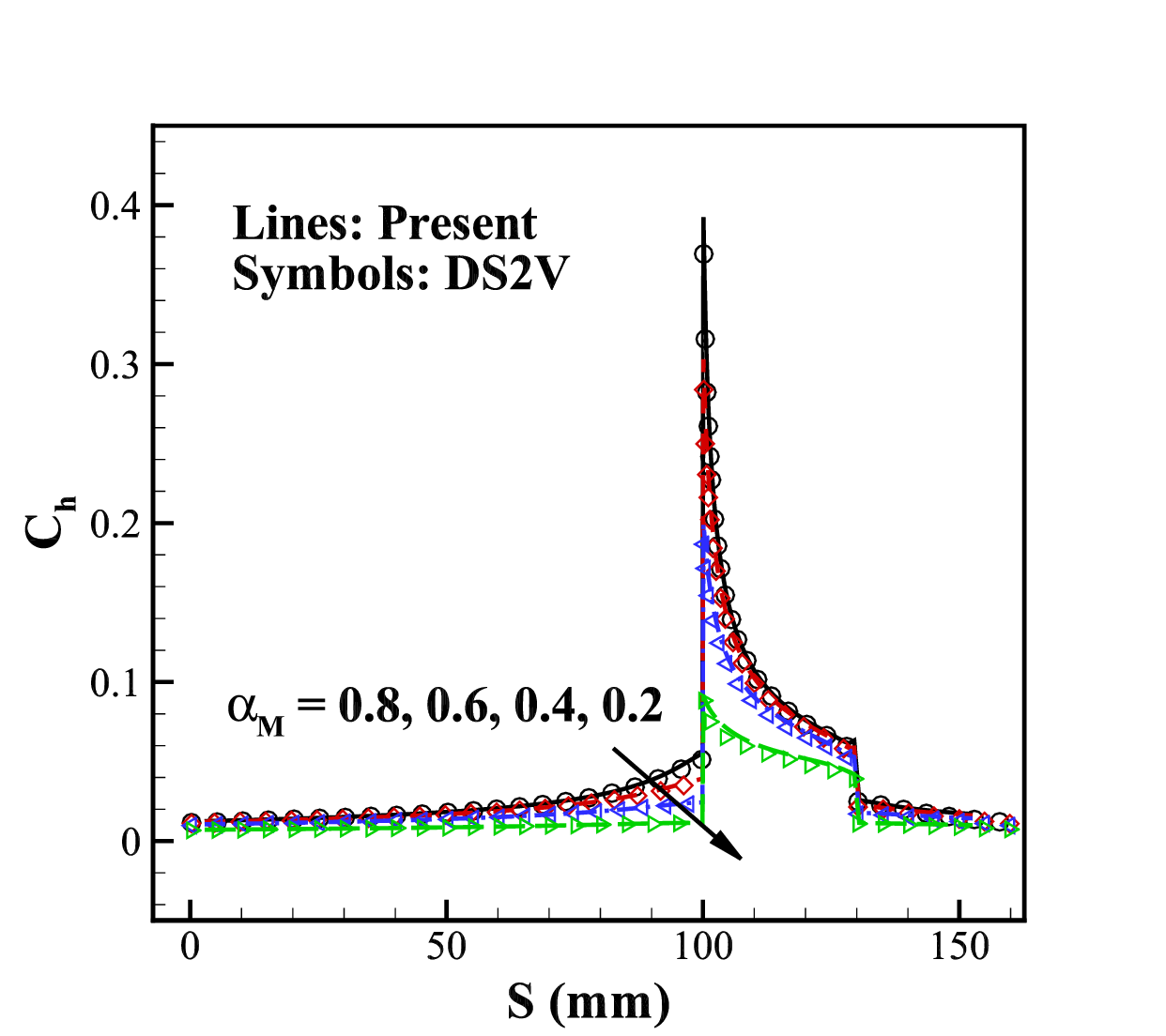}}
		\caption{\label{sharpplate_MAX}{Comparison of the (a) pressure coefficient, (b) skin friction coefficient, and (c) heat transfer coefficient on the surface of flat plate with different $\alpha_{M}$ when employing the Maxwell boundary ($Ma = 4.89$, $Kn = 0.0078$, $T_{\infty} = 116 K$, $T_{w} = 290 K$).}}
	\end{figure}
	
	\begin{figure}[!htp]
		\centering
		\subfigure[]{\label{sharpplate_C_P_a}\includegraphics[width=0.32\textwidth]{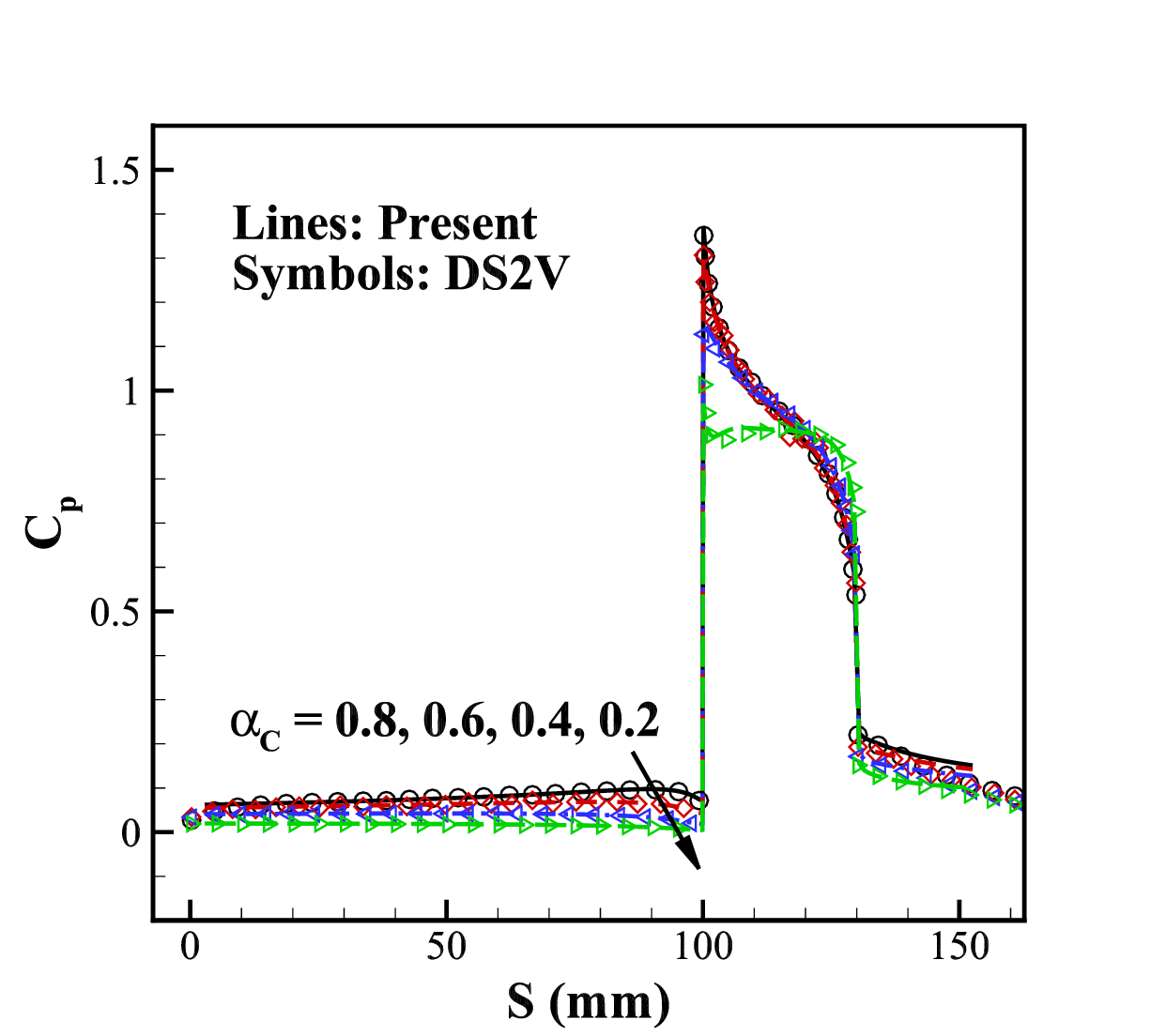}}
		\subfigure[]{\label{sharpplate_C_S_a}\includegraphics[width=0.32\textwidth]{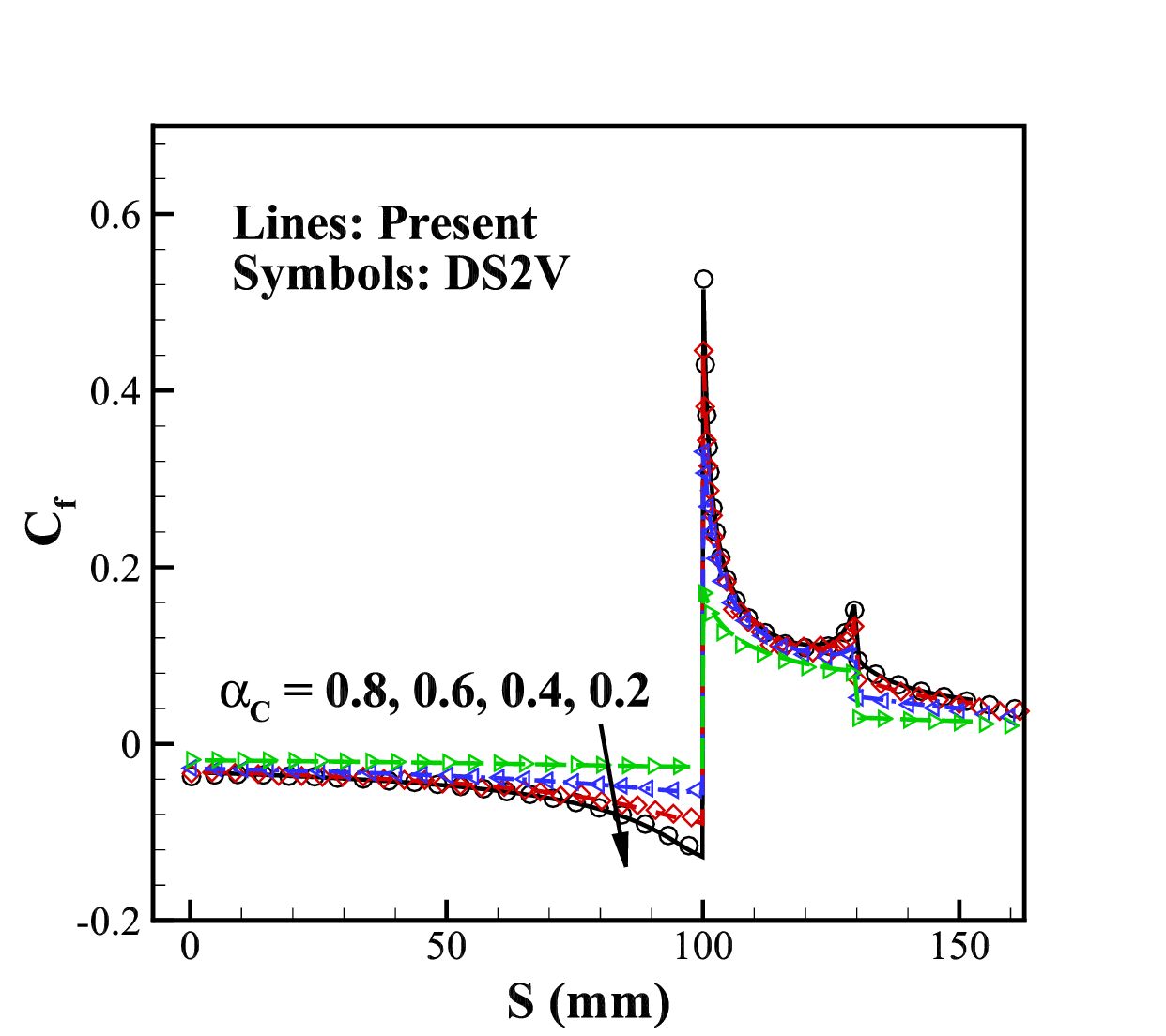}}
		\subfigure[]{\label{sharpplate_C_H_a}\includegraphics[width=0.32\textwidth]{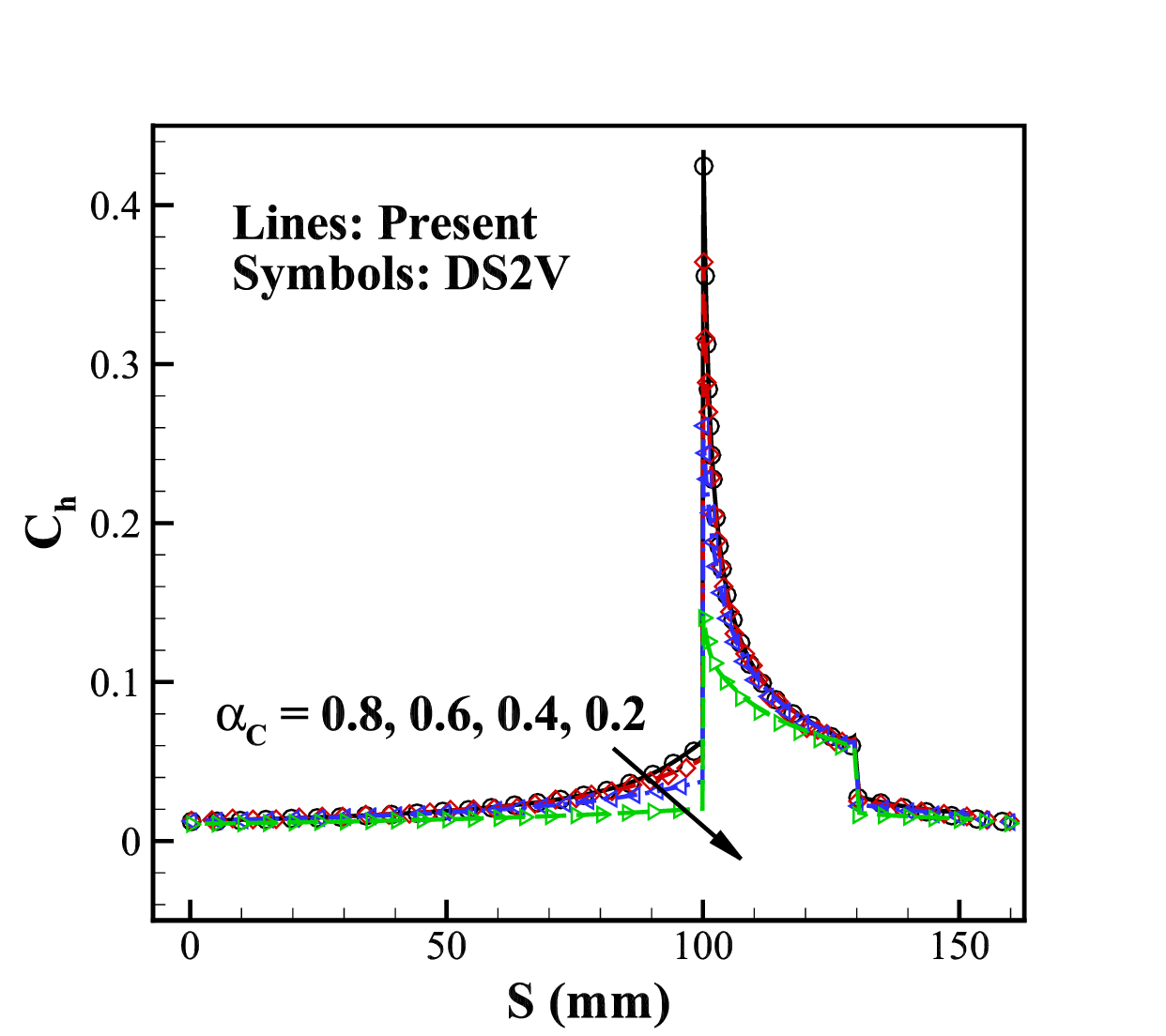}}
		\caption{\label{sharpplate_CLL}{Comparison of the (a) pressure coefficient, (b) skin friction coefficient, and (c) heat transfer coefficient on the surface of flat plate with different $\alpha_{C}$ when employing the CLL boundary ($Ma = 4.89$, $Kn = 0.0078$, $T_{\infty} = 116 K$, $T_{w} = 290 K$).}}
	\end{figure}
	
	\begin{figure}[!htp]
		\centering
		\subfigure[]{\label{sharpplate_M_U_5}\includegraphics[width=0.45\textwidth]{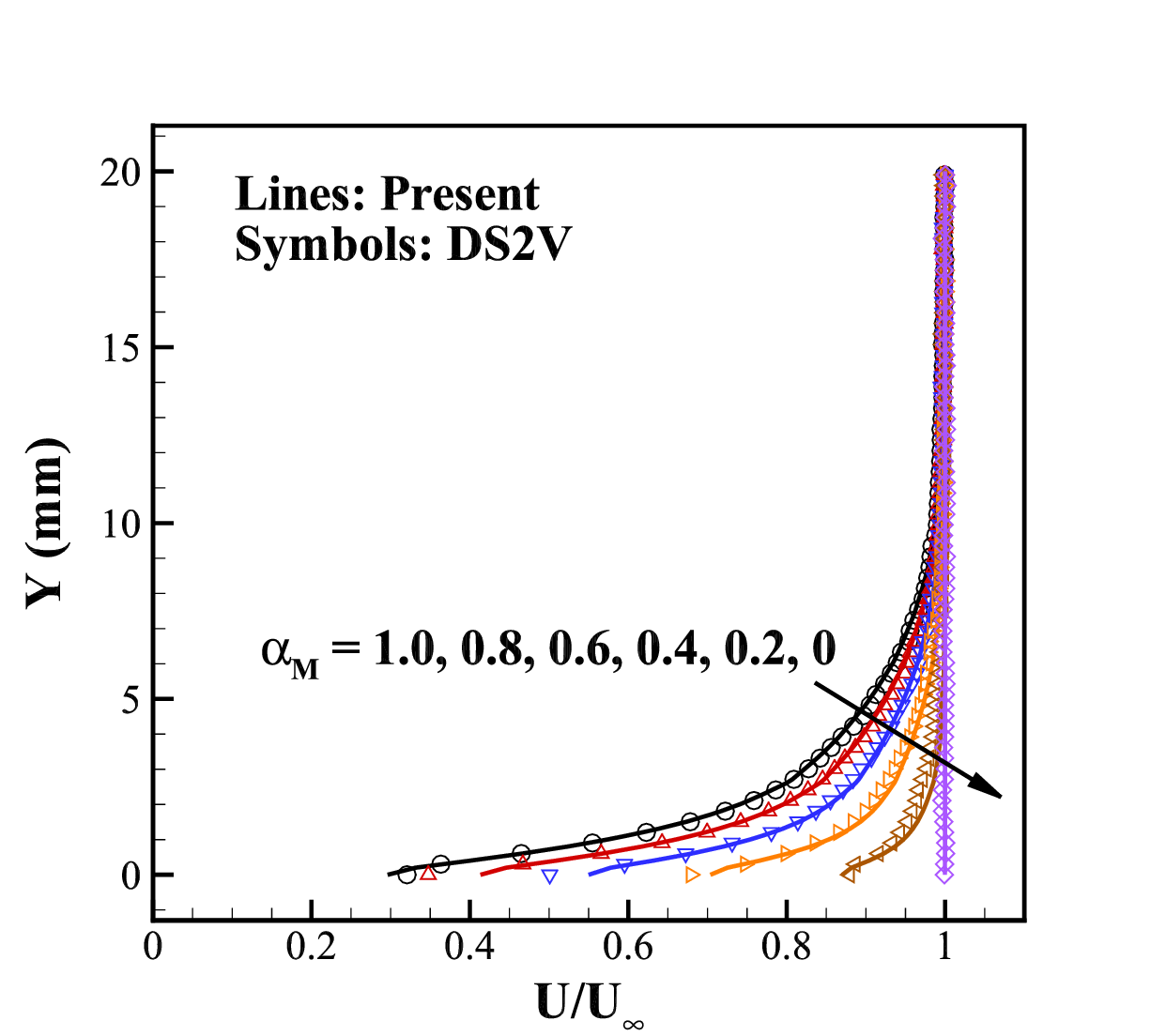}}
		\subfigure[]{\label{sharpplate_M_U_20}\includegraphics[width=0.45\textwidth]{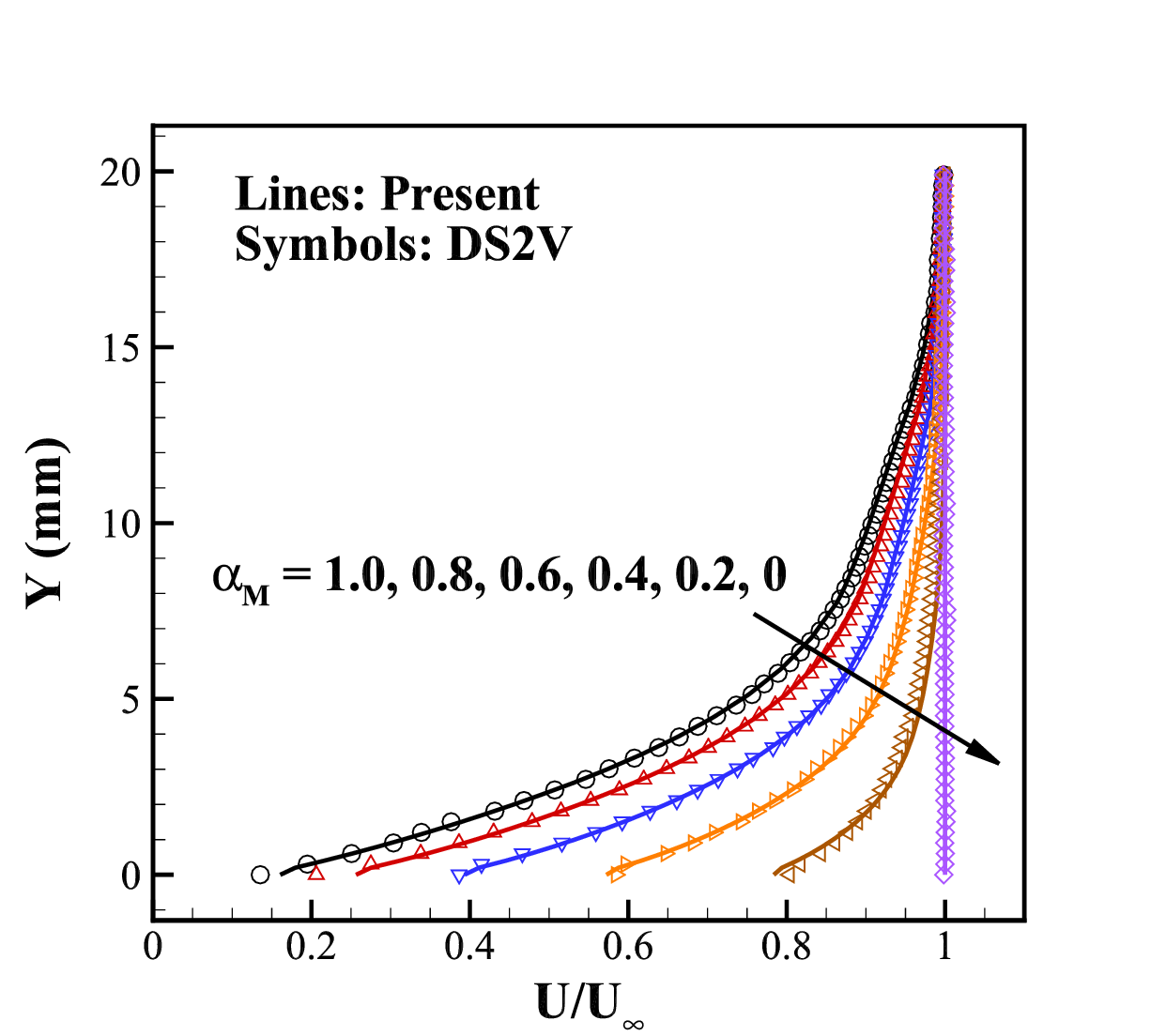}}
		\caption{\label{sharpplate_M_U}{Horizontal velocity profiles over the flat plate along the vertical line (a) X = 5 mm, and (b) X = 20 mm with different $\alpha_{M}$ when employing the Maxwell boundary ($Ma = 4.89$, $Kn = 0.0078$, $T_{\infty} = 116 K$, $T_{w} = 290 K$).}}
		
	\end{figure}
	
	\begin{figure}[!htp]
		\centering
		\subfigure[]{\label{sharpplate_M_T_5}\includegraphics[width=0.45\textwidth]{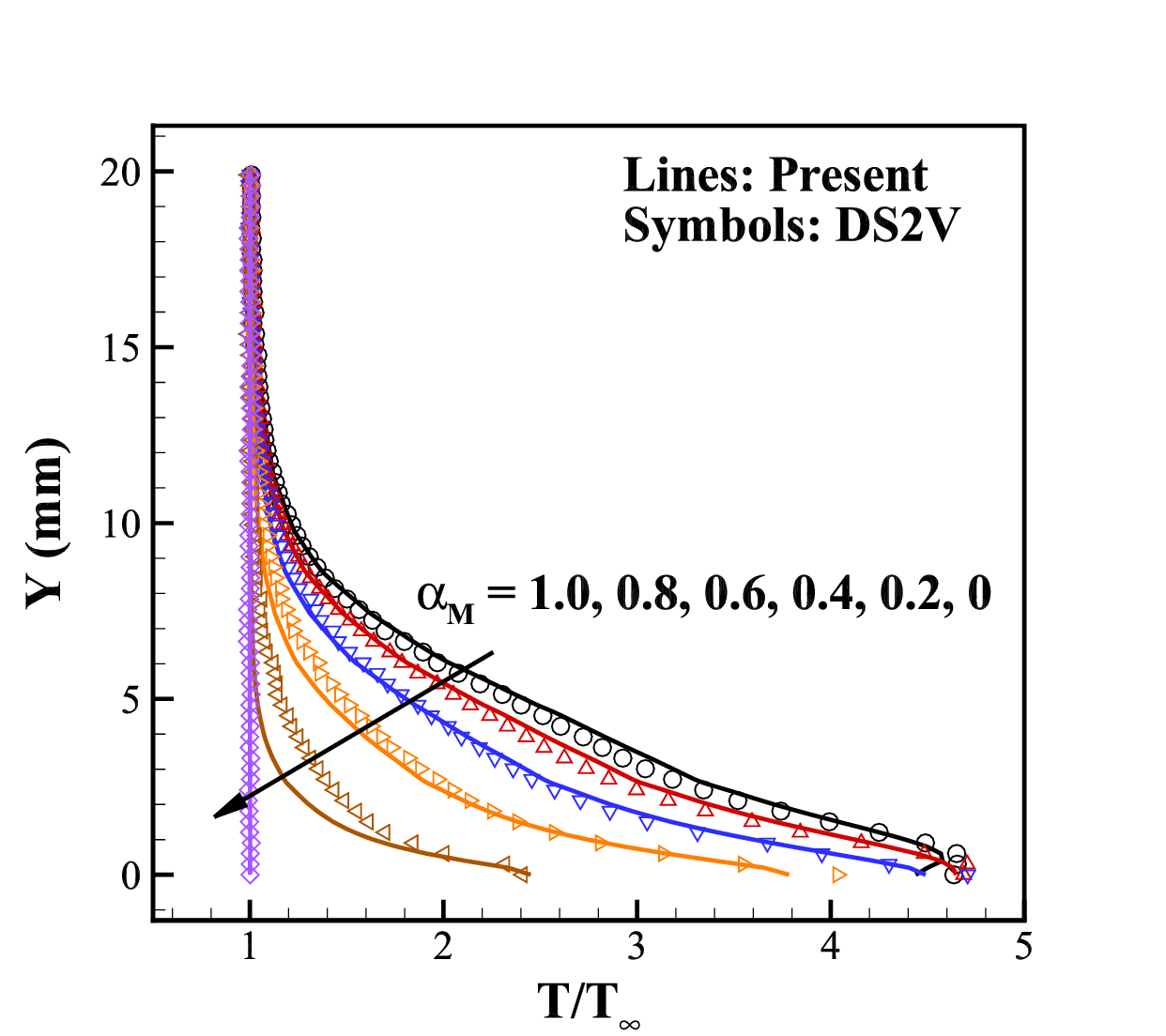}}
		\subfigure[]{\label{sharpplate_M_T_20}\includegraphics[width=0.45\textwidth]{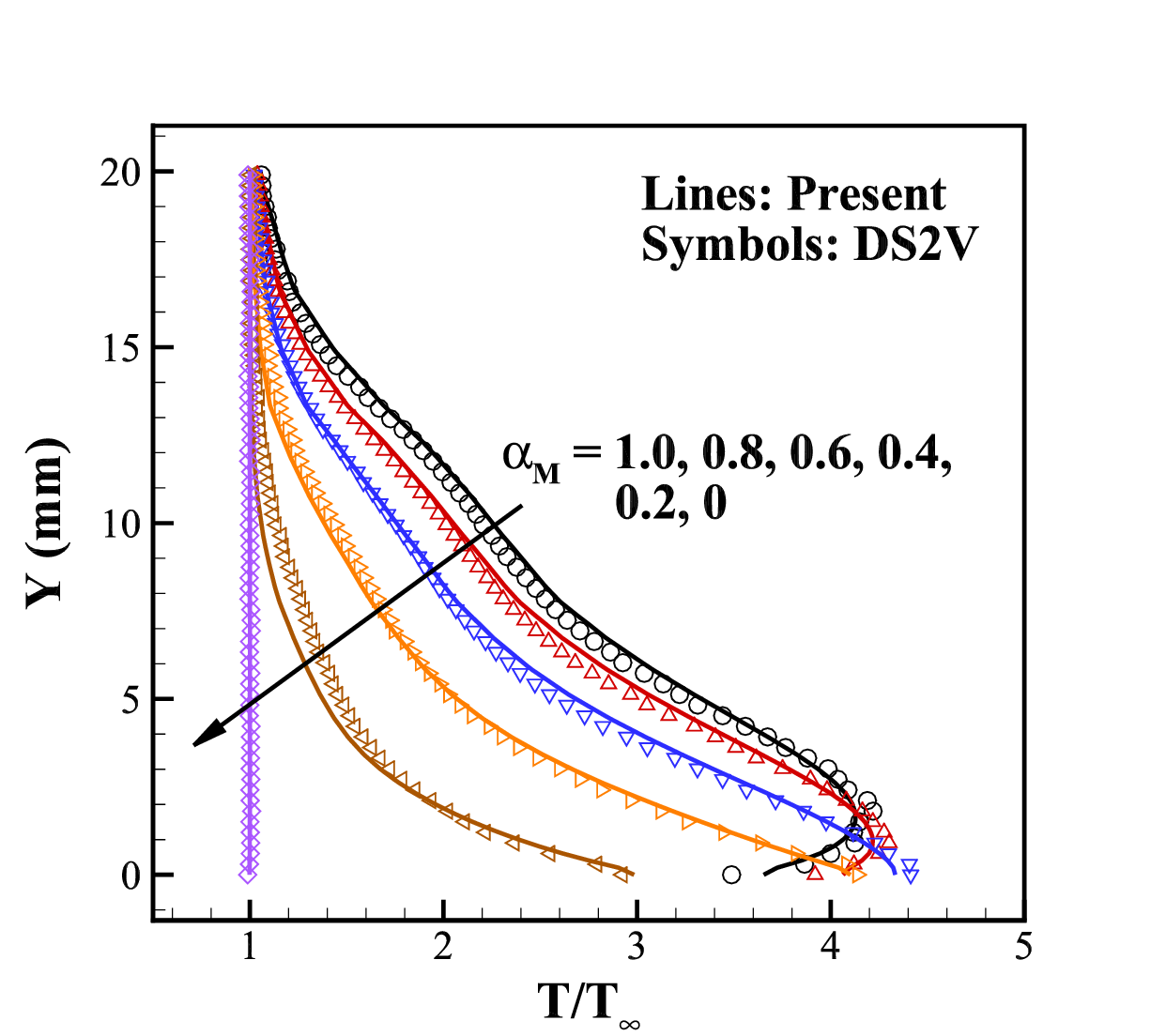}}
		\caption{\label{sharpplate_M_T}{Temperature profiles over the flat plate along the vertical line (a) X = 5 mm, and (b) X = 20 mm with different $\alpha_{M}$ when employing the Maxwell boundary ($Ma = 4.89$, $Kn = 0.0078$, $T_{\infty} = 116 K$, $T_{w} = 290 K$).}}
	\end{figure}
	
	\begin{figure}[!htp]
		\centering
		\subfigure[]{\label{sharpplate_C_U_5}\includegraphics[width=0.45\textwidth]{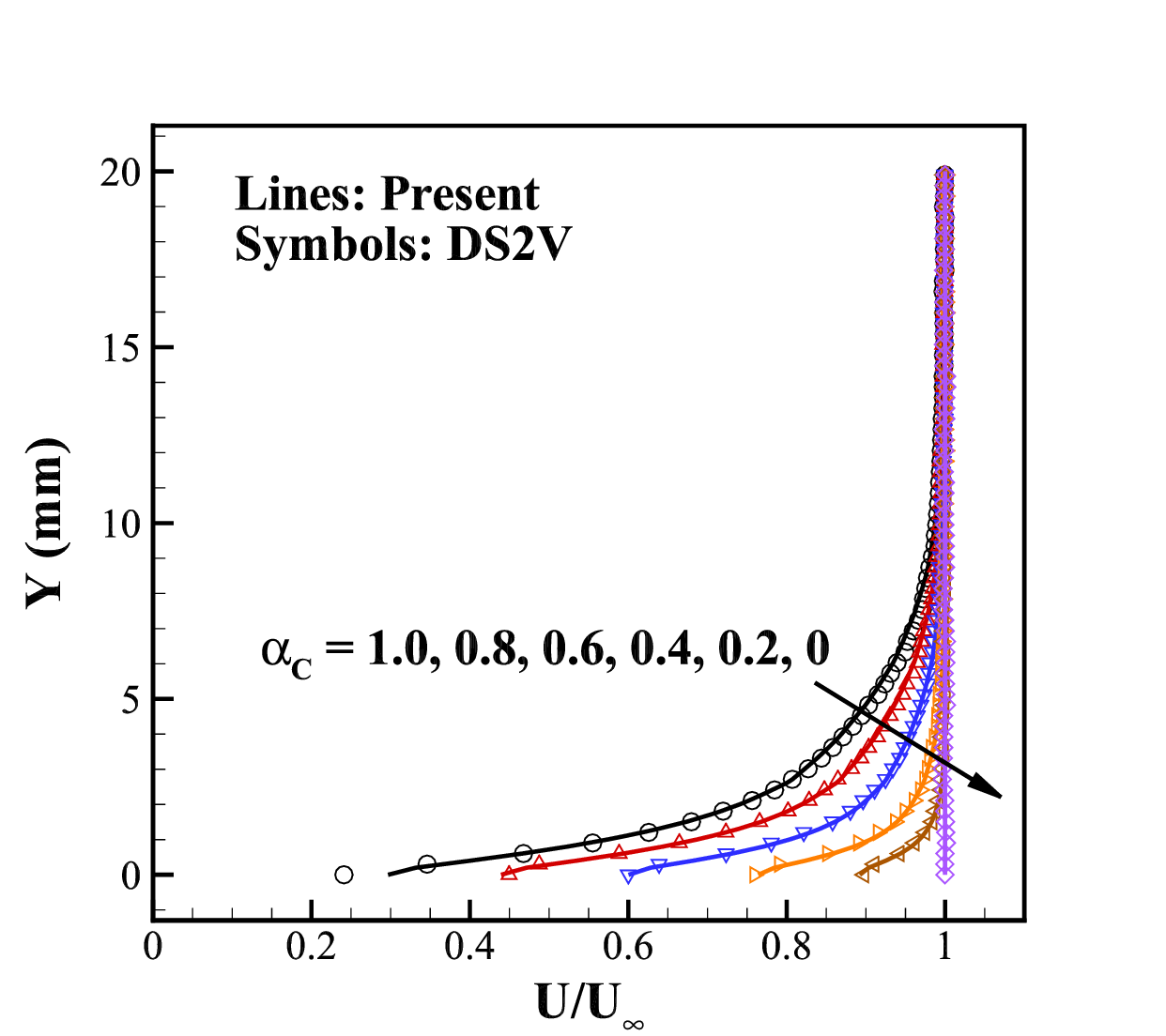}}
		\subfigure[]{\label{sharpplate_C_U_20}\includegraphics[width=0.45\textwidth]{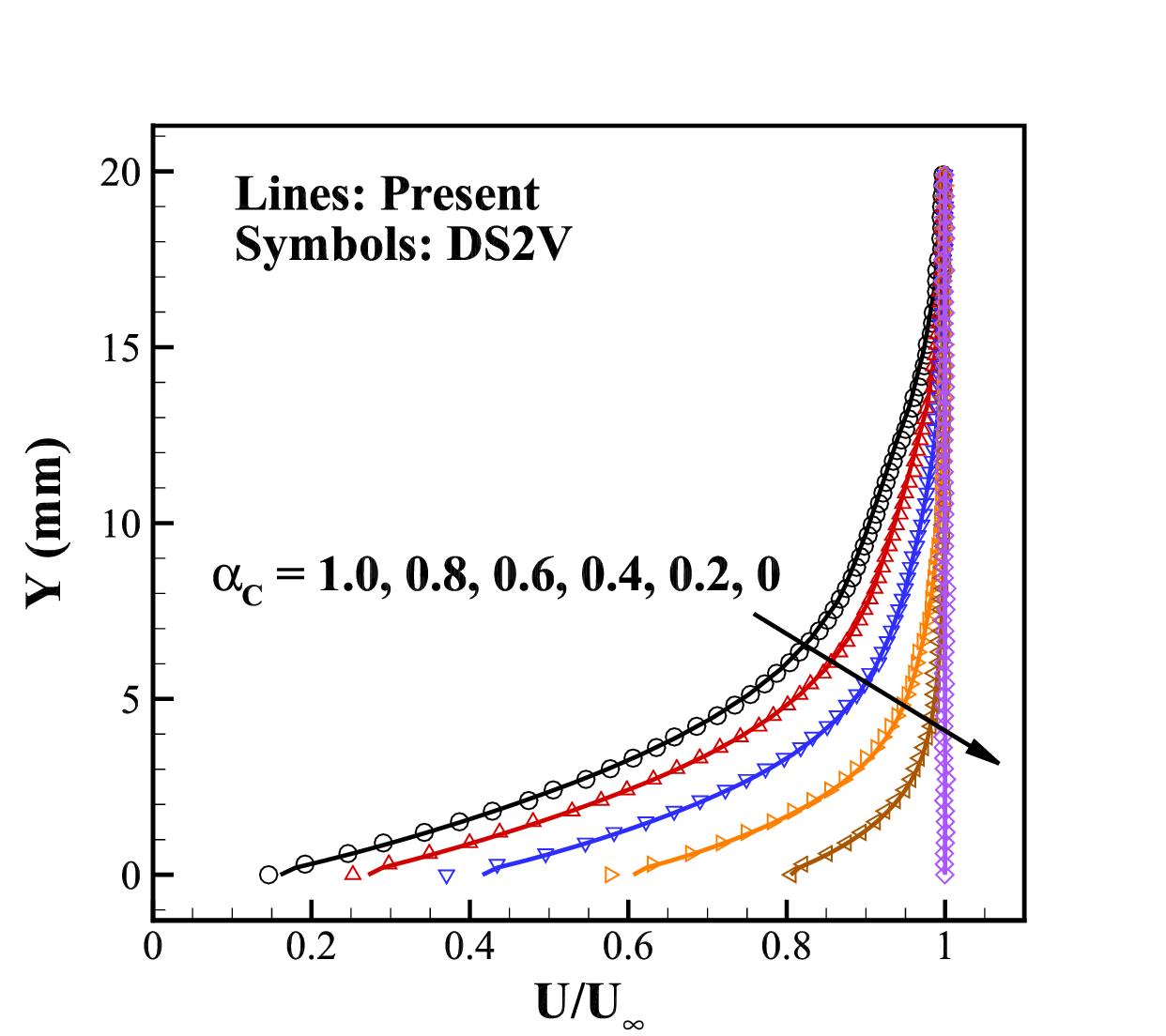}}
		\caption{\label{sharpplate_C_U}{Horizontal velocity profiles over the flat plate along the vertical line (a) X = 5 mm, and (b) X = 20 mm with different $\alpha_{C}$ when employing the CLL boundary ($Ma = 4.89$, $Kn = 0.0078$, $T_{\infty} = 116 K$, $T_{w} = 290 K$).}}
	\end{figure}
	
	\begin{figure}[!htp]
		\centering
		\subfigure[]{\label{sharpplate_C_T_5}\includegraphics[width=0.45\textwidth]{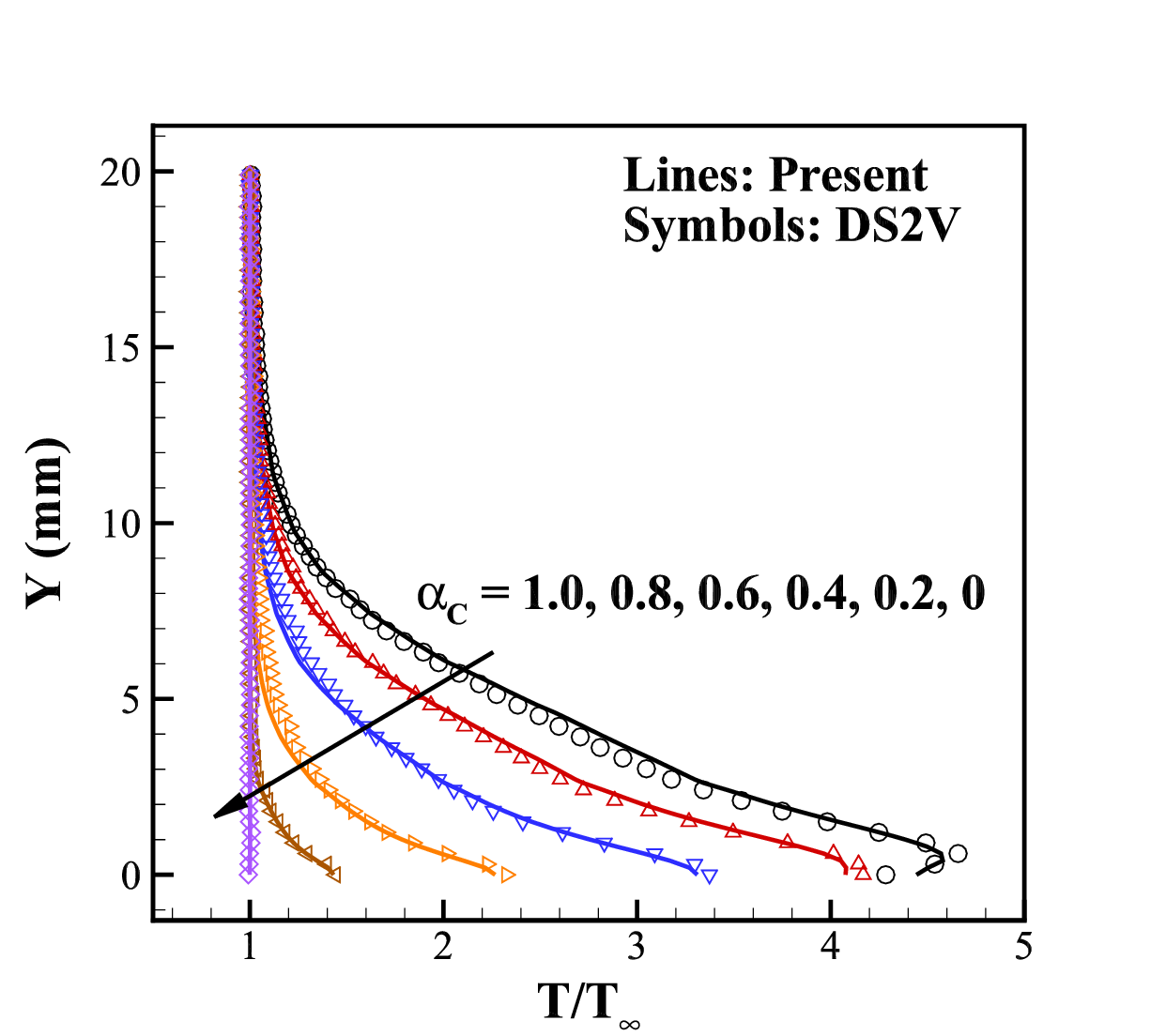}}
		\subfigure[]{\label{sharpplate_C_T_20}\includegraphics[width=0.45\textwidth]{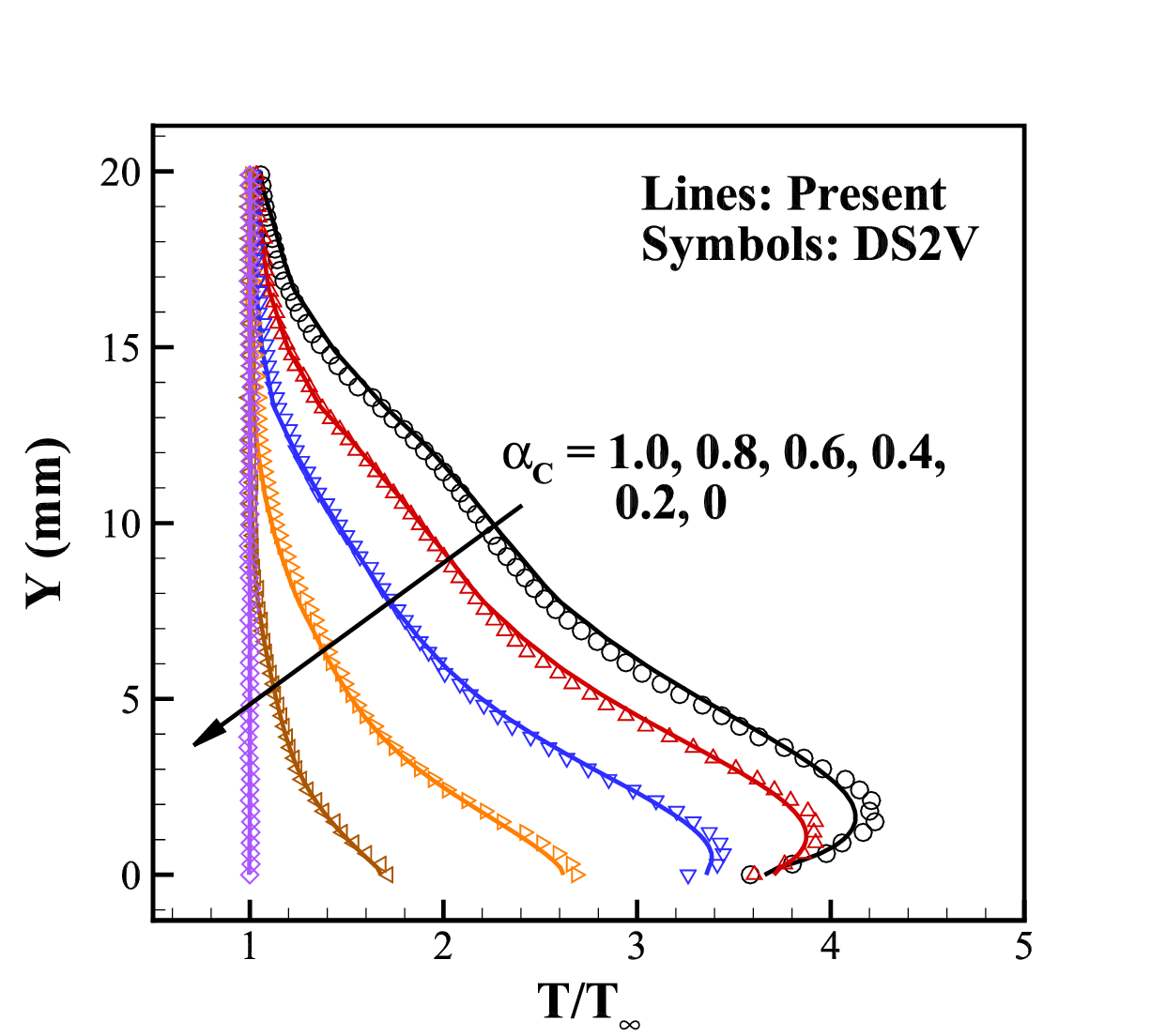}}
		\caption{\label{sharpplate_C_T}{Temperature profiles over the flat plate along the vertical line (a) X = 5 mm, and (b) X = 20 mm with different $\alpha_{C}$ when employing the CLL boundary ($Ma = 4.89$, $Kn = 0.0078$, $T_{\infty} = 116 K$, $T_{w} = 290 K$).}}
	\end{figure}
	
	\begin{figure}[!htp]
		\centering
		\includegraphics[width=0.7\textwidth]{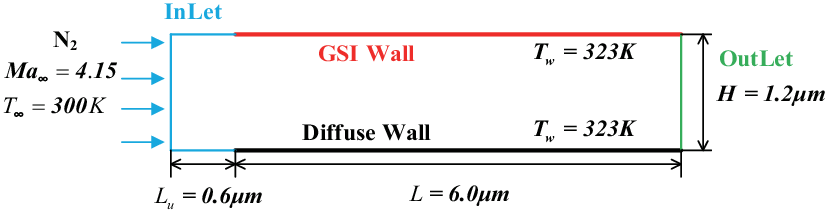}
		\caption{\label{microchannel_geo}{Schematic view of the computational domain.}}
	\end{figure}
	
	\begin{figure}[!htp]
		\centering
		\includegraphics[width=0.6\textwidth]{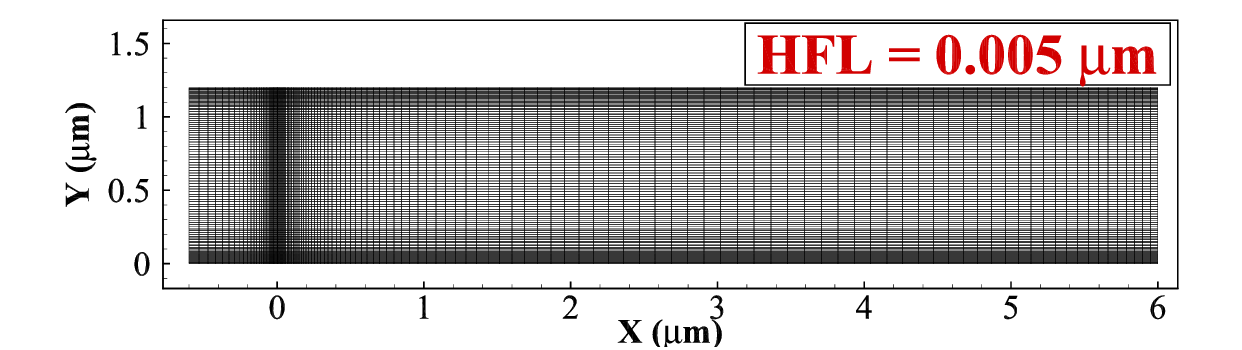}
		\caption{\label{microchannel_pmesh}{The physical mesh for the high speed microchannel flow ($Ma = 4.15$, $Kn = 0.062$, $T_{\infty} = 300 K$, $T_{w} = 323 K$).}}
	\end{figure}
	
	\begin{figure}[!htp]
		\centering
		\includegraphics[width=0.5\textwidth]{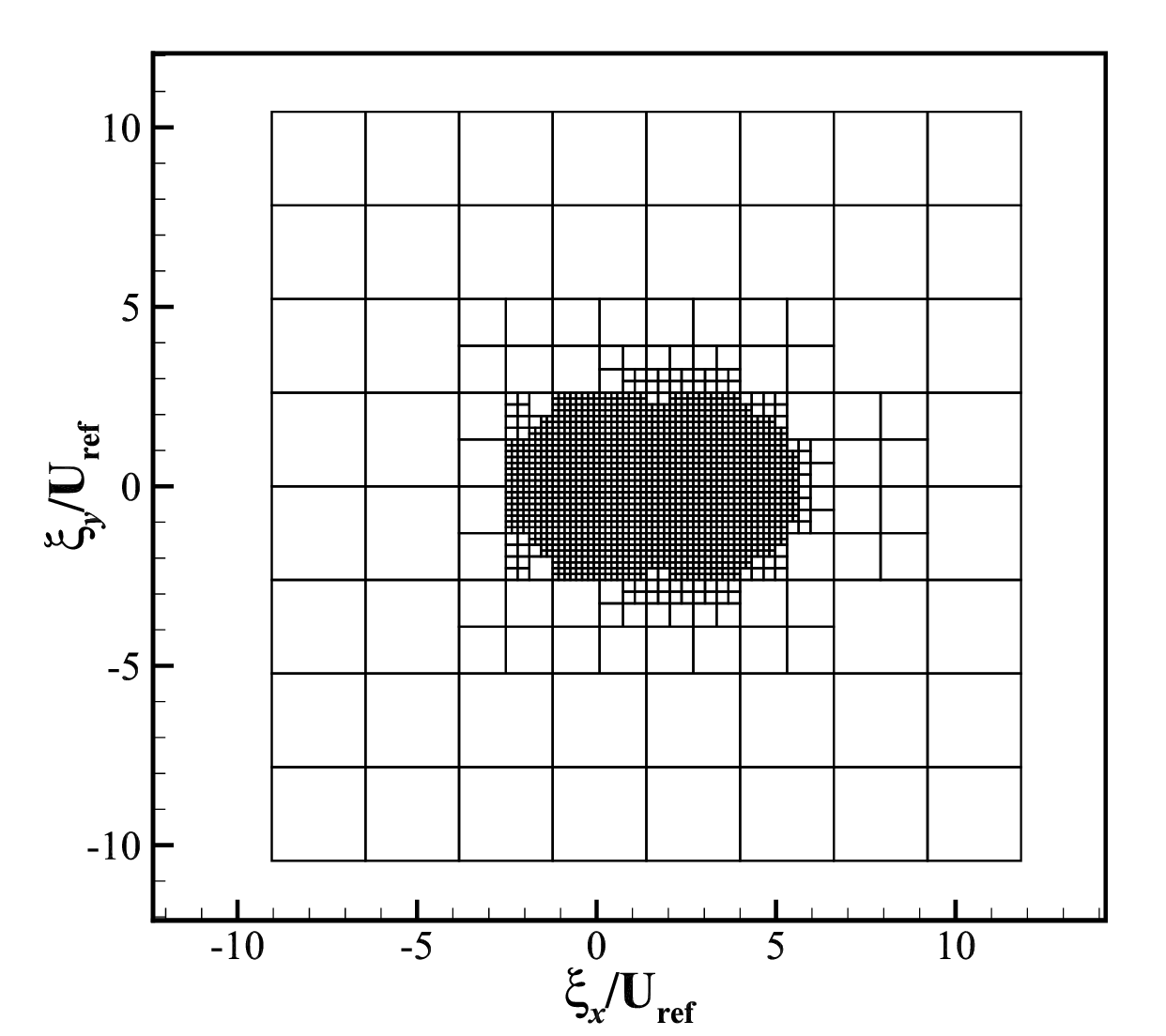}
		\caption{\label{microchannel_dvmesh}{The velocity mesh for the high speed microchannel flow ($Ma = 4.15$, $Kn = 0.062$, $T_{\infty} = 300 K$, $T_{w} = 323 K$).}}
	\end{figure}
	
	\begin{figure}[!htp]
		\centering
		\subfigure[]{\label{microchannel_contour_D}\includegraphics[width=0.45\textwidth]{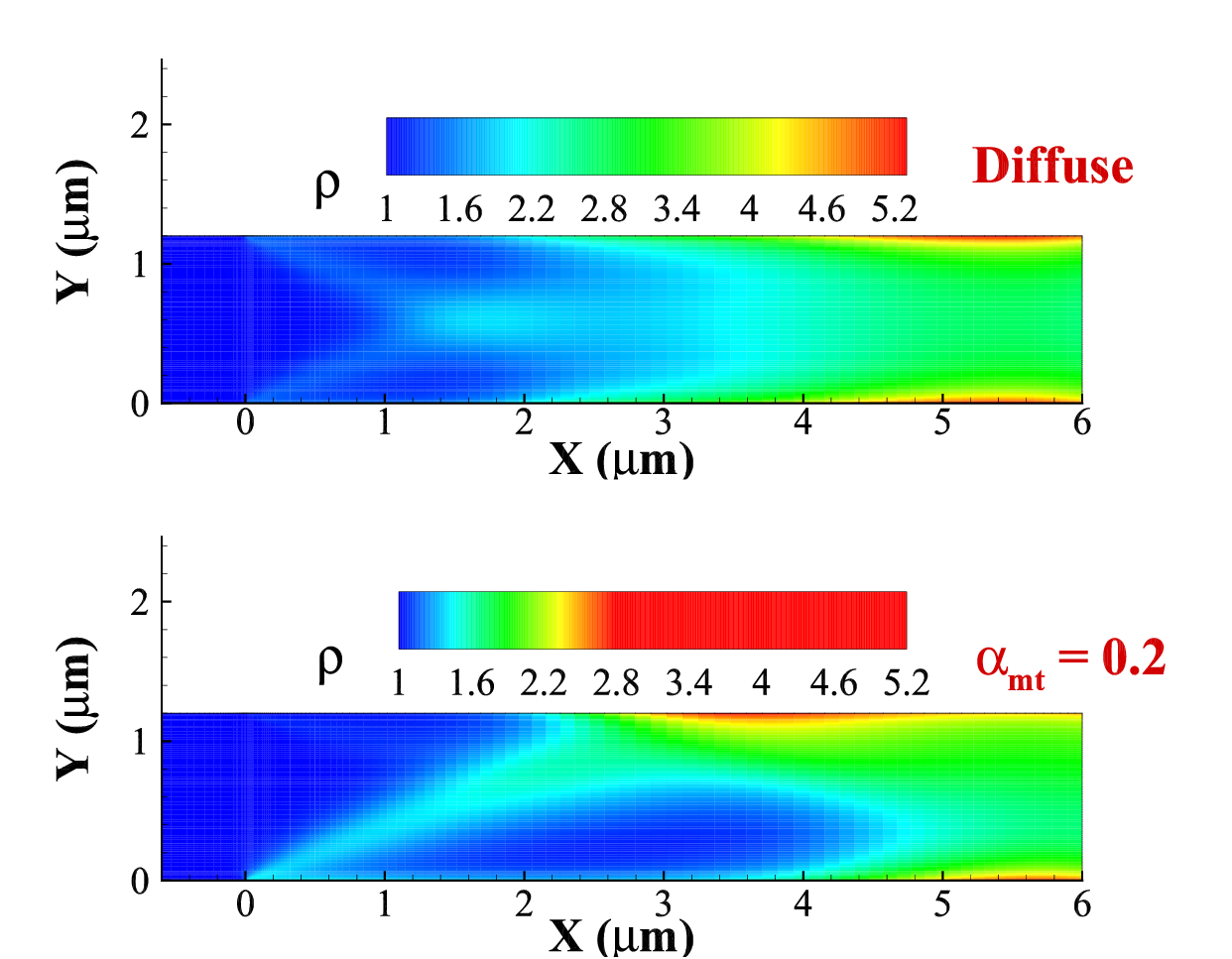}}
		\subfigure[]{\label{microchannel_contour_P}\includegraphics[width=0.45\textwidth]{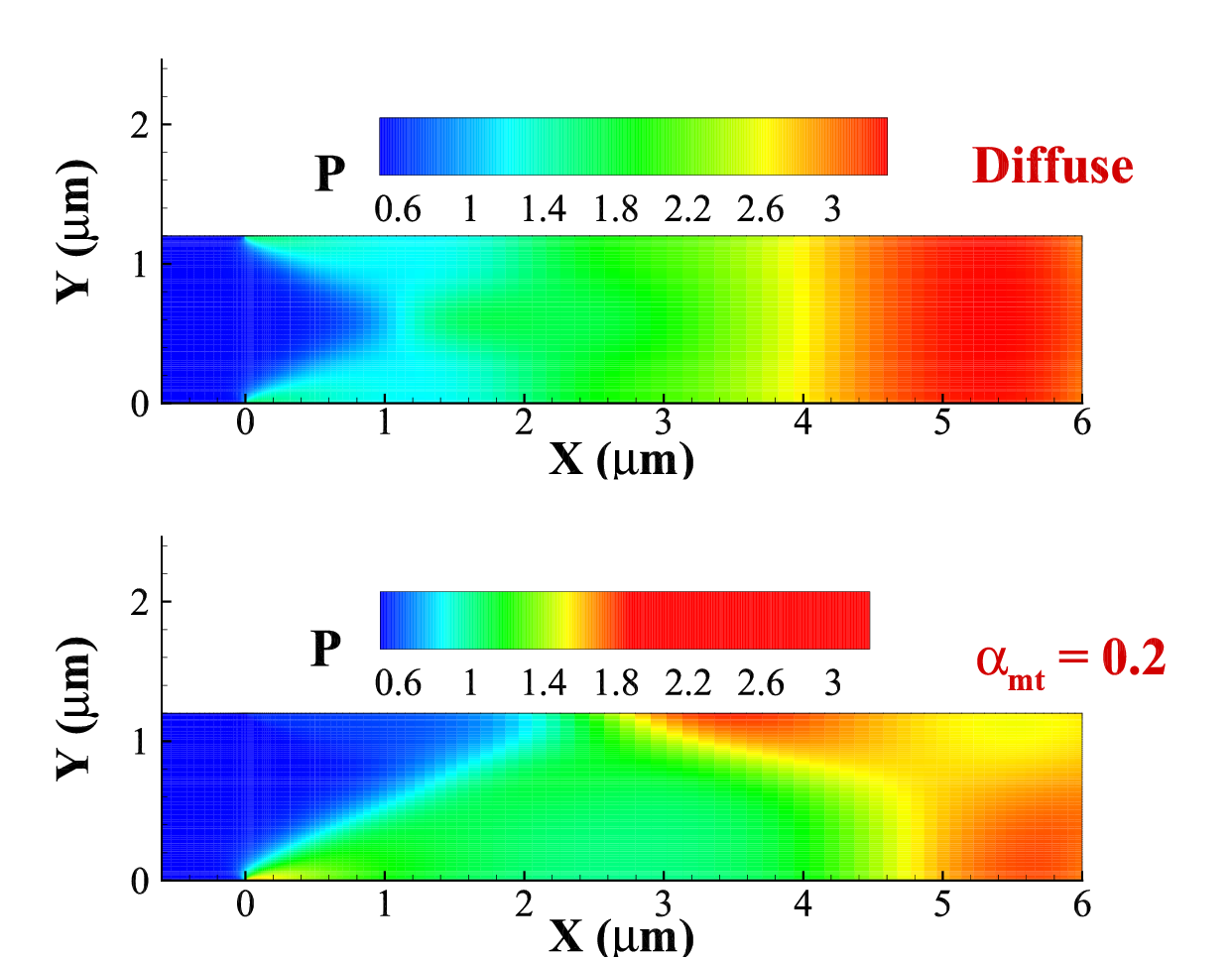}}
		\subfigure[]{\label{microchannel_contour_U}\includegraphics[width=0.45\textwidth]{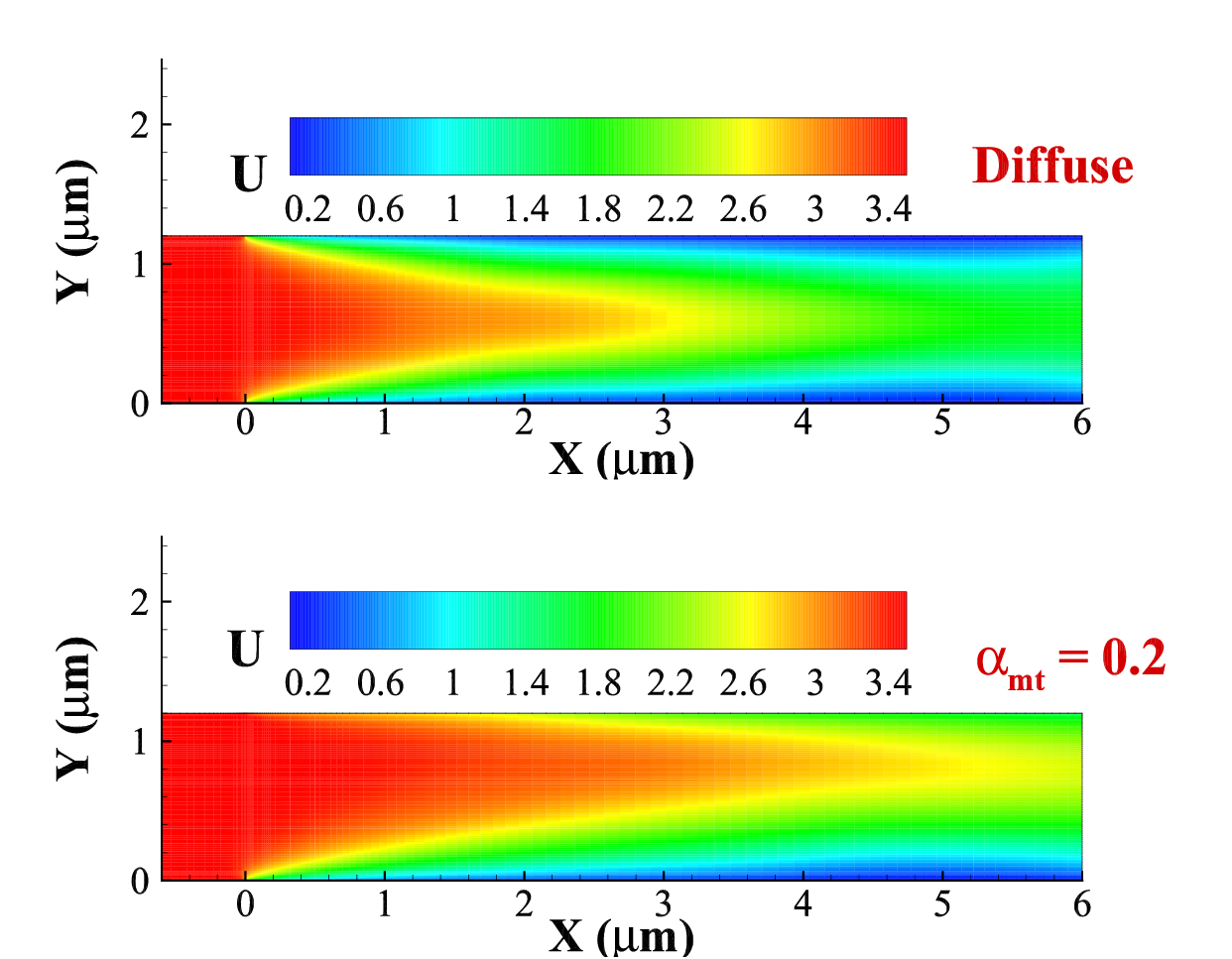}}
		\subfigure[]{\label{microchannel_contour_Tt}\includegraphics[width=0.45\textwidth]{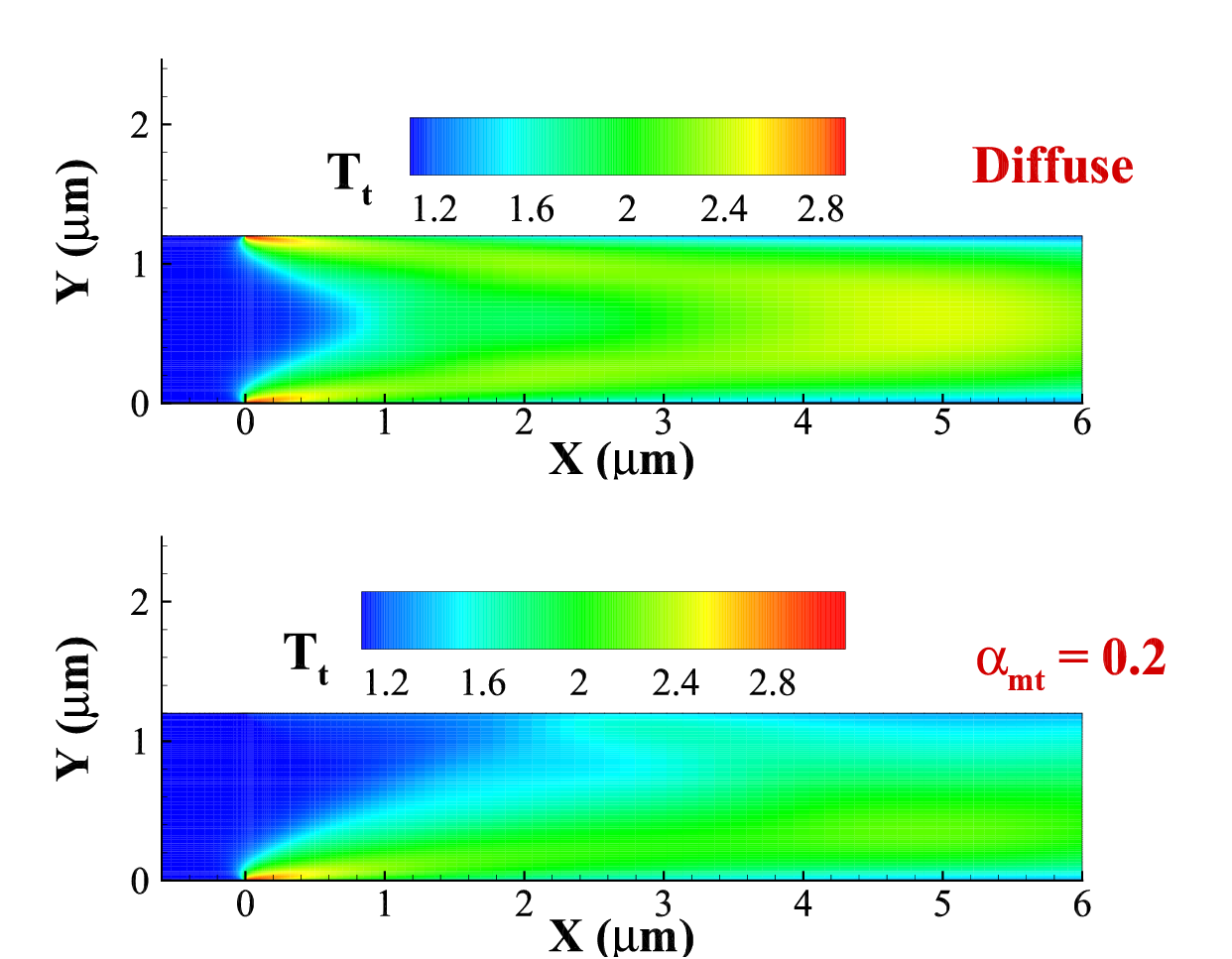}}
		\caption{\label{microchannel_contour}{Comparison of (a) density, (b) pressure, (c) horizontal velocity, and (d) translational temperature contours along the microchannel for $\alpha_{mt}$  = 1.0 (top) and 0.2 (bottom).}}
	\end{figure}	
	
	\begin{figure}[!htp]
		\centering
		\subfigure[]{\label{microchannel_X_04_U}\includegraphics[width=0.45\textwidth]{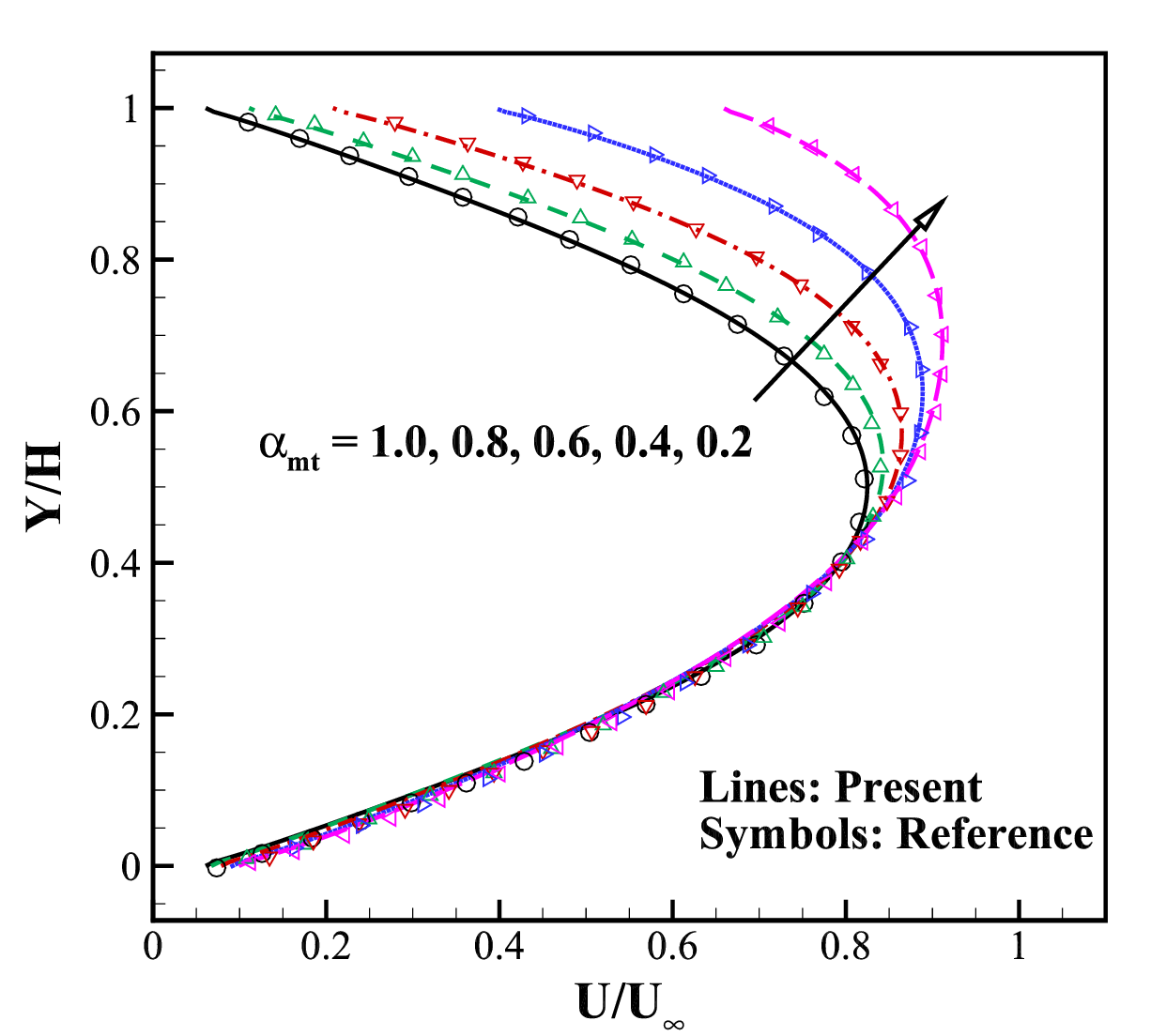}}
		\subfigure[]{\label{microchannel_X_06_U}\includegraphics[width=0.45\textwidth]{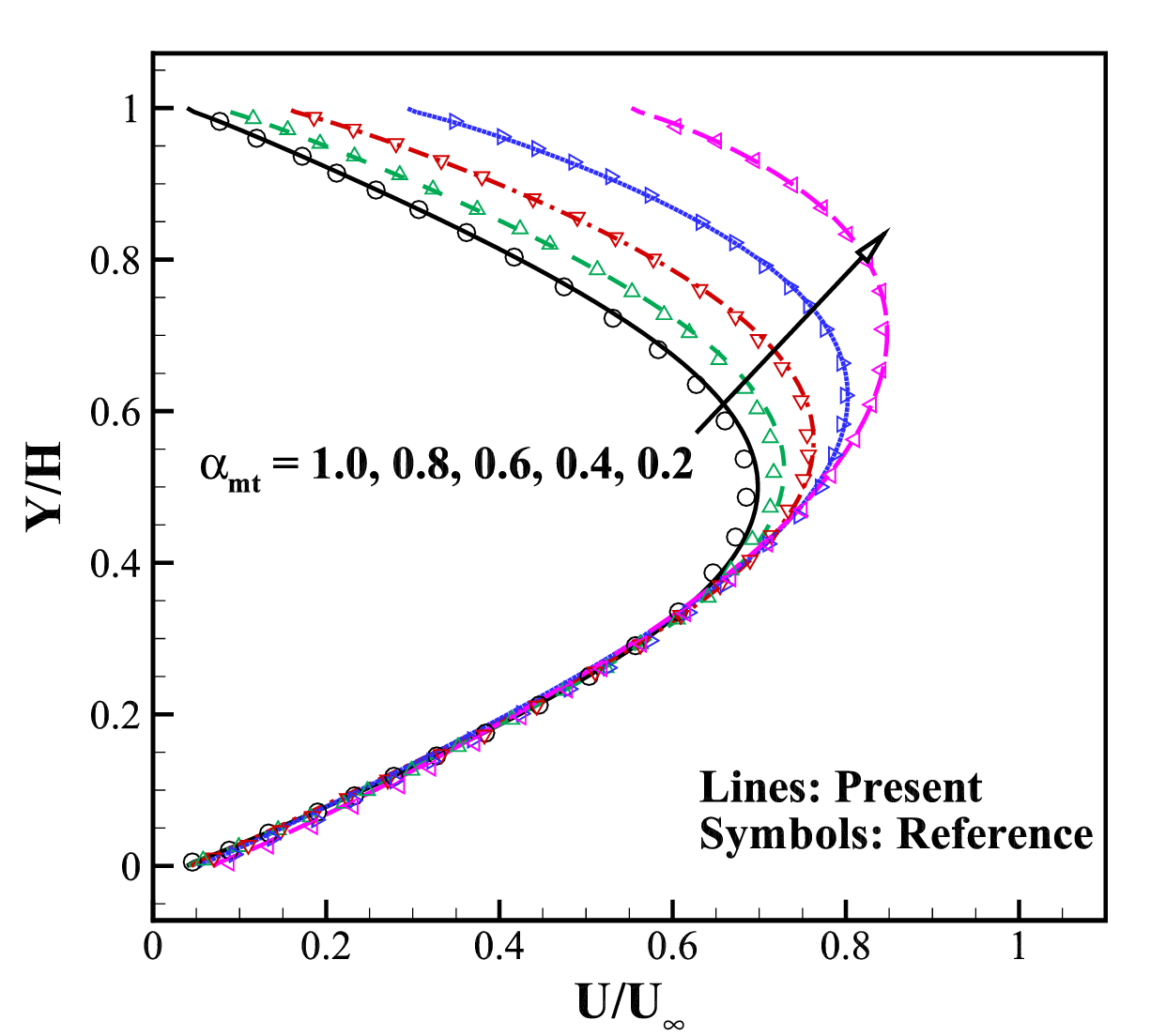}}
		\caption{\label{microchannel_X_U}{Horizontal velocity profiles along the microchannel at (a) X = 2.4 $\mu m$ and (b) X = 3.6 $\mu m$ with different $\alpha_{mt}$.}}
	\end{figure}	
	
	\begin{figure}[!htp]
		\centering
		\subfigure[]{\label{microchannel_X_04_rho}\includegraphics[width=0.45\textwidth]{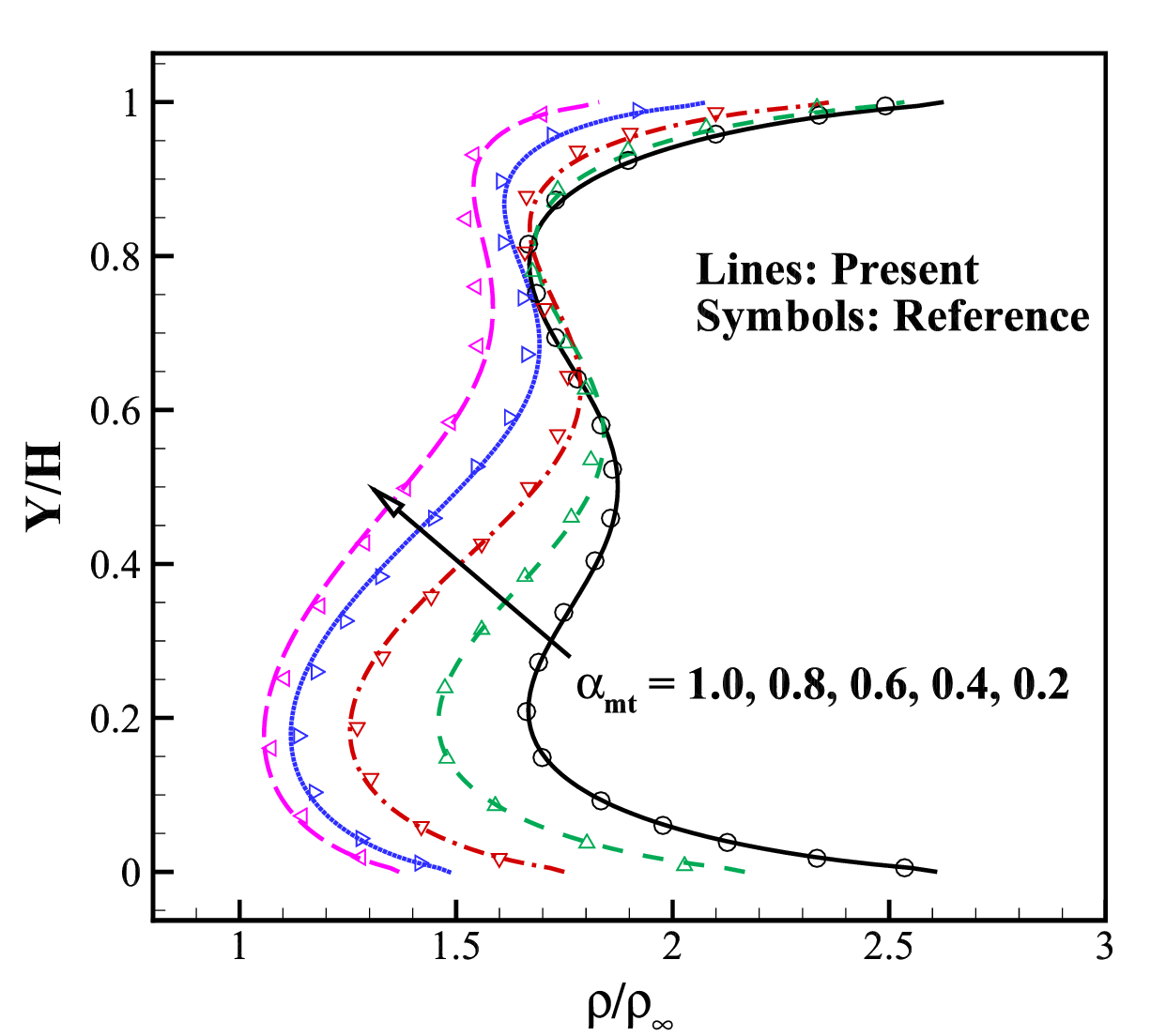}}
		\subfigure[]{\label{microchannel_X_06_rho}\includegraphics[width=0.45\textwidth]{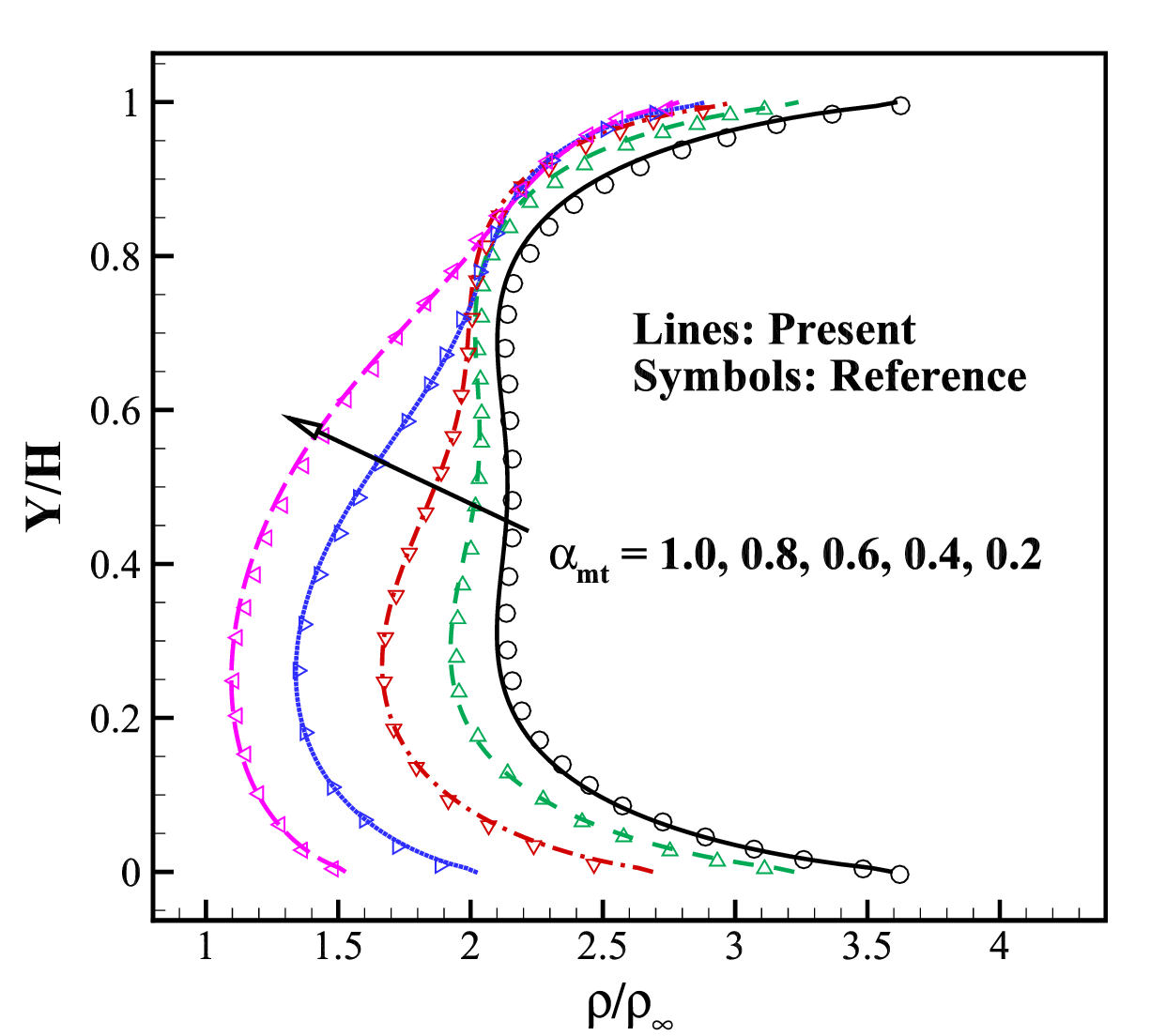}}
		\caption{\label{microchannel_X_rho}{Density profiles along the microchannel at (a) X = 2.4 $\mu m$ and (b) X = 3.6 $\mu m$ with different $\alpha_{mt}$.}}
	\end{figure}
	
	\begin{figure}[!htp]
		\centering
		\subfigure[]{\label{microchannel_X_04_Tt}\includegraphics[width=0.45\textwidth]{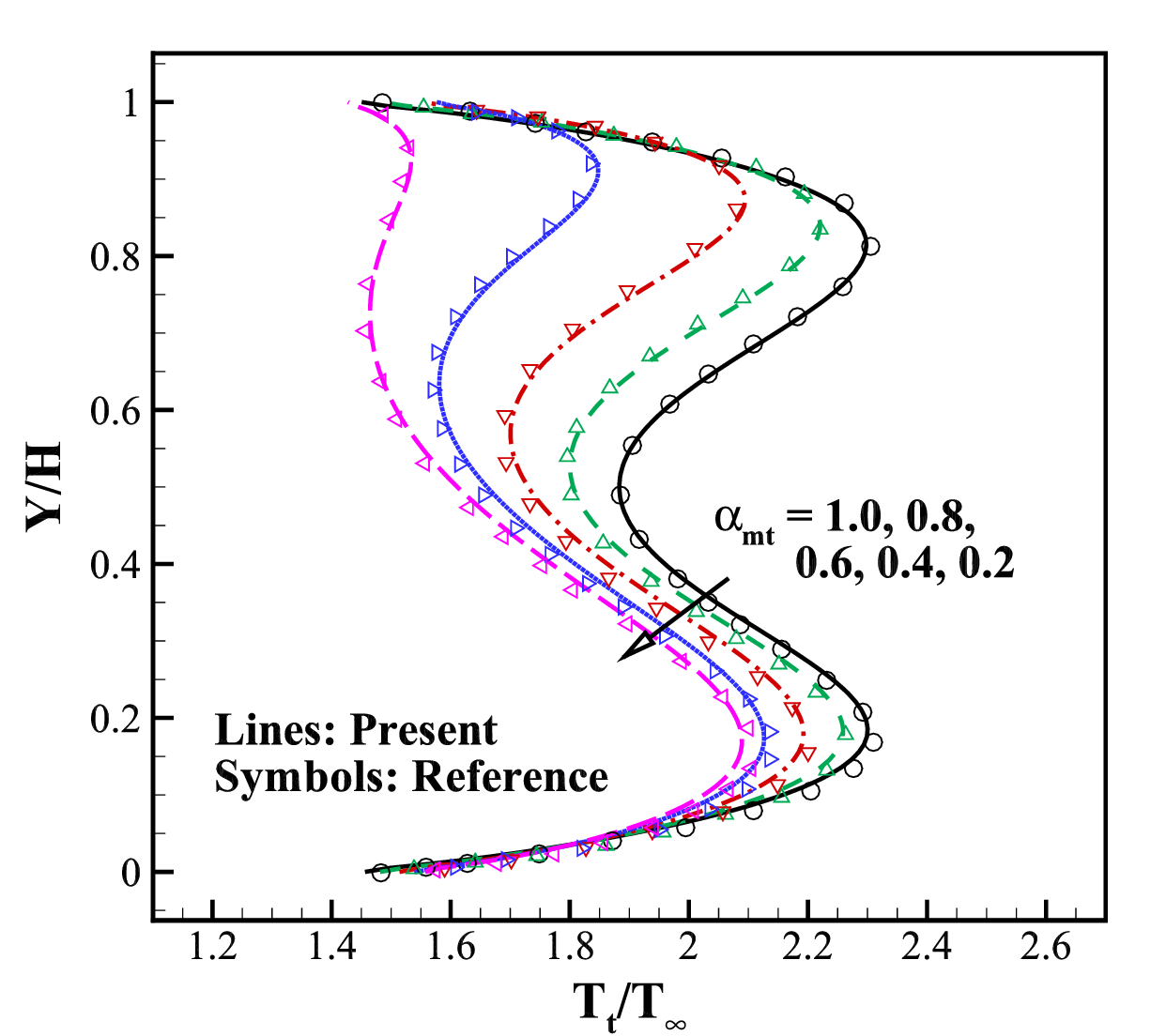}}
		\subfigure[]{\label{microchannel_X_06_Tt}\includegraphics[width=0.45\textwidth]{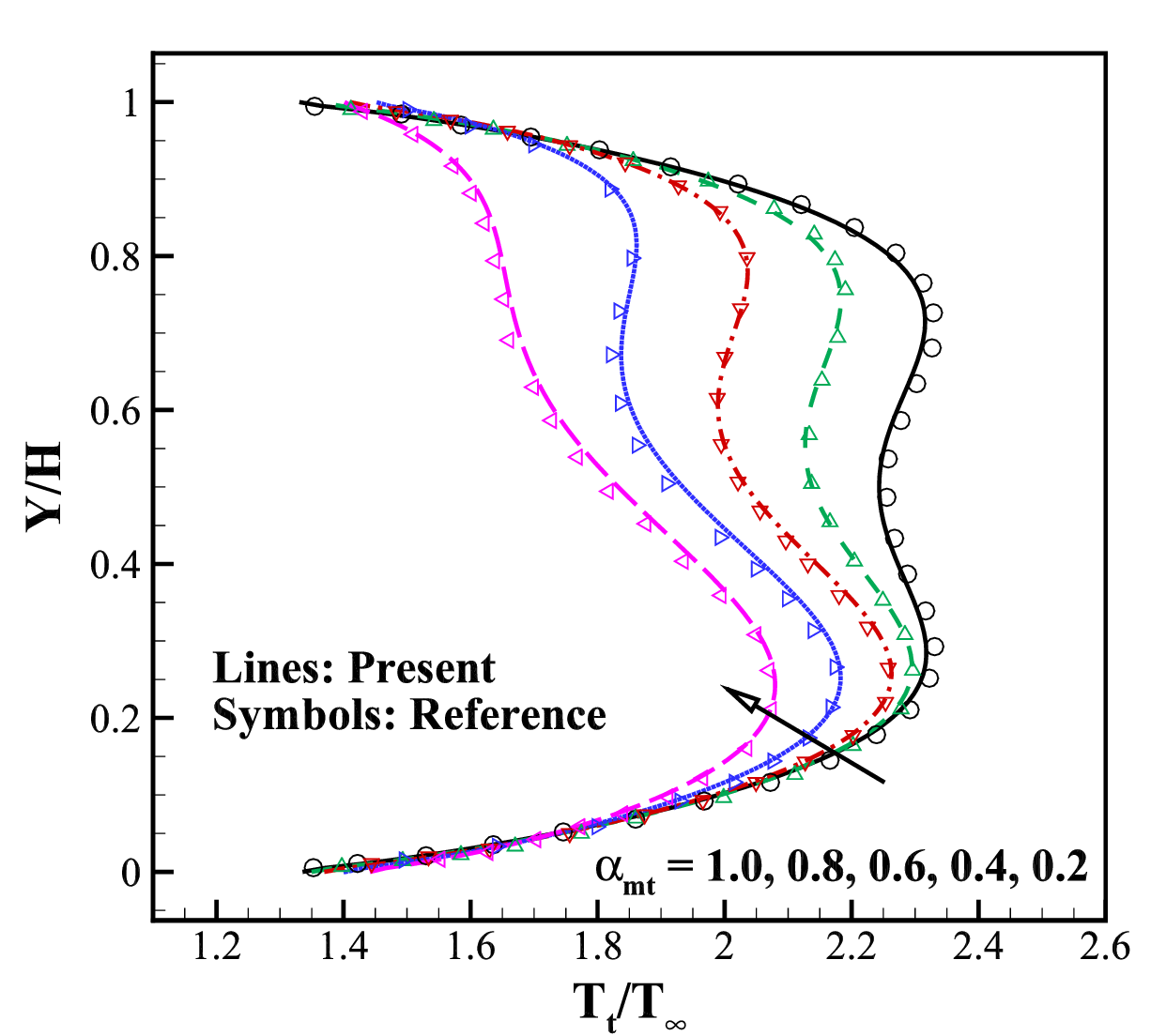}}
		\caption{\label{microchannel_X_Tt}{Translational temperature profiles along the microchannel at (a) X = 2.4 $\mu m$ and (b) X = 3.6 $\mu m$ with different $\alpha_{mt}$.}}
	\end{figure}
	
	\begin{figure}[!htp]
		\centering
		\includegraphics[width=0.6\textwidth]{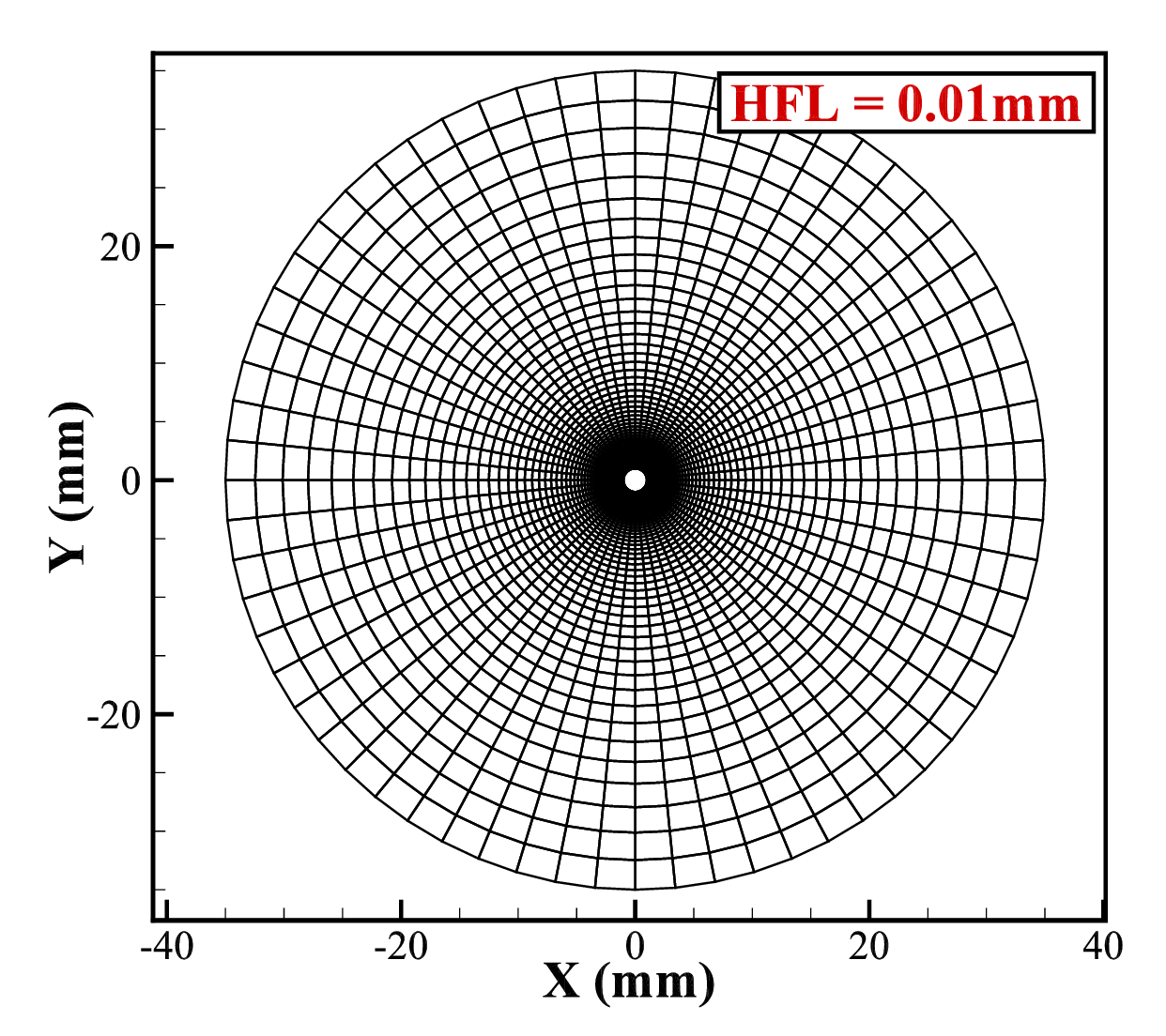}
		\caption{\label{cylinder_pmesh_10_10}{The physical mesh for the supersonic flow over a cylinder ($Ma = 10.0$, $Kn = 10.0$, $T_{\infty} = 273 K$, $T_{w} = 273 K$).}}
	\end{figure}
	
	\begin{figure}[!htp]
		\centering
		\subfigure[]{\label{cylinder_dvmesh_5}\includegraphics[width=0.32\textwidth]{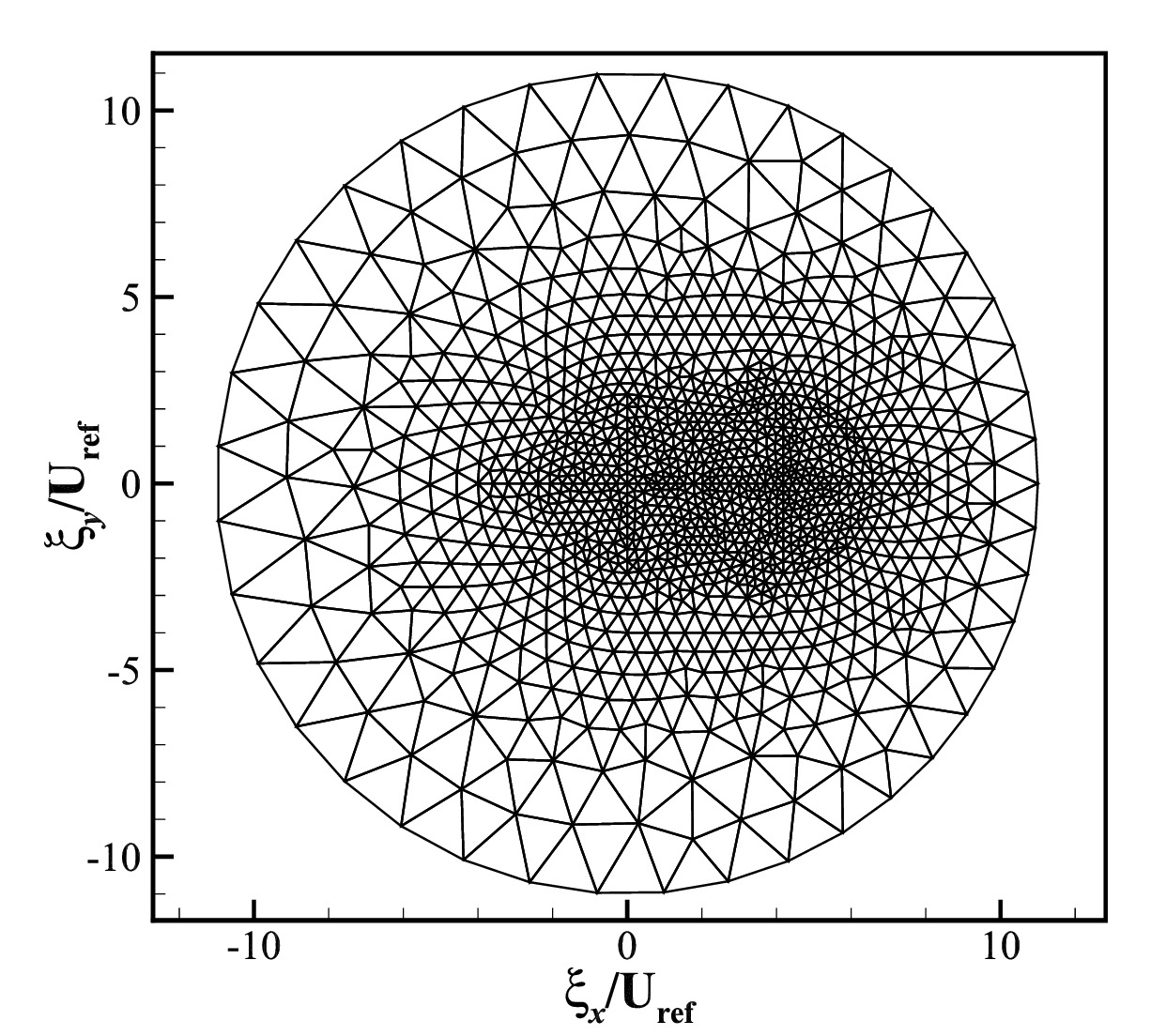}}
		\subfigure[]{\label{cylinder_dvmesh_10}\includegraphics[width=0.32\textwidth]{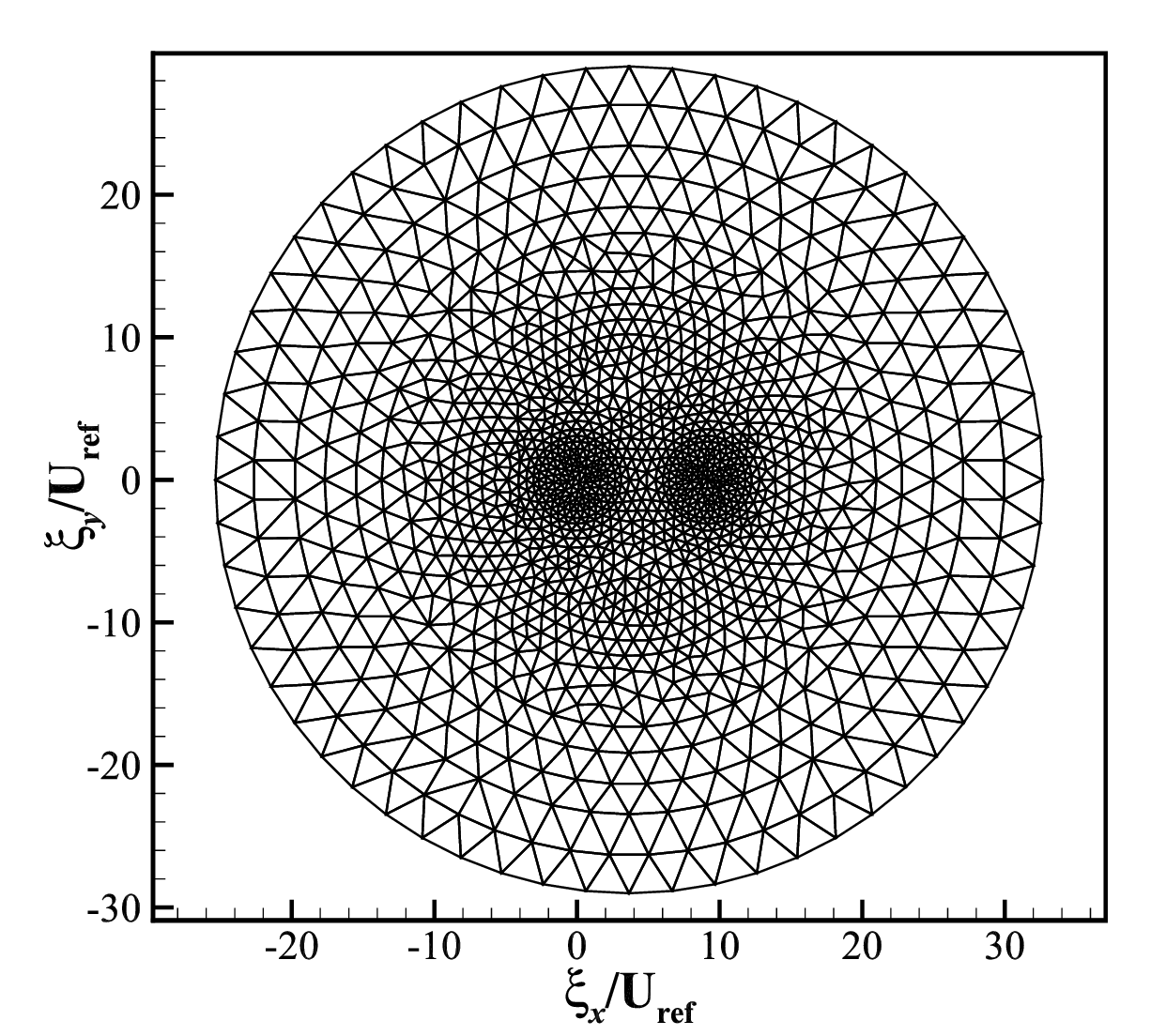}}
		\subfigure[]{\label{cylinder_dvmesh_15}\includegraphics[width=0.32\textwidth]{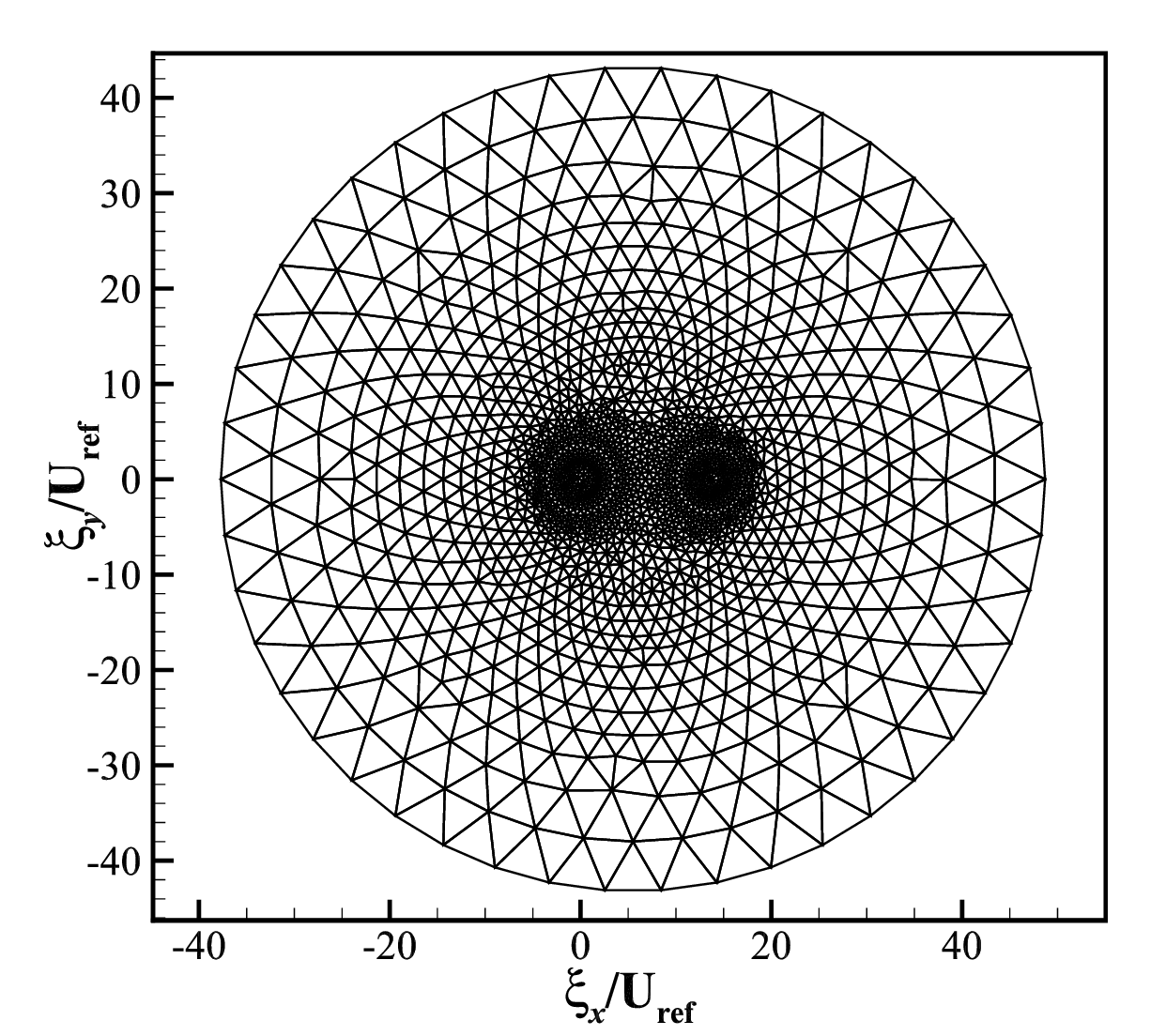}}
		\caption{\label{cylinder_dvmesh}{The velocity mesh for the supersonic flow over a cylinder: (a) Ma = 5.0, (b) Ma = 10.0, and (c) Ma = 15.0.}}
	\end{figure}
	
	\begin{figure}[!htp]
		\centering
		\subfigure[]{\label{cylinder_MAX_5_10_P}\includegraphics[width=0.32\textwidth]{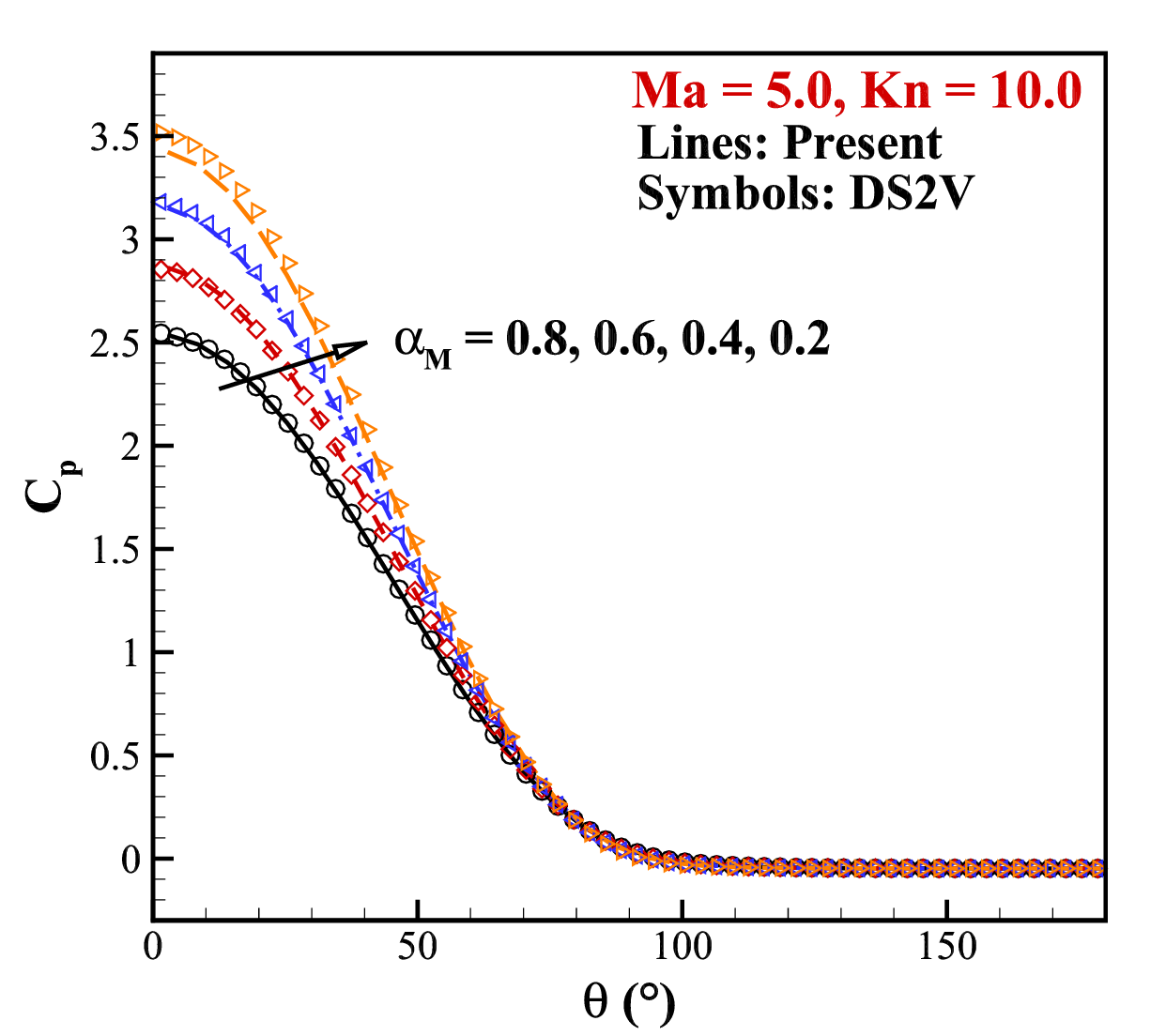}}
		\subfigure[]{\label{cylinder_MAX_5_10_S}\includegraphics[width=0.32\textwidth]{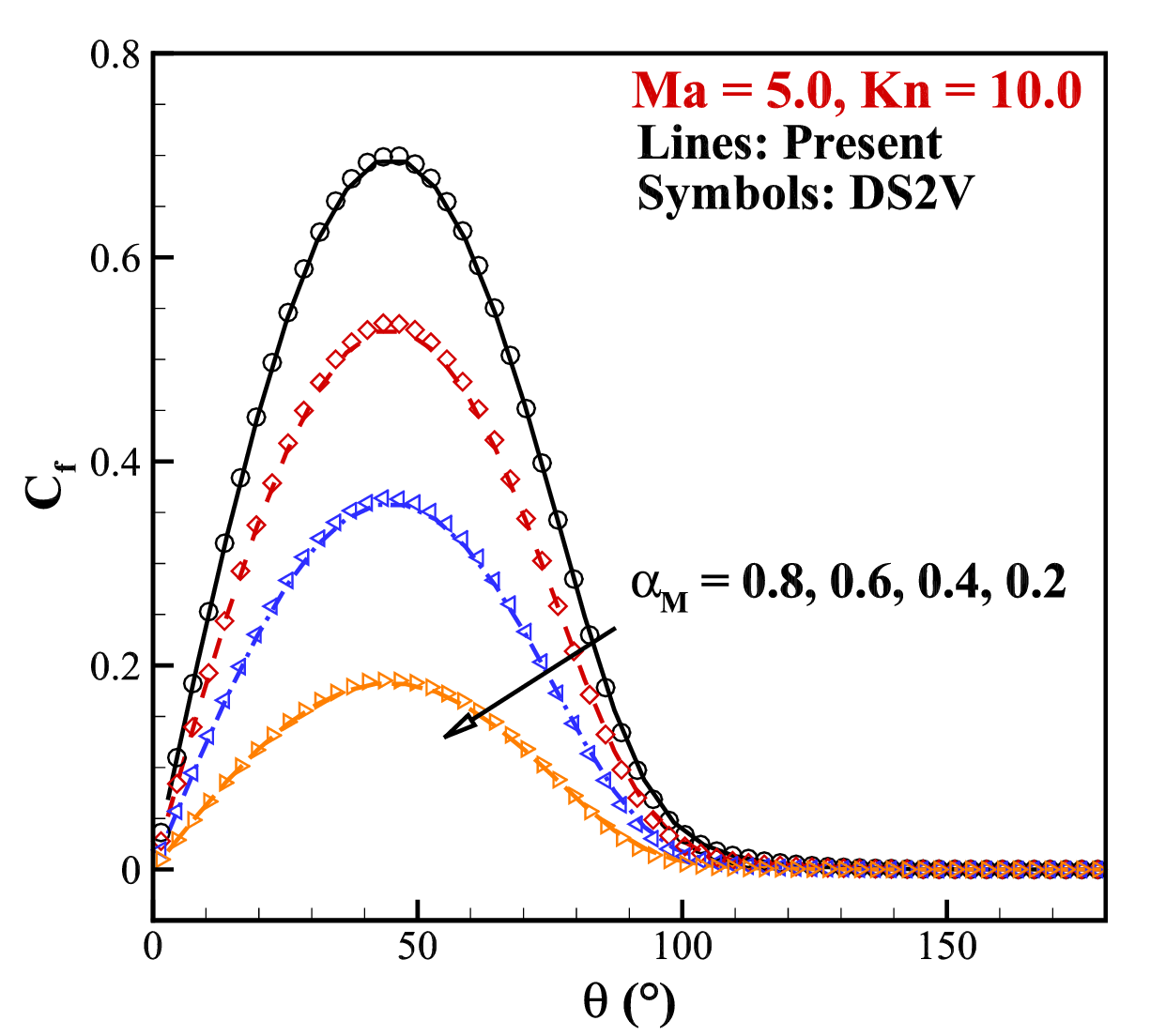}}
		\subfigure[]{\label{cylinder_MAX_5_10_H}\includegraphics[width=0.32\textwidth]{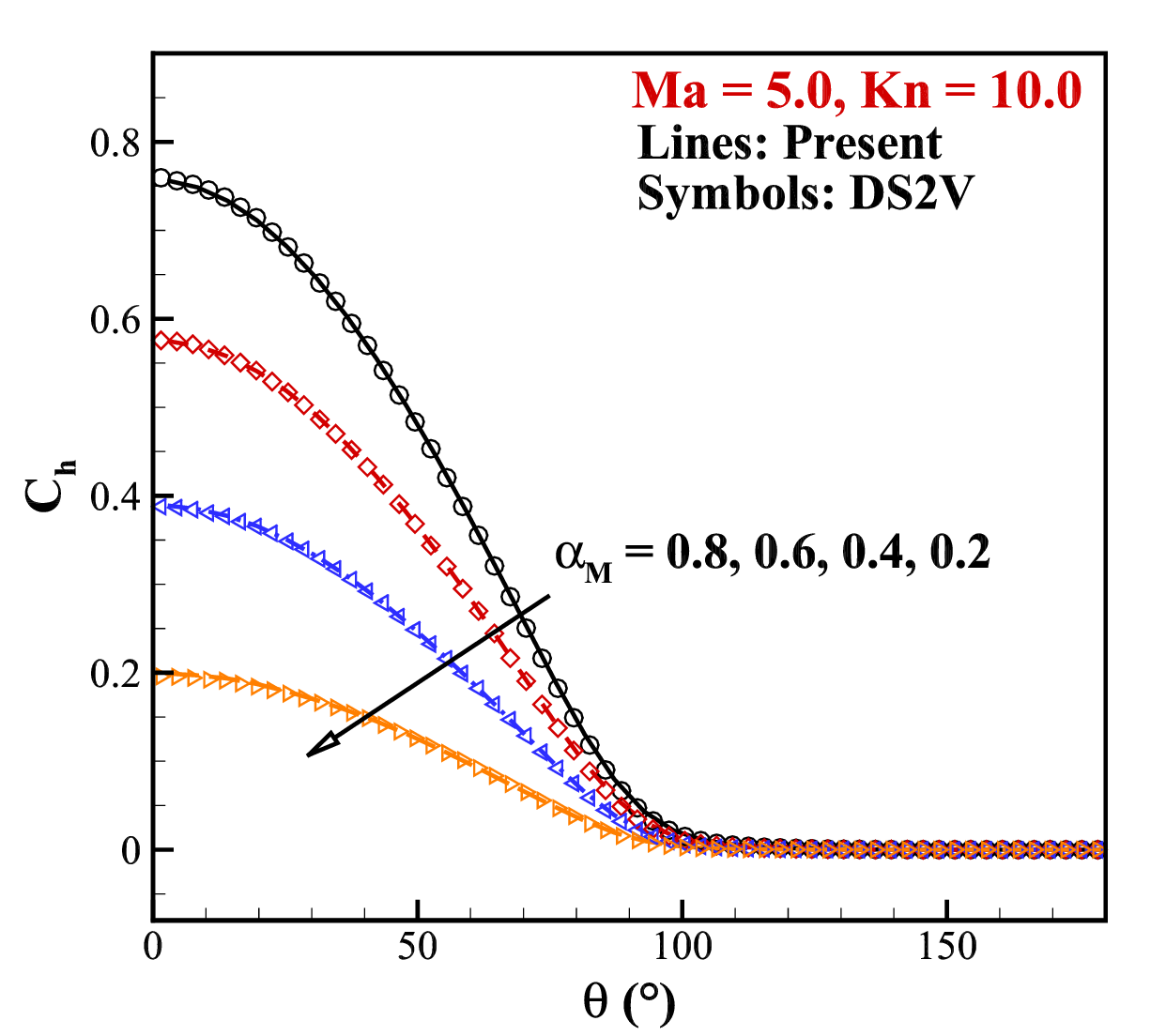}}
		\caption{\label{cylinder_MAX_5_10}{Comparison of the (a) pressure coefficient, (b) skin friction coefficient, and (c) heat transfer coefficient on the surface of cylinder with different $\alpha_{M}$ when employing the Maxwell boundary ($Ma = 5.0$, $Kn = 10.0$, $T_{\infty} = 273 K$, $T_{w} = 273 K$).}}
	\end{figure}
	
	\begin{figure}[!htp]
		\centering
		\subfigure[]{\label{cylinder_MAX_5_1_P}\includegraphics[width=0.32\textwidth]{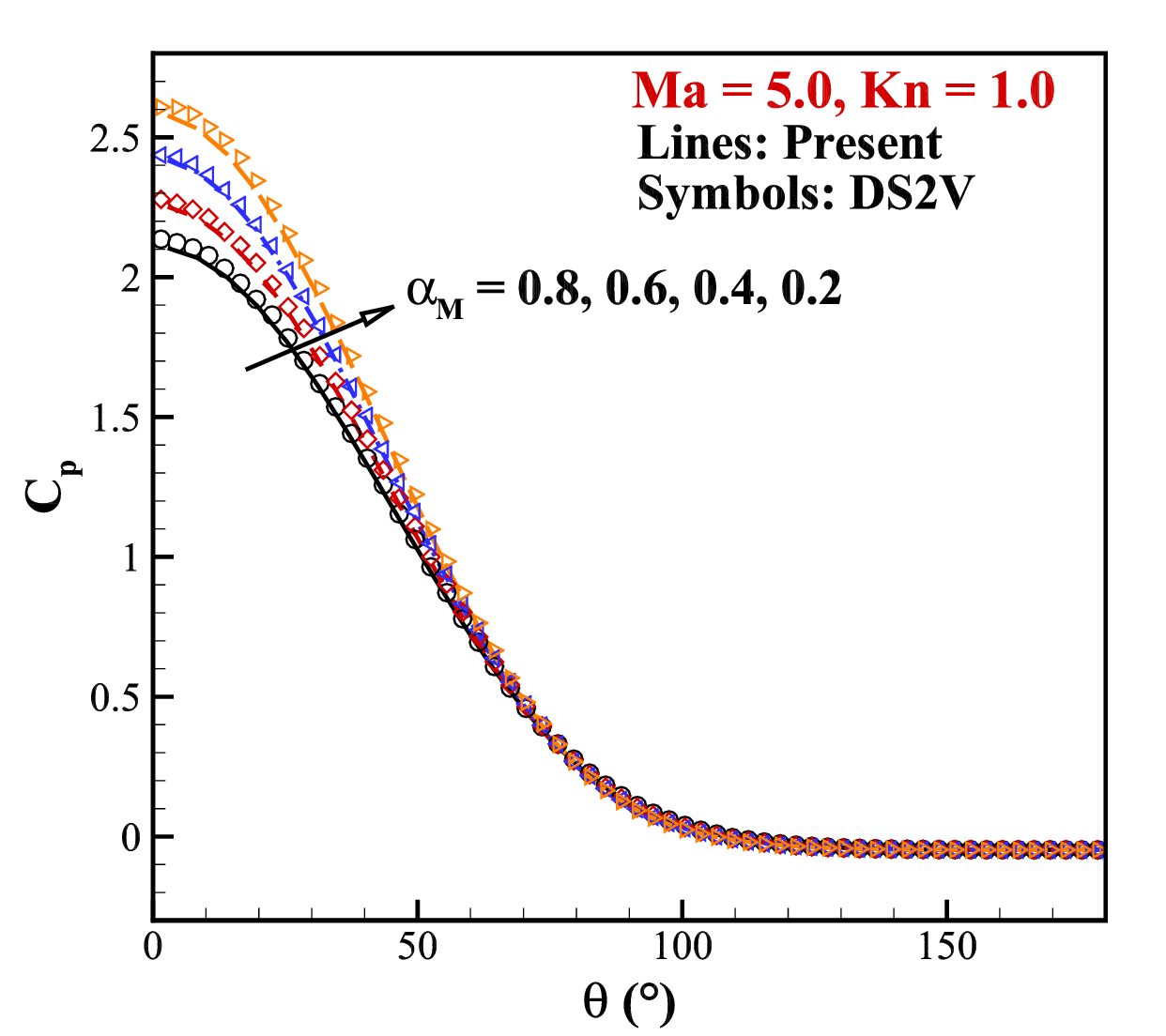}}
		\subfigure[]{\label{cylinder_MAX_5_1_S}\includegraphics[width=0.32\textwidth]{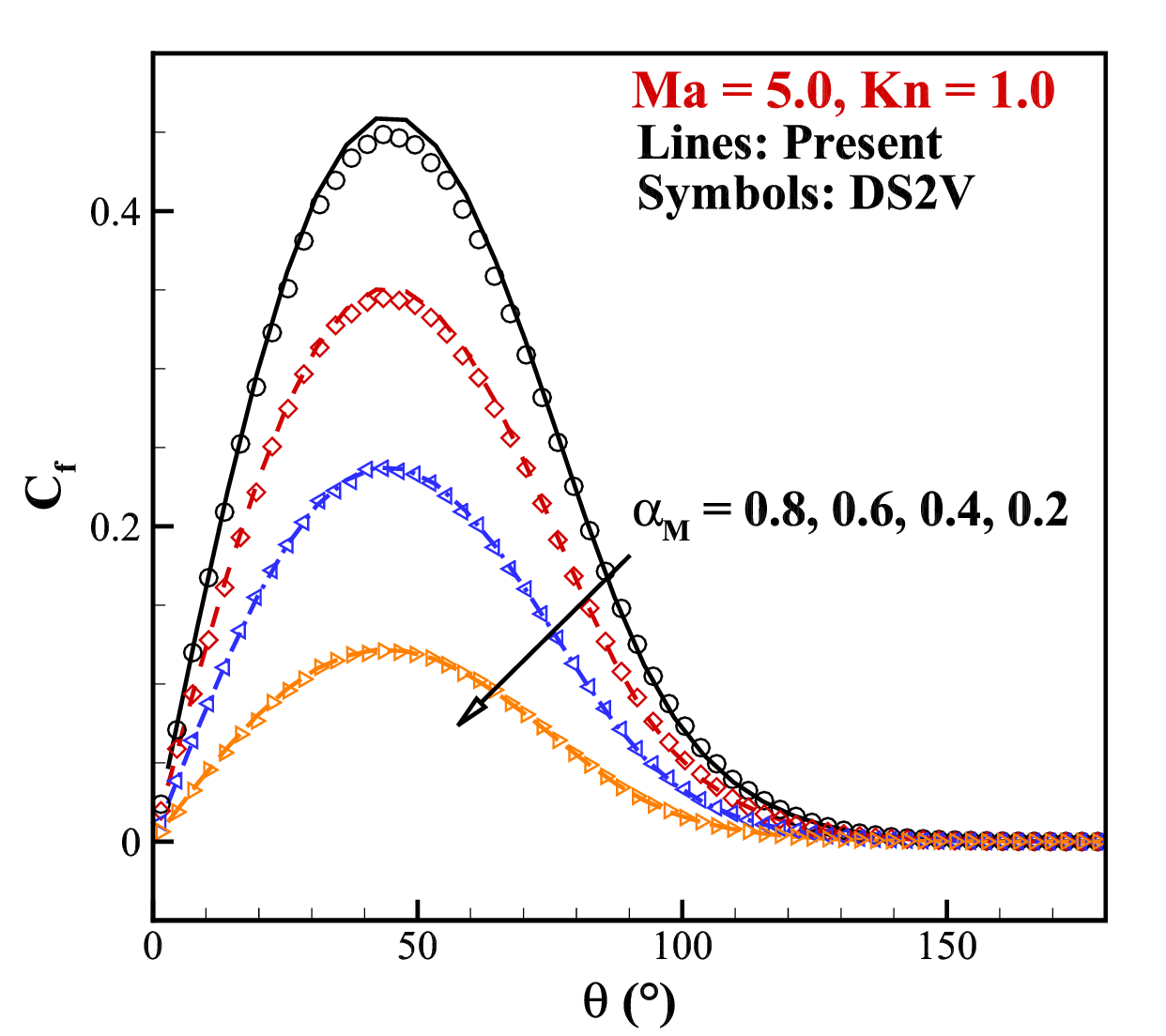}}
		\subfigure[]{\label{cylinder_MAX_5_1_H}\includegraphics[width=0.32\textwidth]{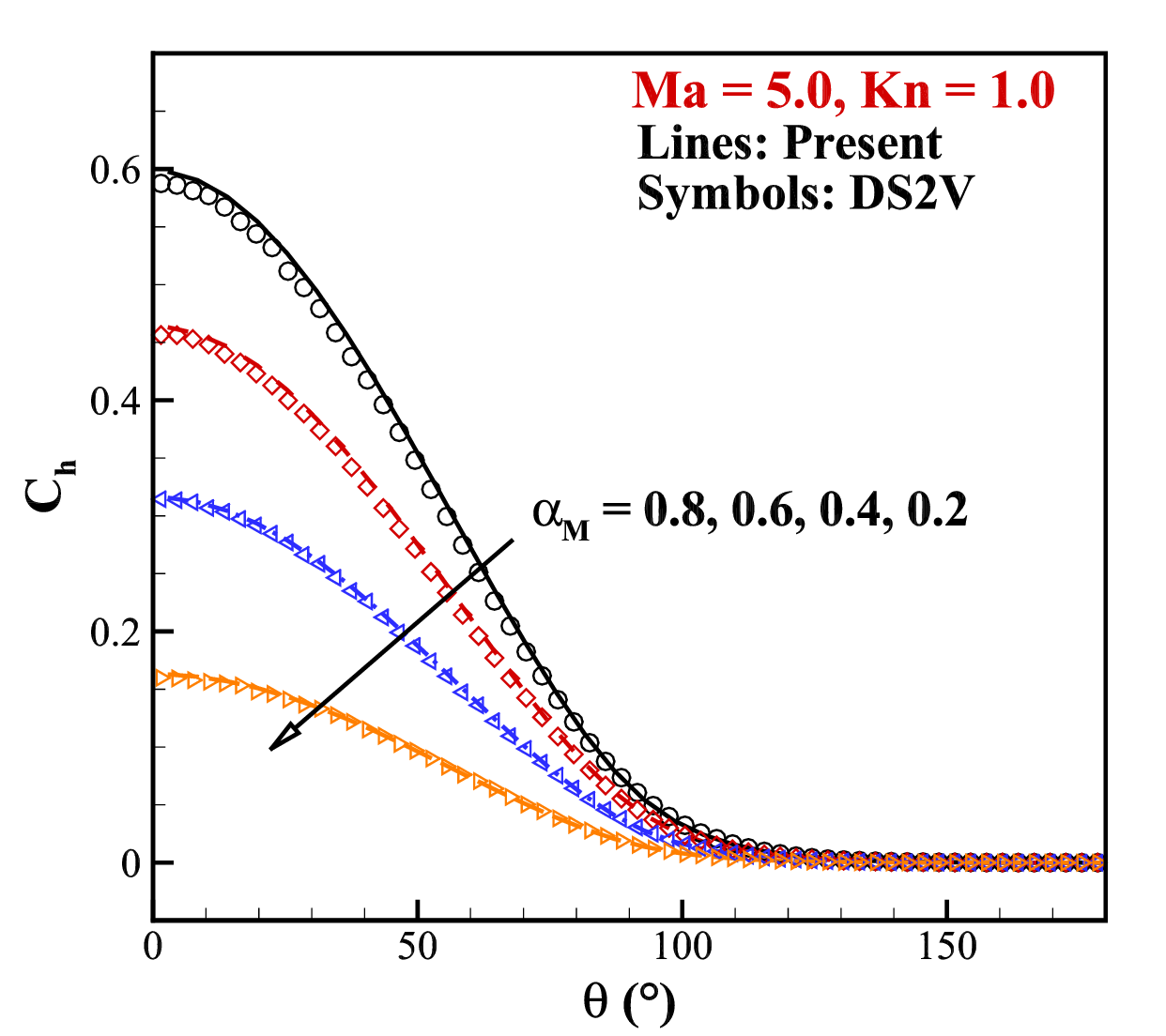}}
		\caption{\label{cylinder_MAX_5_1}{Comparison of the (a) pressure coefficient, (b) skin friction coefficient, and (c) heat transfer coefficient on the surface of cylinder with different $\alpha_{M}$ when employing the Maxwell boundary ($Ma = 5.0$, $Kn = 1.0$, $T_{\infty} = 273 K$, $T_{w} = 273 K$).}}
	\end{figure}
	
	\begin{figure}[!htp]
		\centering
		\subfigure[]{\label{cylinder_MAX_5_01_P}\includegraphics[width=0.32\textwidth]{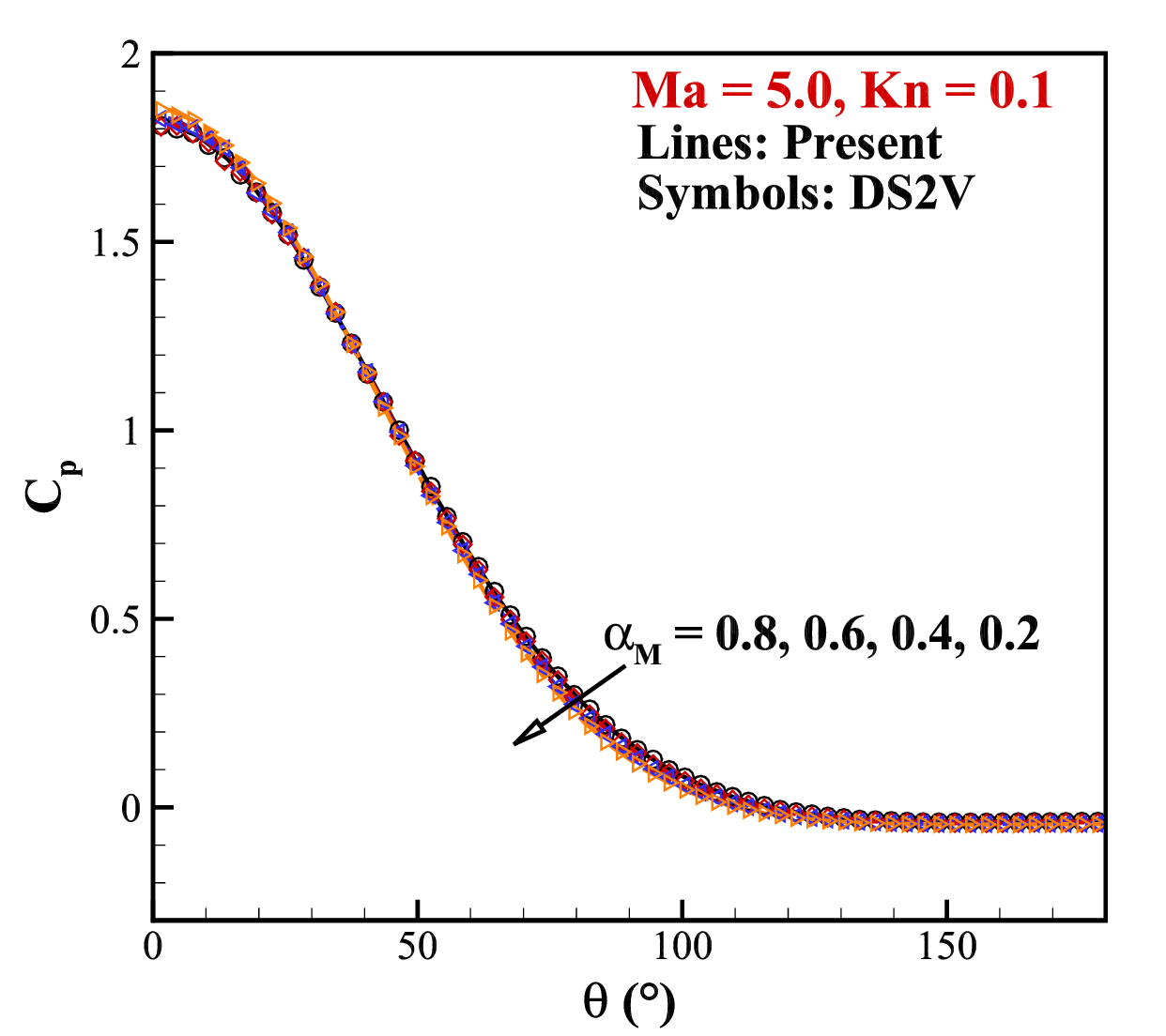}}
		\subfigure[]{\label{cylinder_MAX_5_01_S}\includegraphics[width=0.32\textwidth]{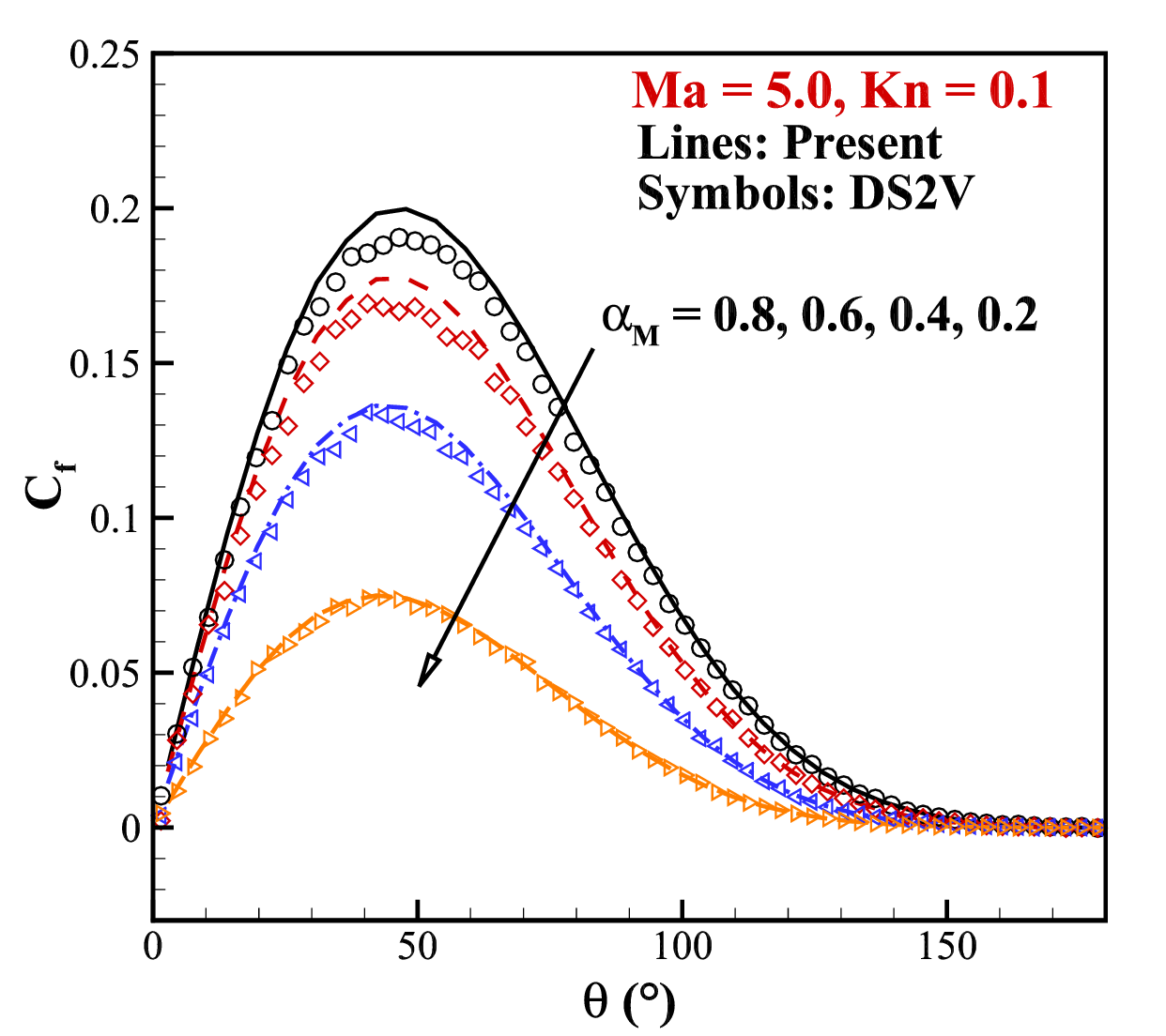}}
		\subfigure[]{\label{cylinder_MAX_5_01_H}\includegraphics[width=0.32\textwidth]{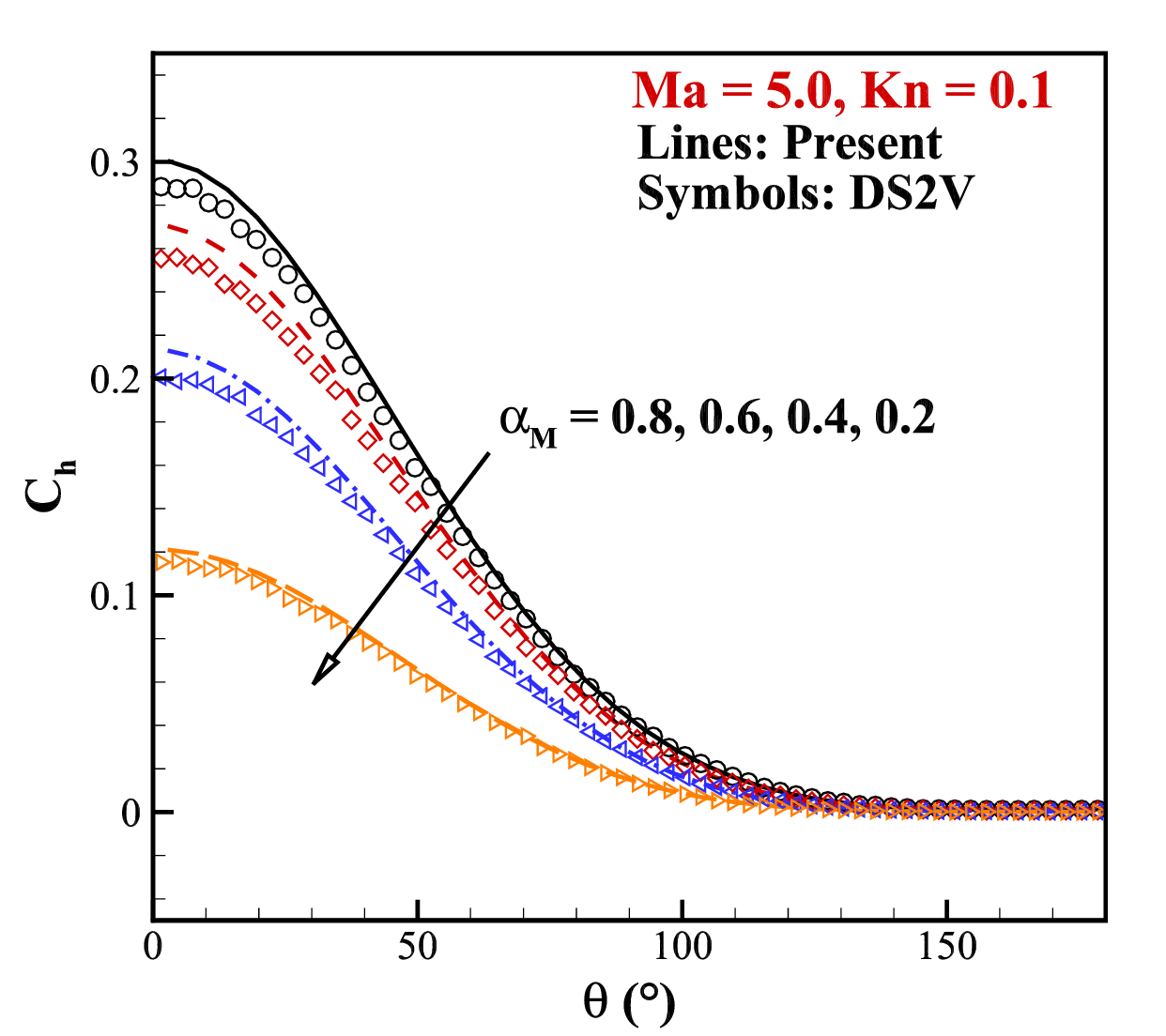}}
		\caption{\label{cylinder_MAX_5_01}{Comparison of the (a) pressure coefficient, (b) skin friction coefficient, and (c) heat transfer coefficient on the surface of cylinder with different $\alpha_{M}$ when employing the Maxwell boundary ($Ma = 5.0$, $Kn = 0.1$, $T_{\infty} = 273 K$, $T_{w} = 273 K$).}}
	\end{figure}
	
	\begin{figure}[!htp]
		\centering
		\subfigure[]{\label{cylinder_MAX_10_10_P}\includegraphics[width=0.32\textwidth]{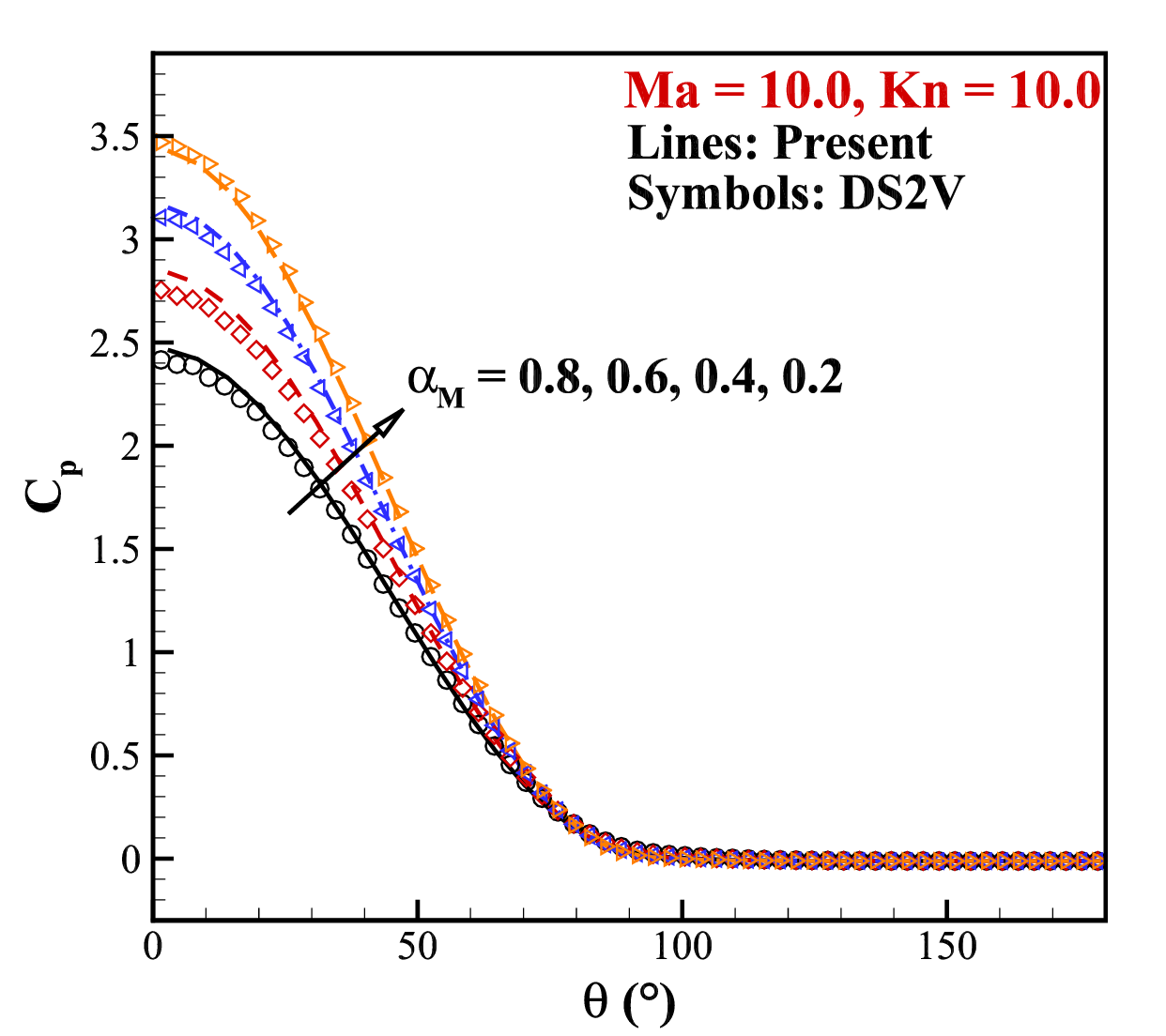}}
		\subfigure[]{\label{cylinder_MAX_10_10_S}\includegraphics[width=0.32\textwidth]{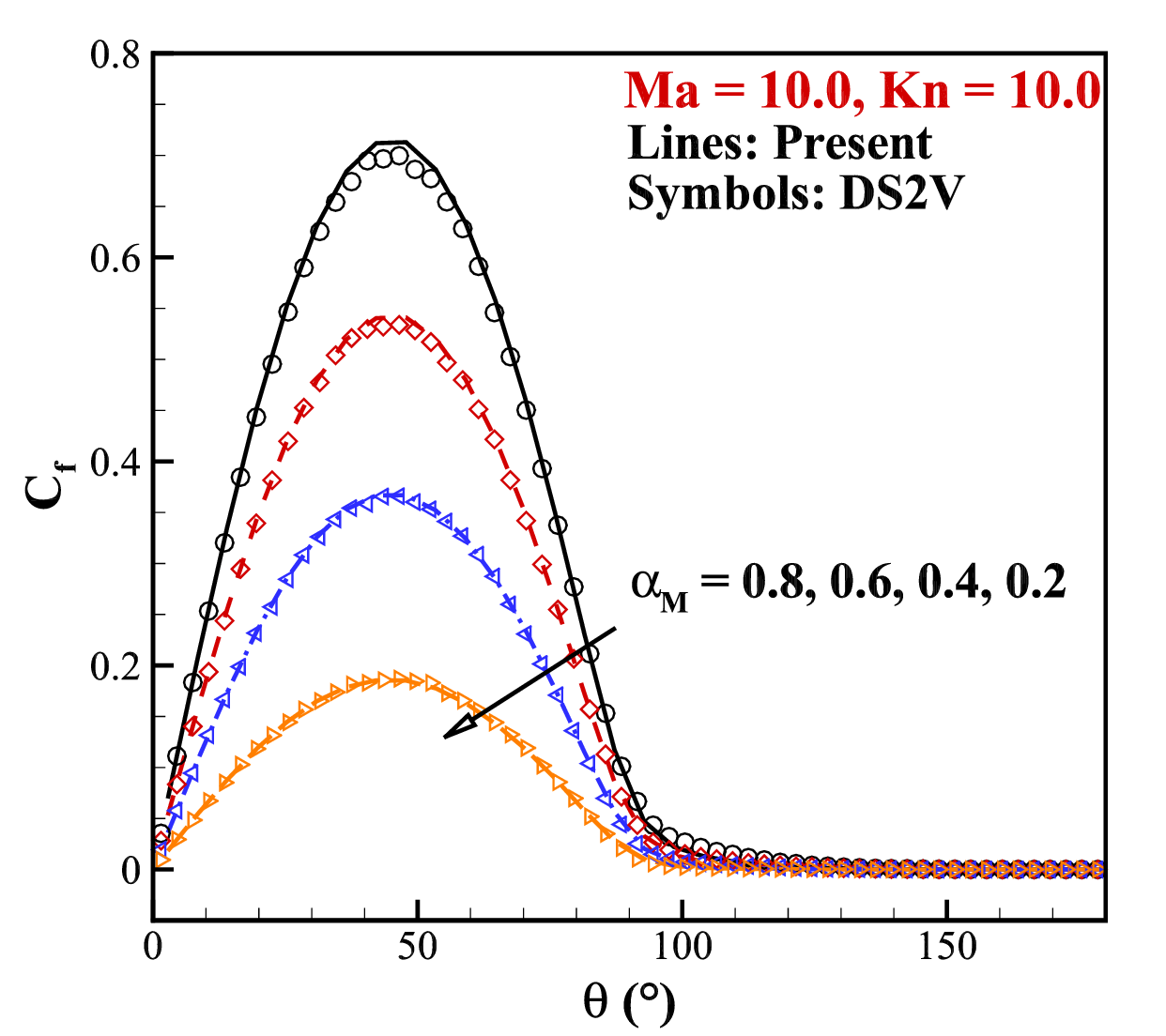}}
		\subfigure[]{\label{cylinder_MAX_10_10_H}\includegraphics[width=0.32\textwidth]{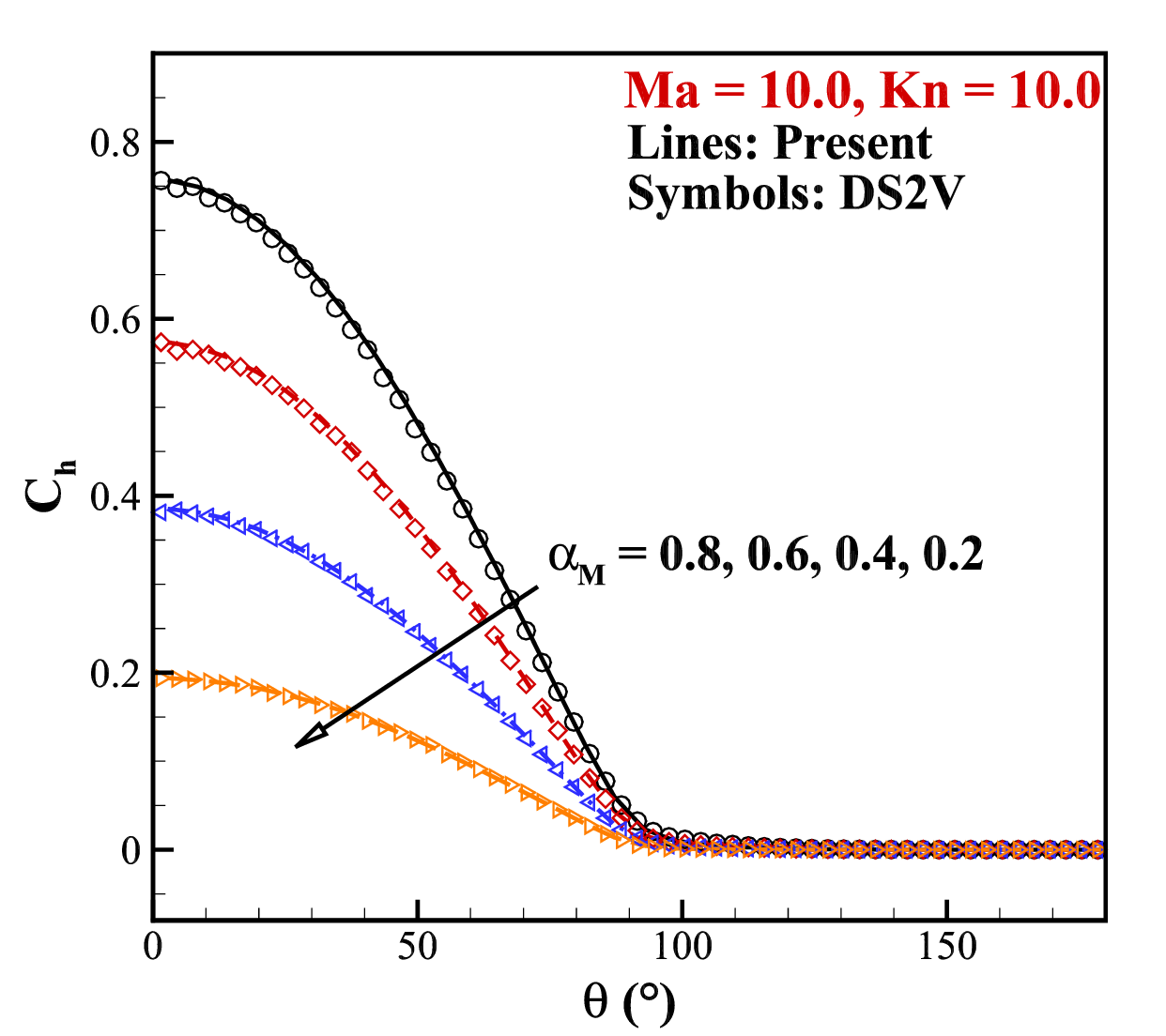}}
		\caption{\label{cylinder_MAX_10_10}{Comparison of the (a) pressure coefficient, (b) skin friction coefficient, and (c) heat transfer coefficient on the surface of cylinder with different $\alpha_{M}$ when employing the Maxwell boundary ($Ma = 10.0$, $Kn = 10.0$, $T_{\infty} = 273 K$, $T_{w} = 273 K$).}}
	\end{figure}
	
	\begin{figure}[!htp]
		\centering
		\subfigure[]{\label{cylinder_MAX_10_1_P}\includegraphics[width=0.32\textwidth]{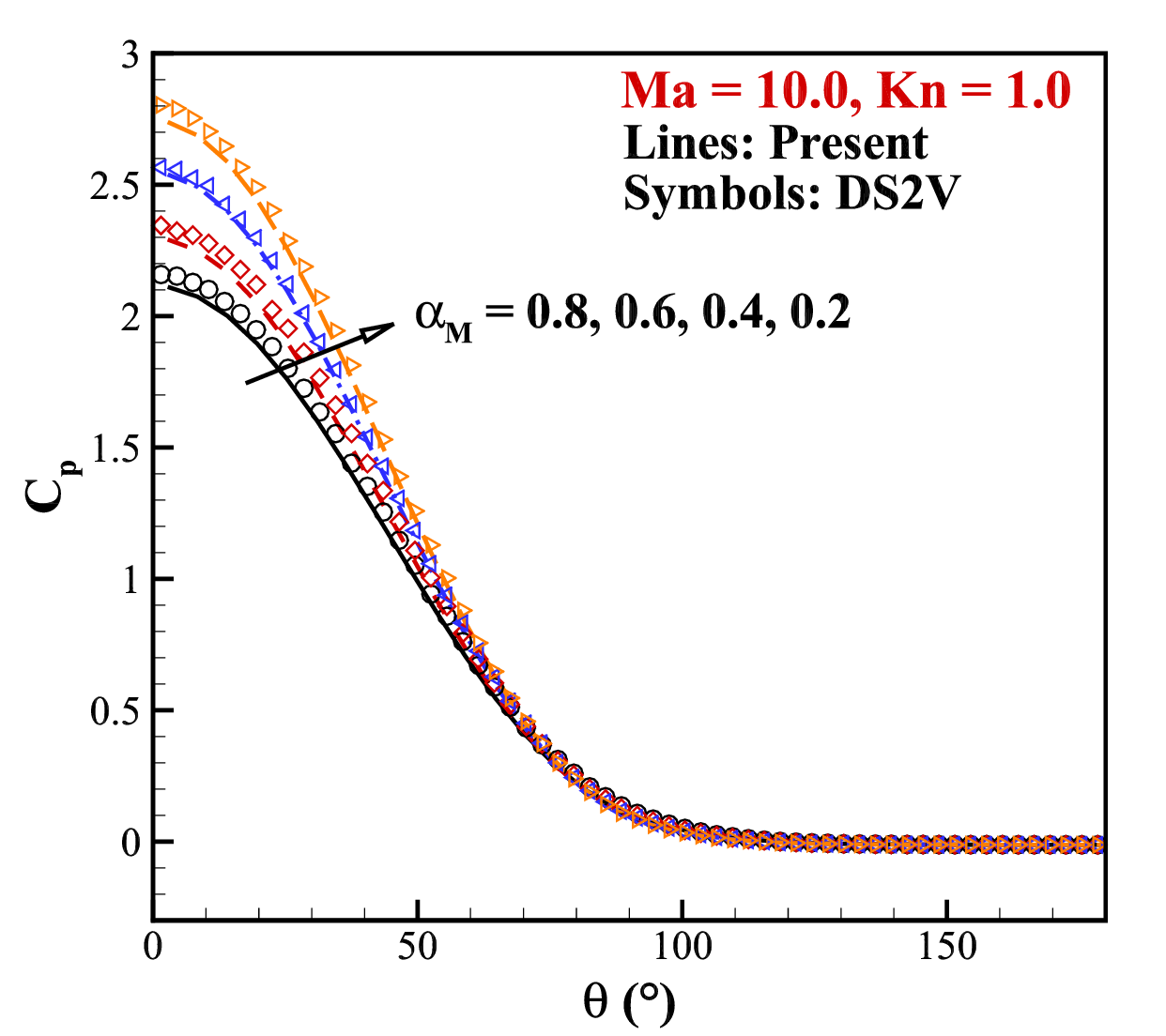}}
		\subfigure[]{\label{cylinder_MAX_10_1_S}\includegraphics[width=0.32\textwidth]{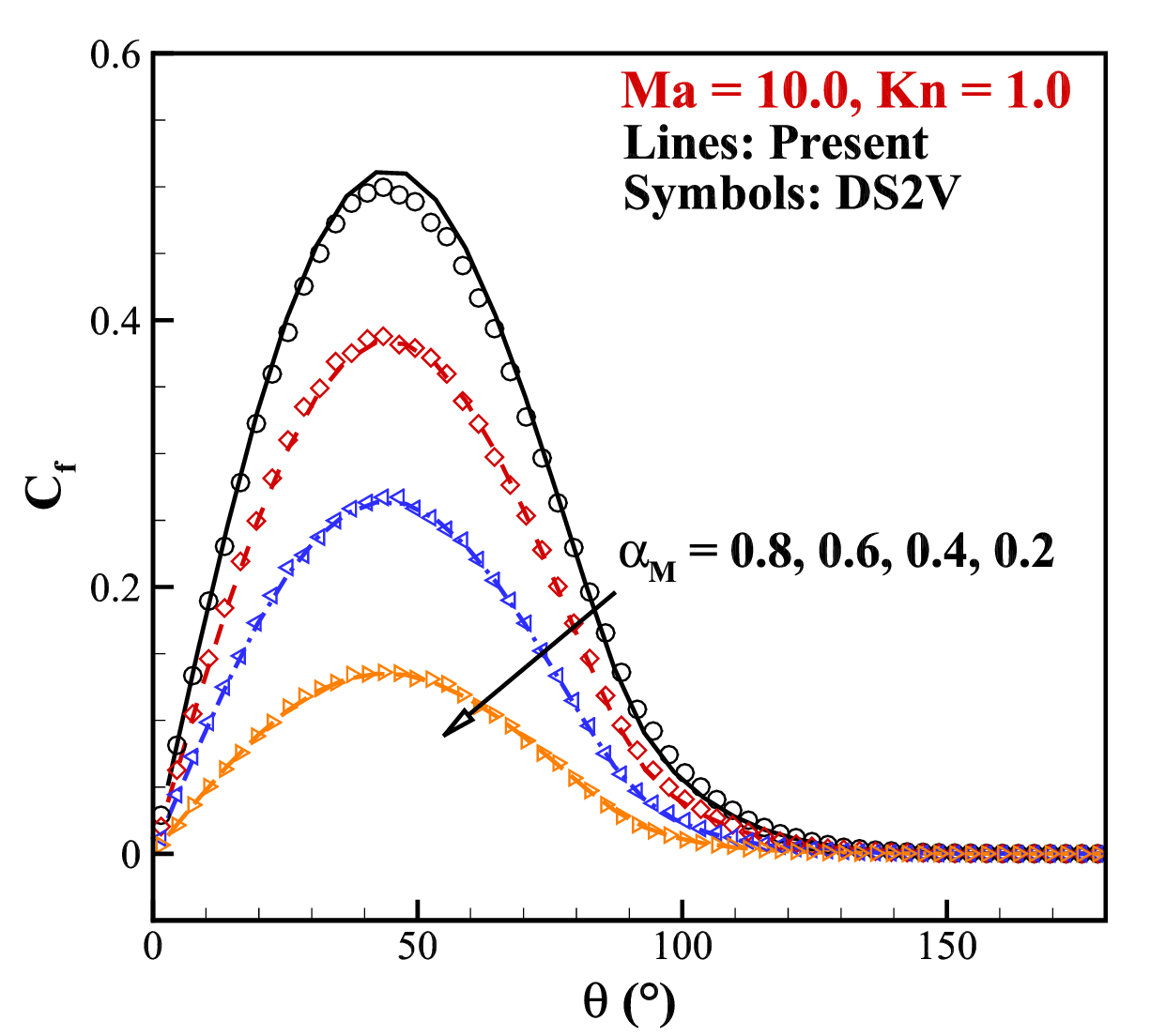}}
		\subfigure[]{\label{cylinder_MAX_10_1_H}\includegraphics[width=0.32\textwidth]{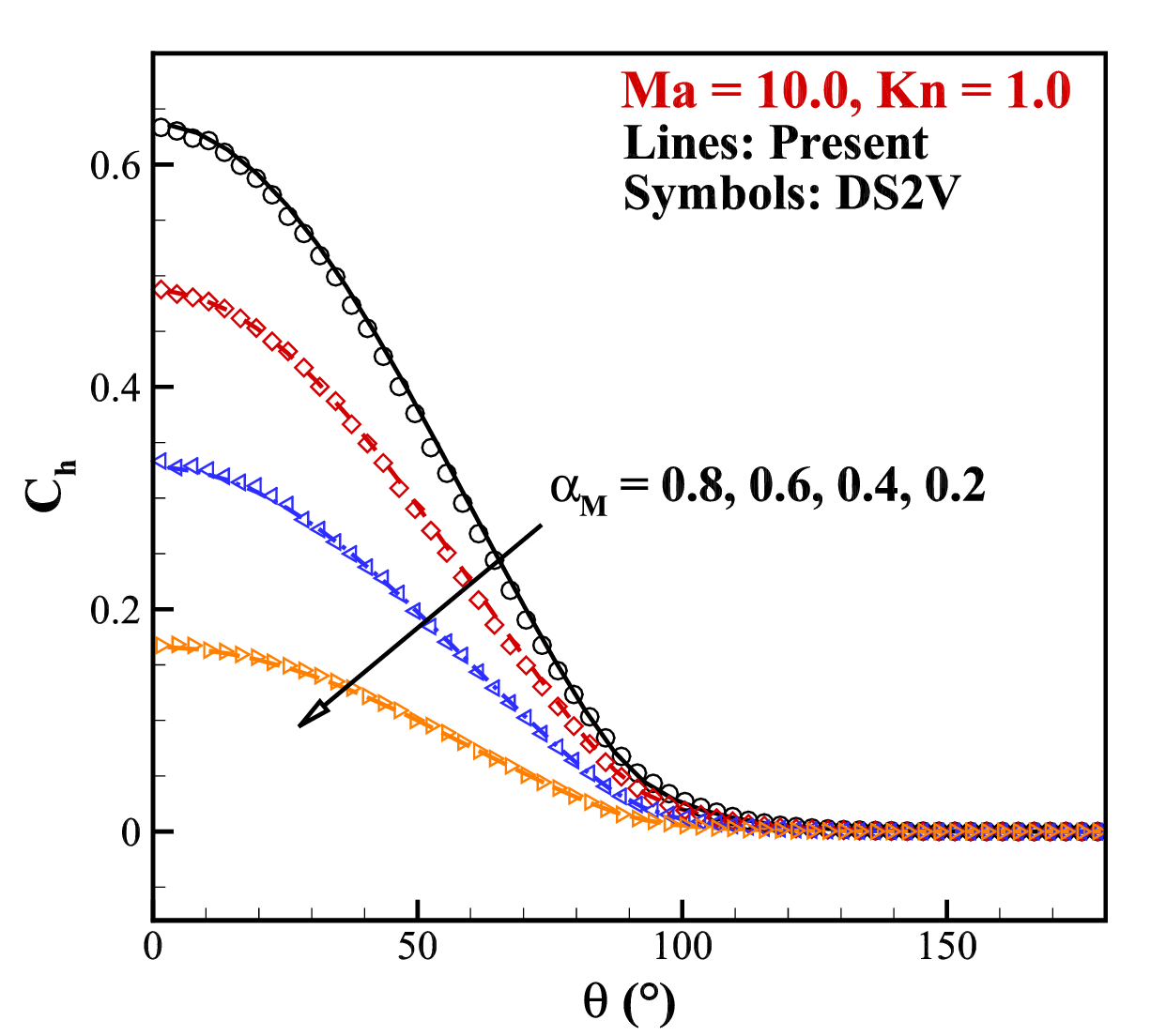}}
		\caption{\label{cylinder_MAX_10_1}{Comparison of the (a) pressure coefficient, (b) skin friction coefficient, and (c) heat transfer coefficient on the surface of cylinder with different $\alpha_{M}$ when employing the Maxwell boundary ($Ma = 10.0$, $Kn = 1.0$, $T_{\infty} = 273 K$, $T_{w} = 273 K$).}}
	\end{figure}
	
	\begin{figure}[!htp]
		\centering
		\subfigure[]{\label{cylinder_MAX_10_01_P}\includegraphics[width=0.32\textwidth]{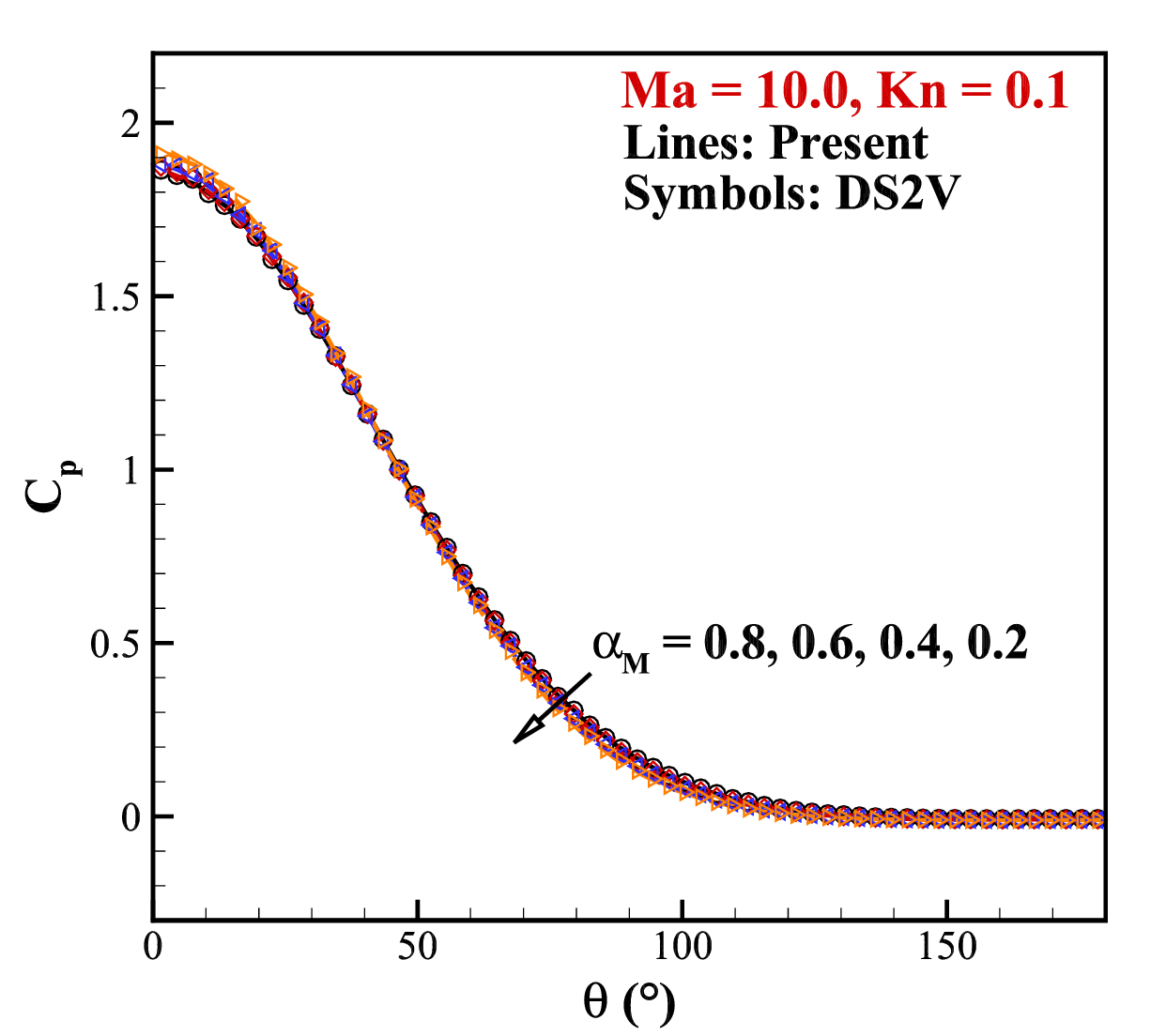}}
		\subfigure[]{\label{cylinder_MAX_10_01_S}\includegraphics[width=0.32\textwidth]{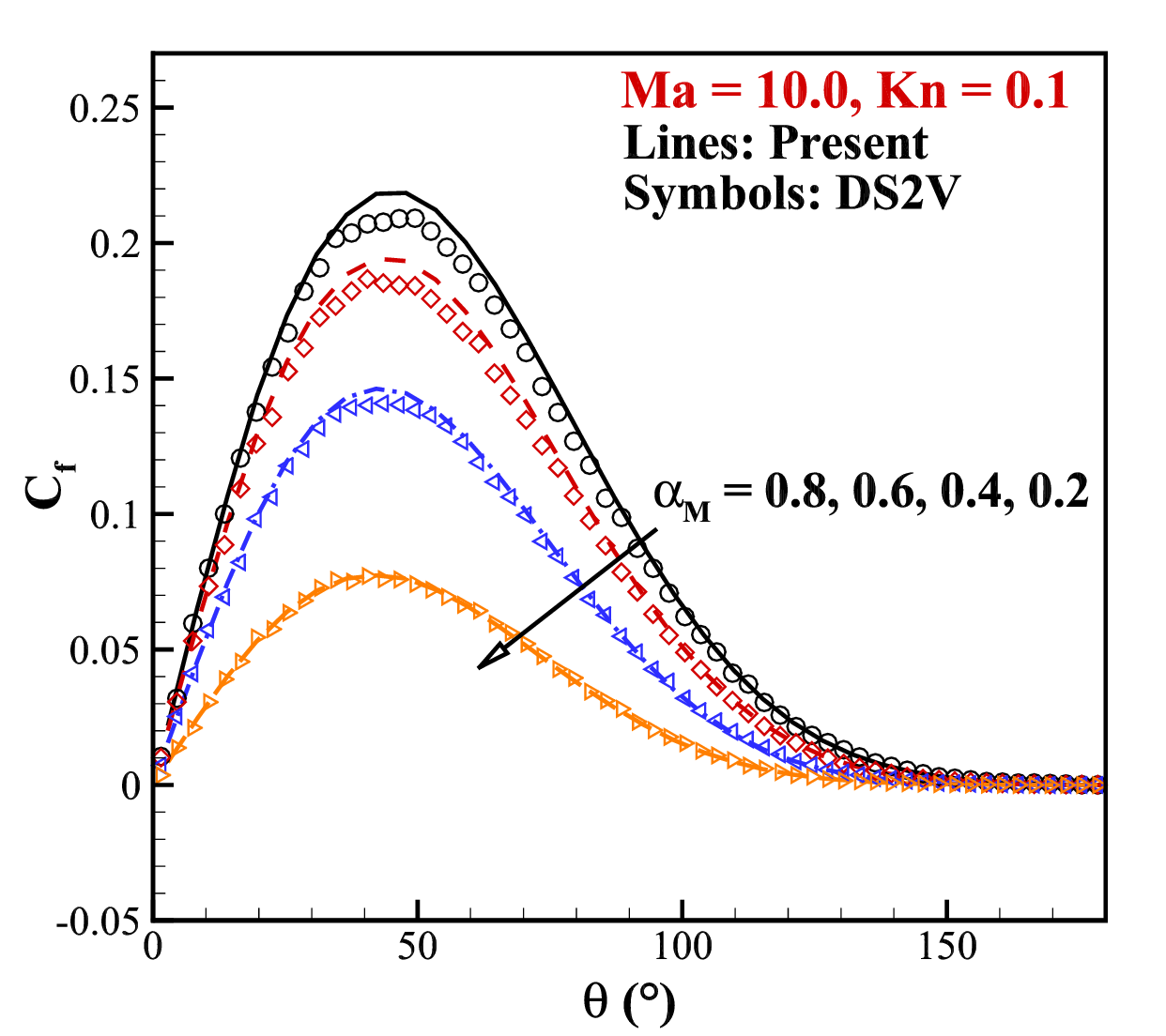}}
		\subfigure[]{\label{cylinder_MAX_10_01_H}\includegraphics[width=0.32\textwidth]{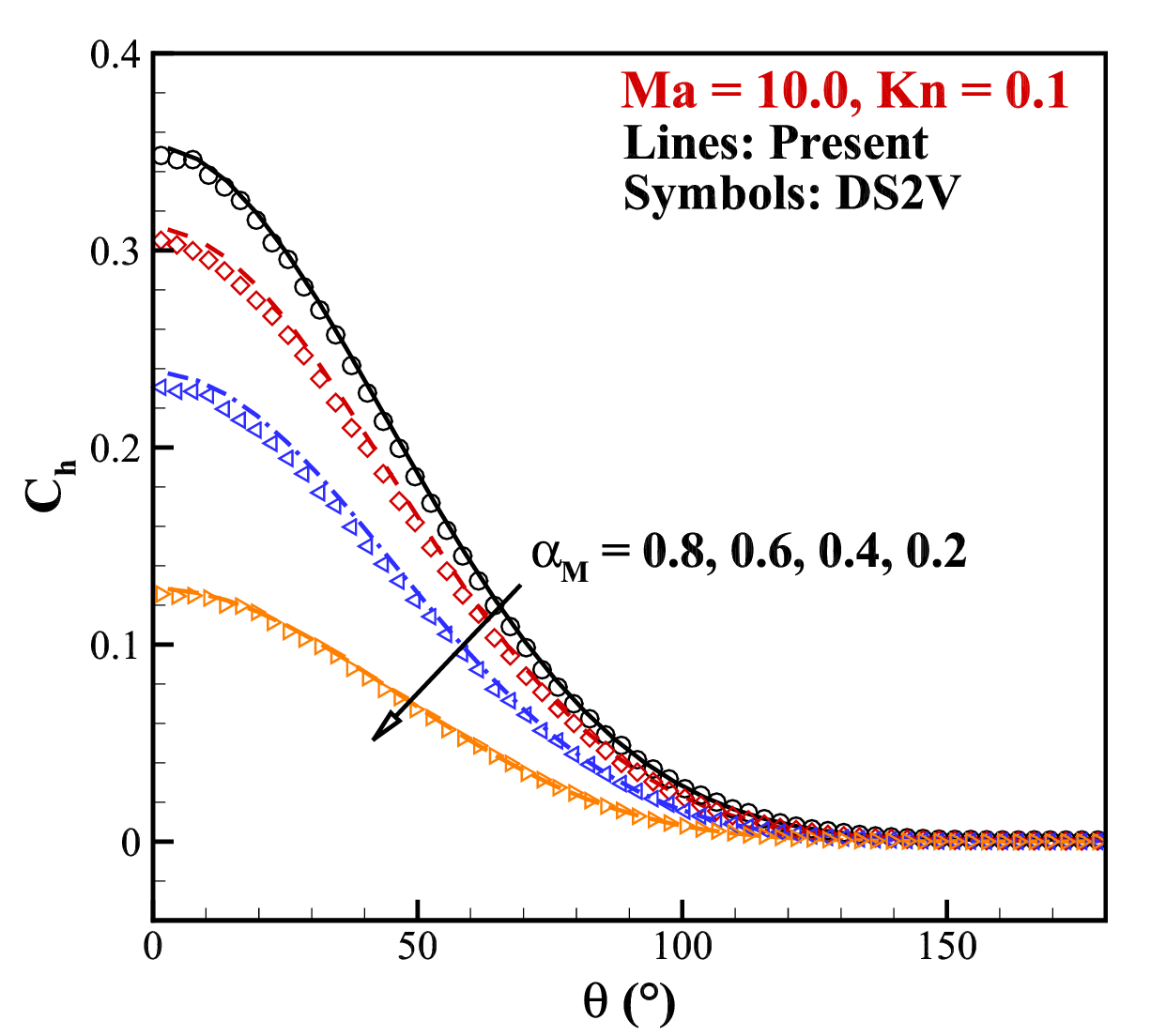}}
		\caption{\label{cylinder_MAX_10_01}{Comparison of the (a) pressure coefficient, (b) skin friction coefficient, and (c) heat transfer coefficient on the surface of cylinder with different $\alpha_{M}$ when employing the Maxwell boundary ($Ma = 10.0$, $Kn = 0.1$, $T_{\infty} = 273 K$, $T_{w} = 273 K$).}}
	\end{figure}
	
	\begin{figure}[!htp]
		\centering
		\subfigure[]{\label{cylinder_MAX_15_10_P}\includegraphics[width=0.32\textwidth]{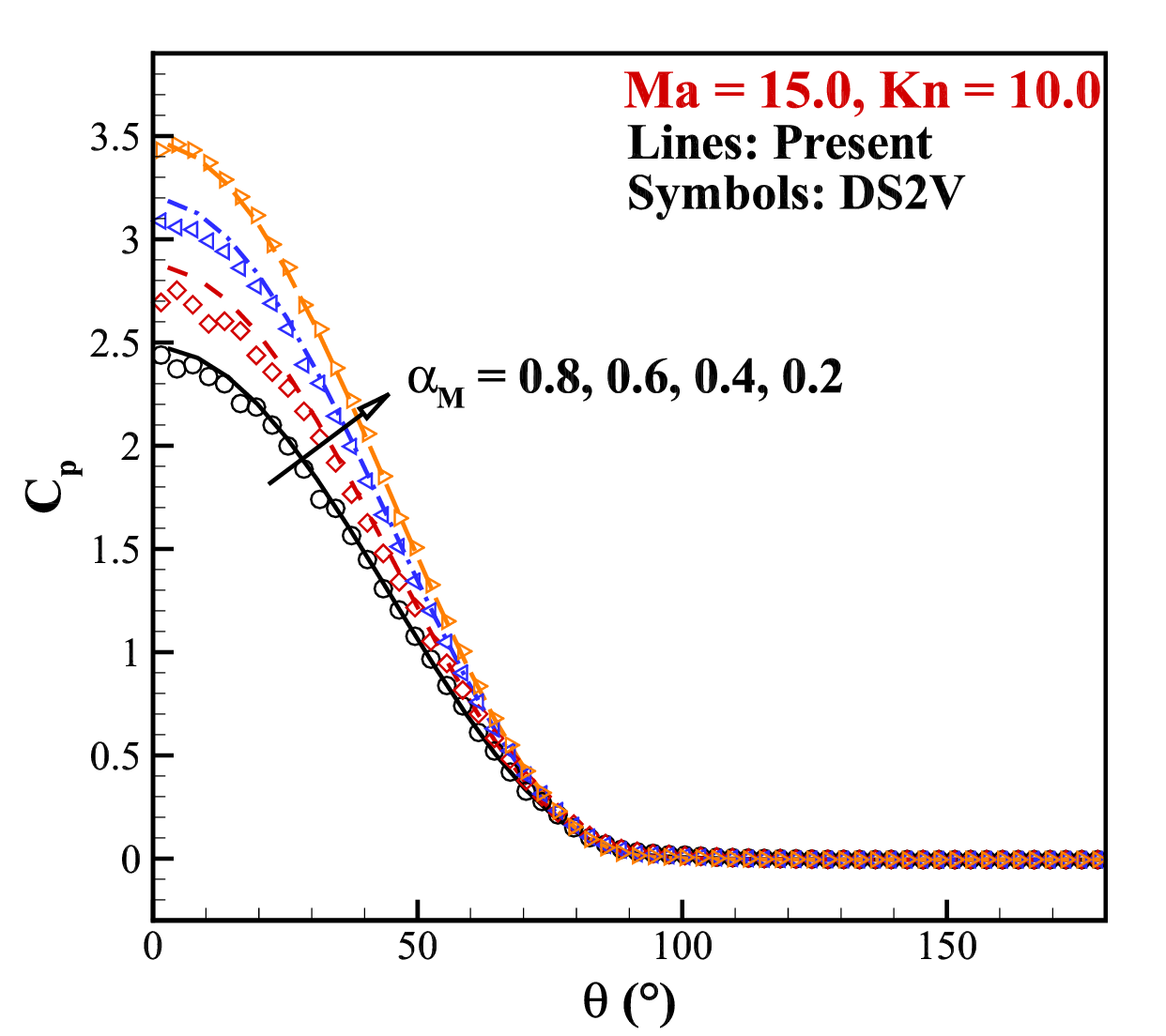}}
		\subfigure[]{\label{cylinder_MAX_15_10_S}\includegraphics[width=0.32\textwidth]{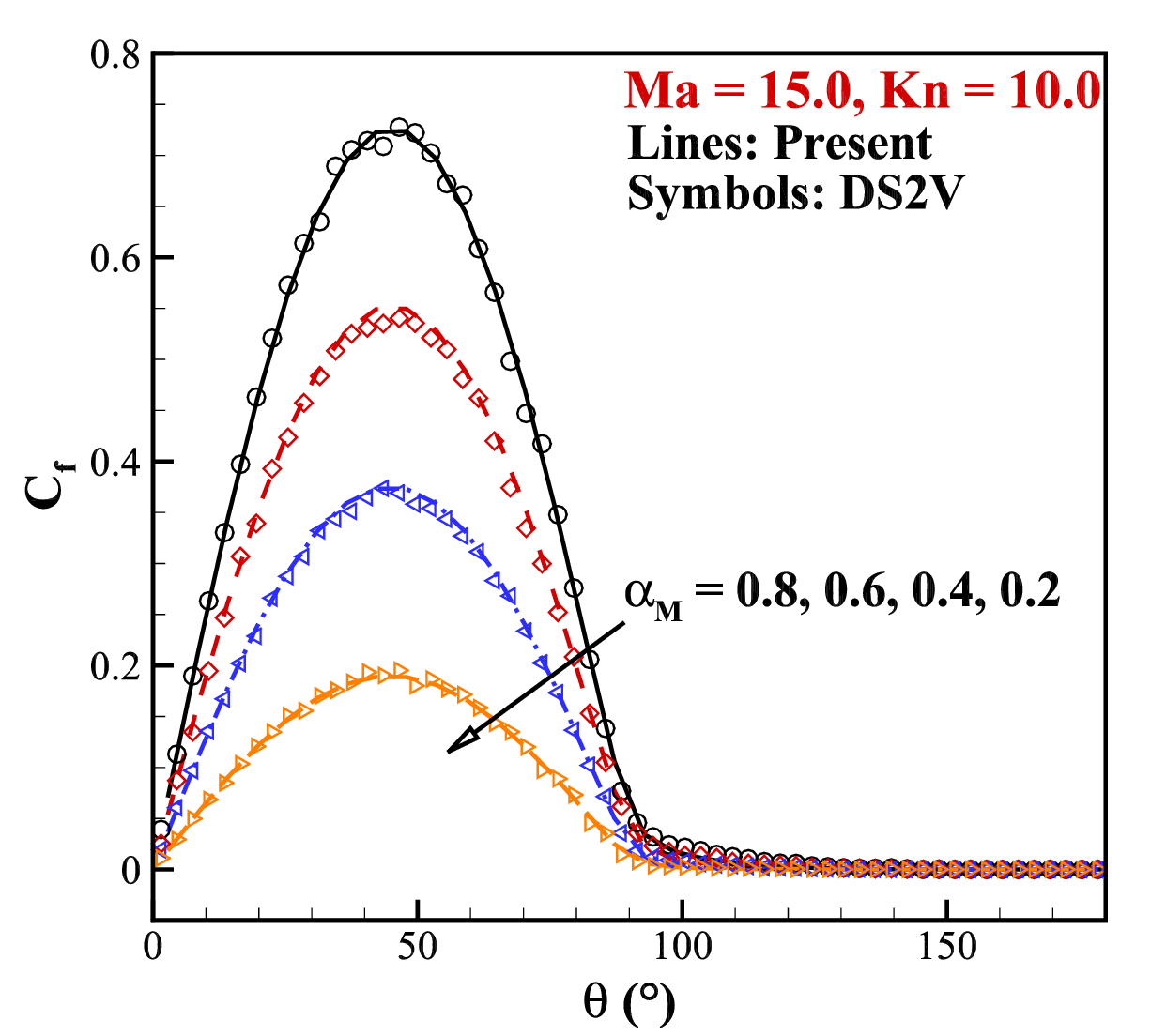}}
		\subfigure[]{\label{cylinder_MAX_15_10_H}\includegraphics[width=0.32\textwidth]{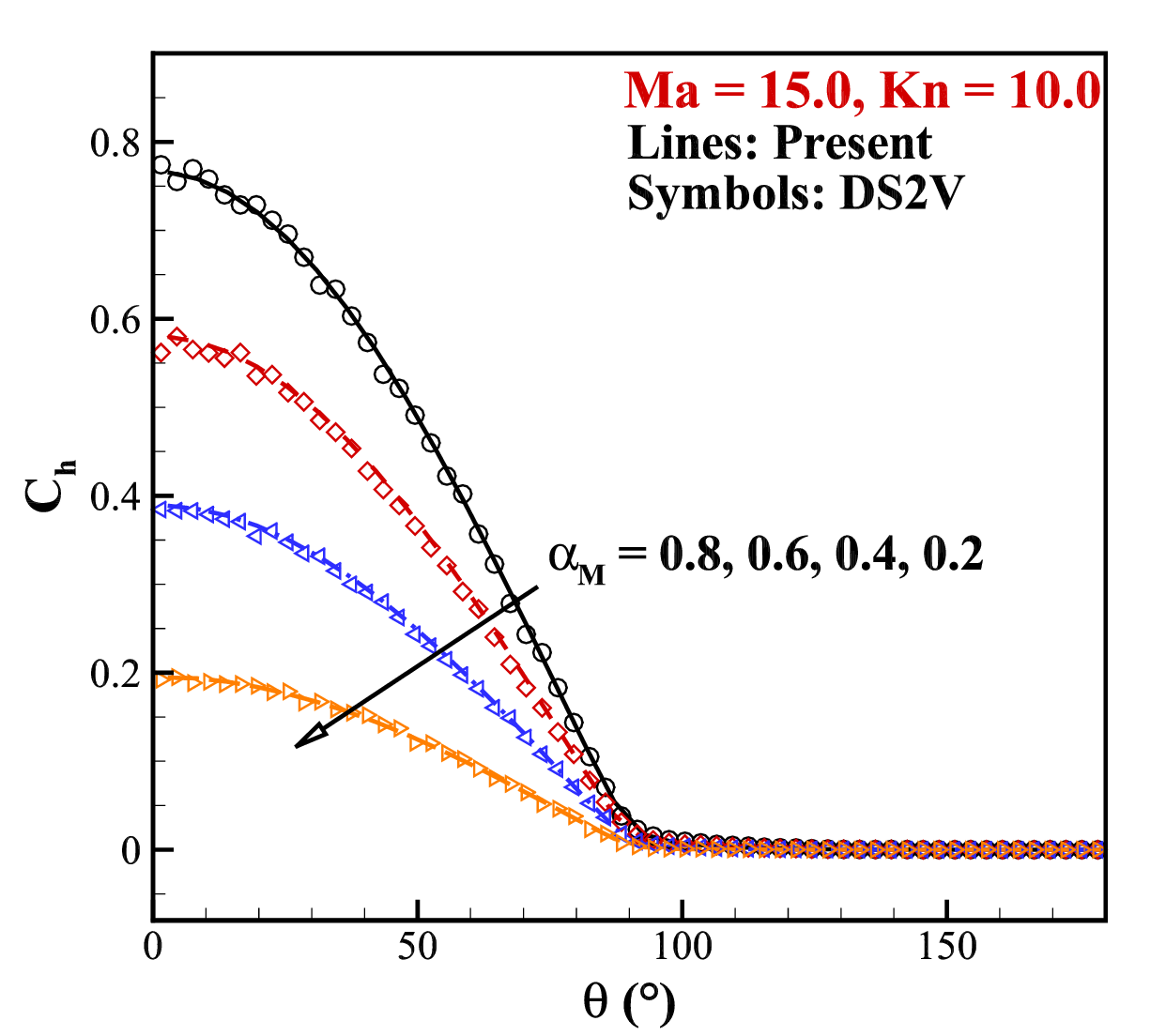}}
		\caption{\label{cylinder_MAX_15_10}{Comparison of the (a) pressure coefficient, (b) skin friction coefficient, and (c) heat transfer coefficient on the surface of cylinder with different $\alpha_{M}$ when employing the Maxwell boundary ($Ma = 15.0$, $Kn = 10.0$, $T_{\infty} = 273 K$, $T_{w} = 273 K$).}}
	\end{figure}
	
	\begin{figure}[!htp]
		\centering
		\subfigure[]{\label{cylinder_CLL_5_10_P}\includegraphics[width=0.32\textwidth]{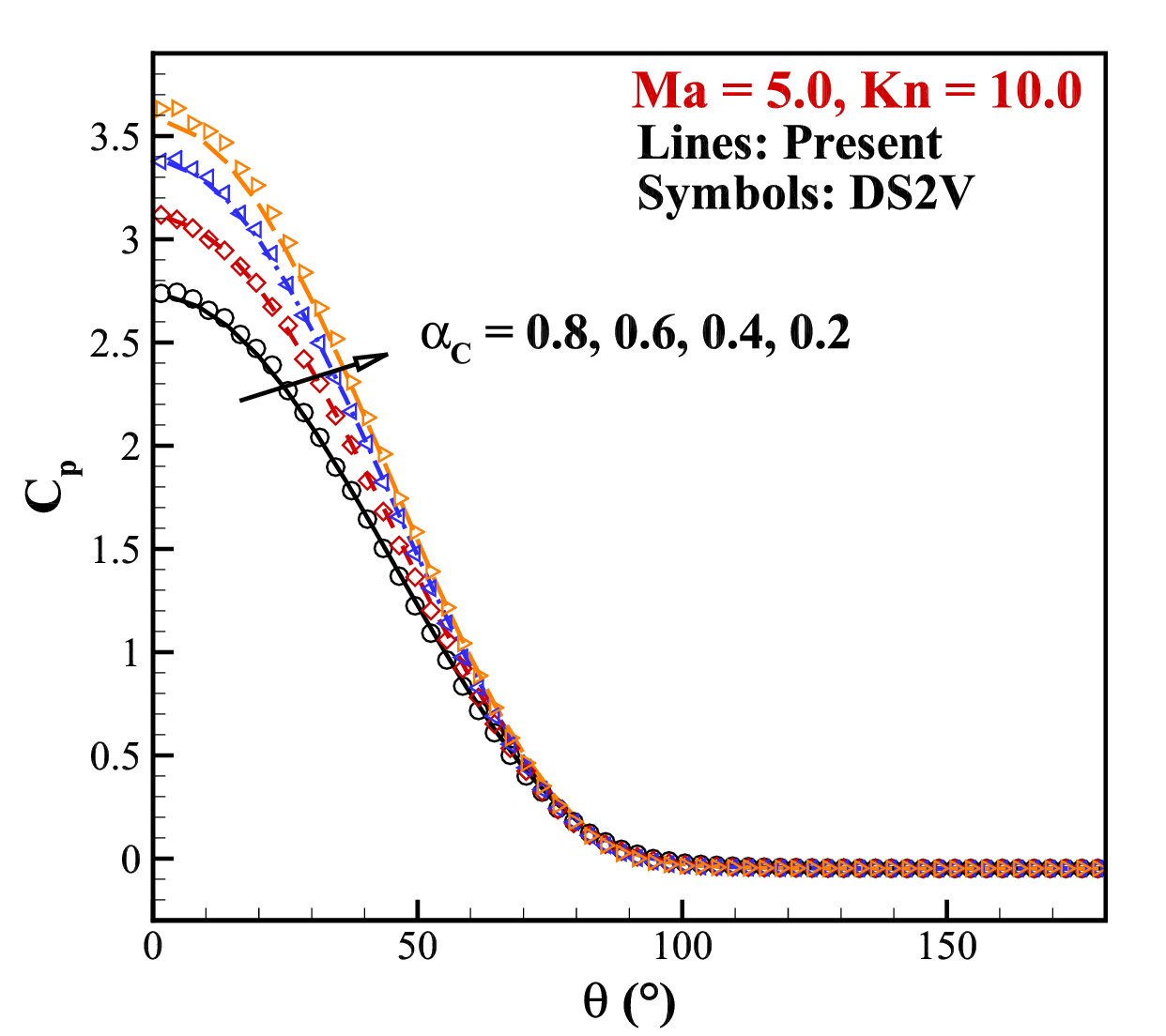}}
		\subfigure[]{\label{cylinder_CLL_5_10_S}\includegraphics[width=0.32\textwidth]{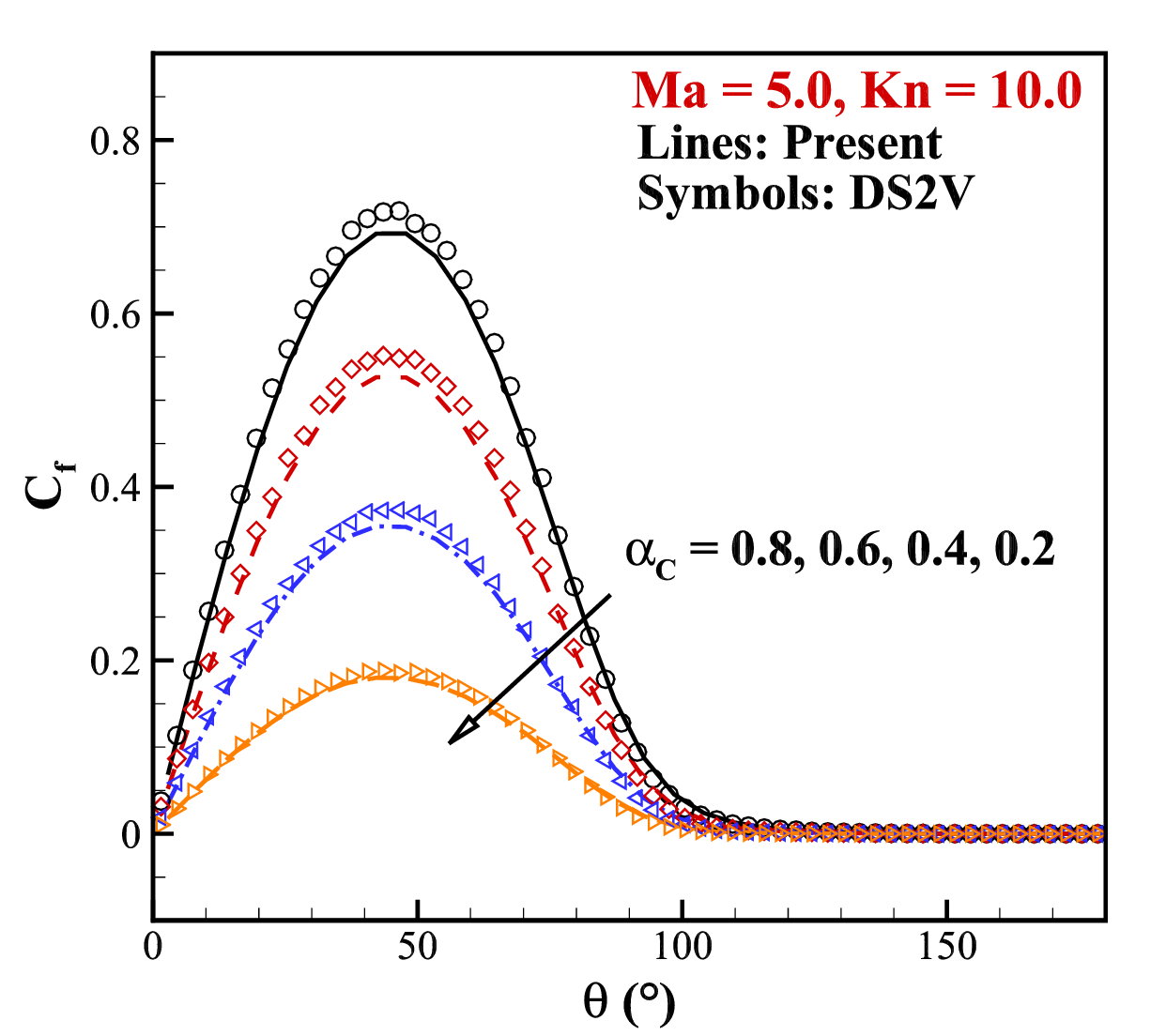}}
		\subfigure[]{\label{cylinder_CLL_5_10_H}\includegraphics[width=0.32\textwidth]{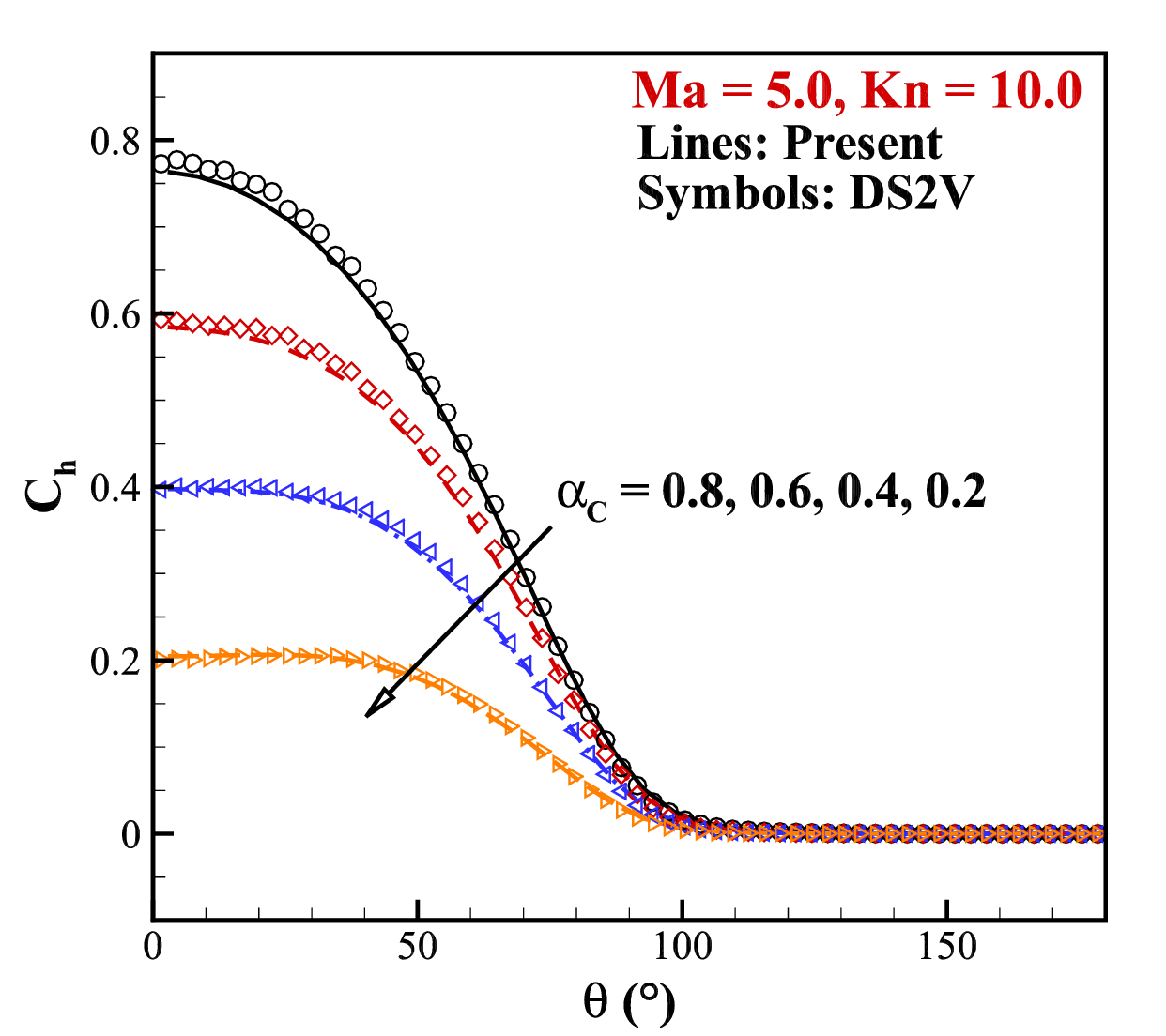}}
		\caption{\label{cylinder_CLL_5_10}{Comparison of the (a) pressure coefficient, (b) skin friction coefficient, and (c) heat transfer coefficient on the surface of cylinder with different $\alpha_{C}$ when employing the CLL boundary ($Ma = 5.0$, $Kn = 10.0$, $T_{\infty} = 273 K$, $T_{w} = 273 K$).}}
	\end{figure}

	\begin{figure}[!htp]
		\centering
		\subfigure[]{\label{cylinder_CLL_5_1_P}\includegraphics[width=0.32\textwidth]{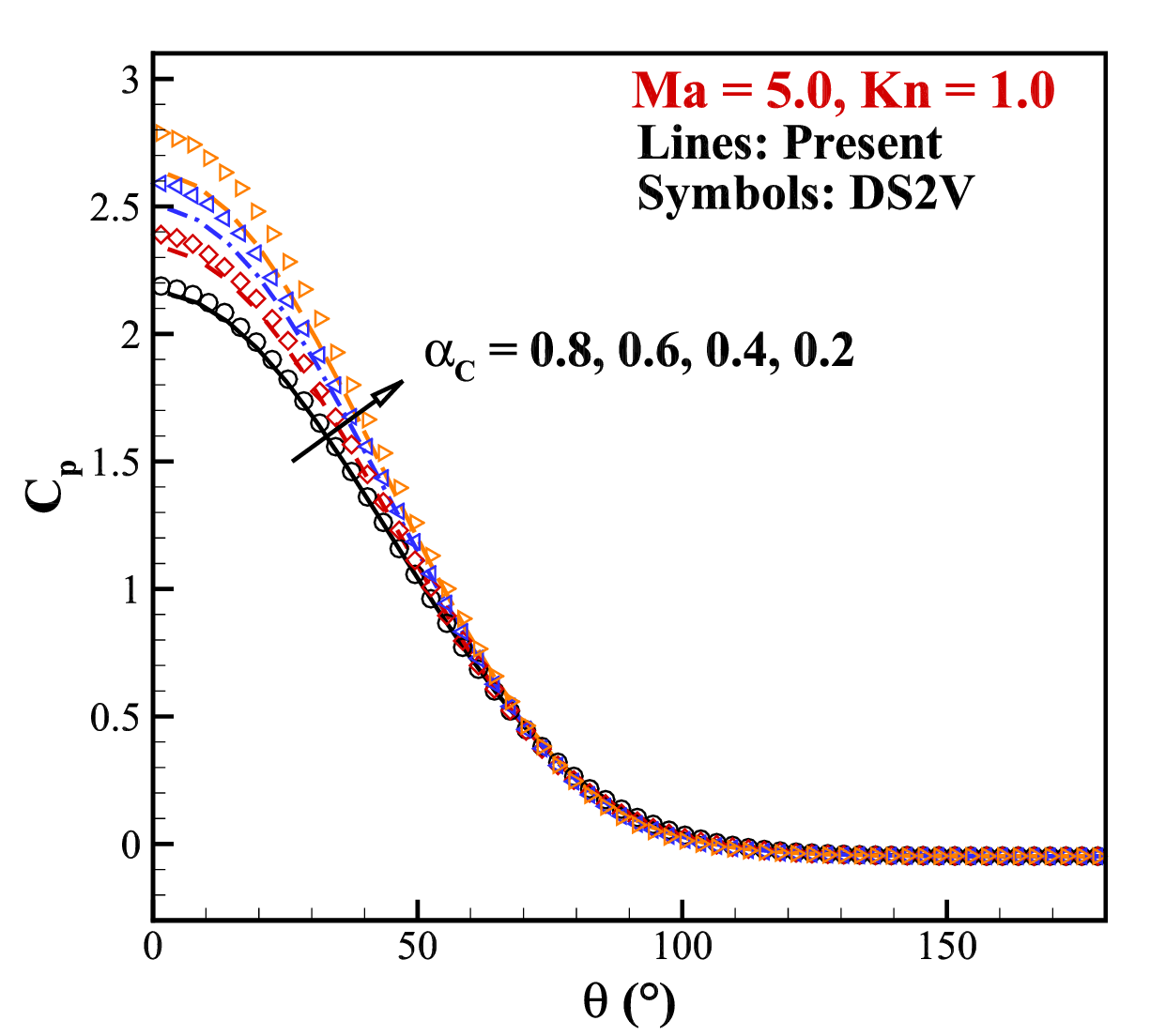}}
		\subfigure[]{\label{cylinder_CLL_5_1_S}\includegraphics[width=0.32\textwidth]{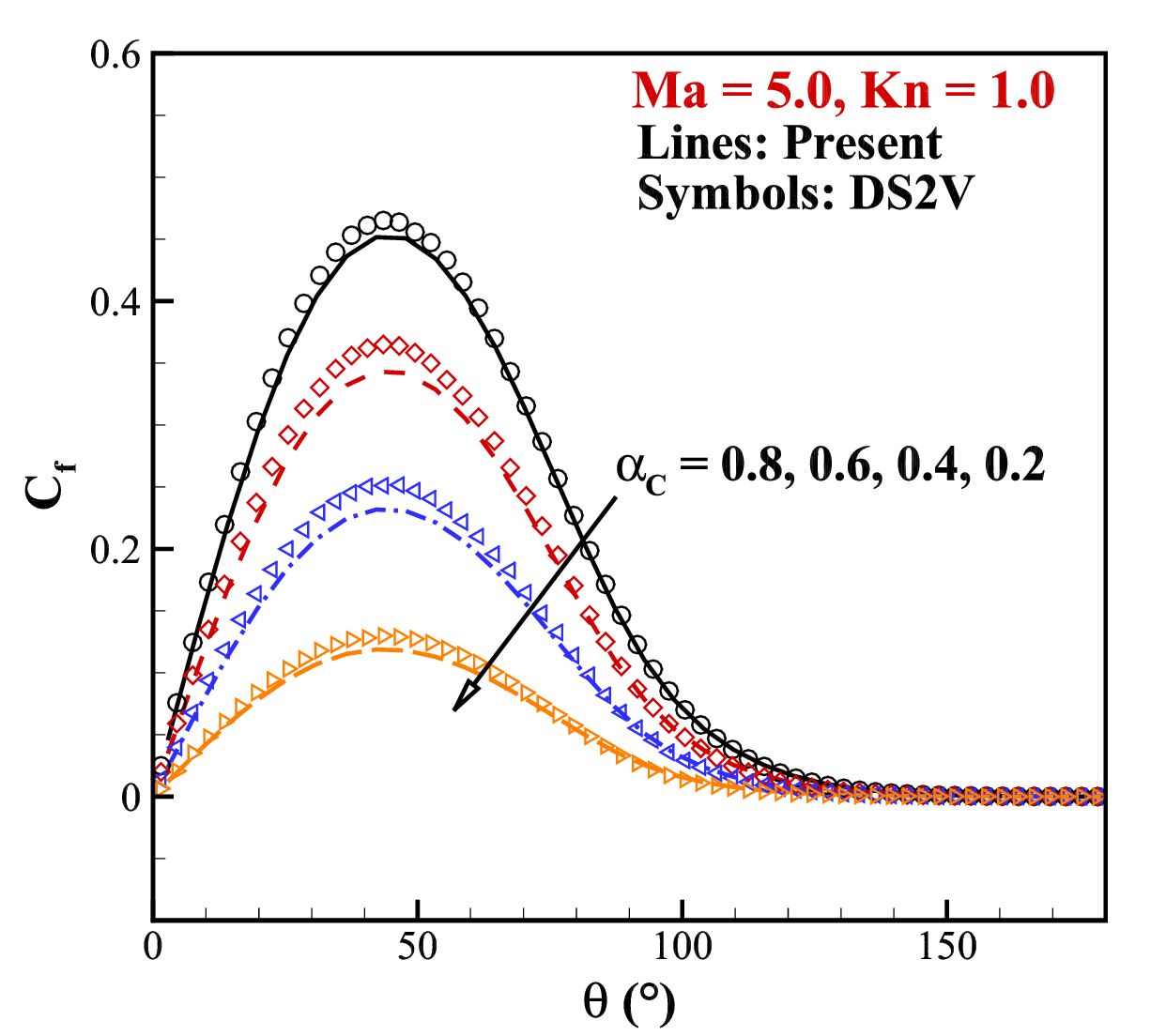}}
		\subfigure[]{\label{cylinder_CLL_5_1_H}\includegraphics[width=0.32\textwidth]{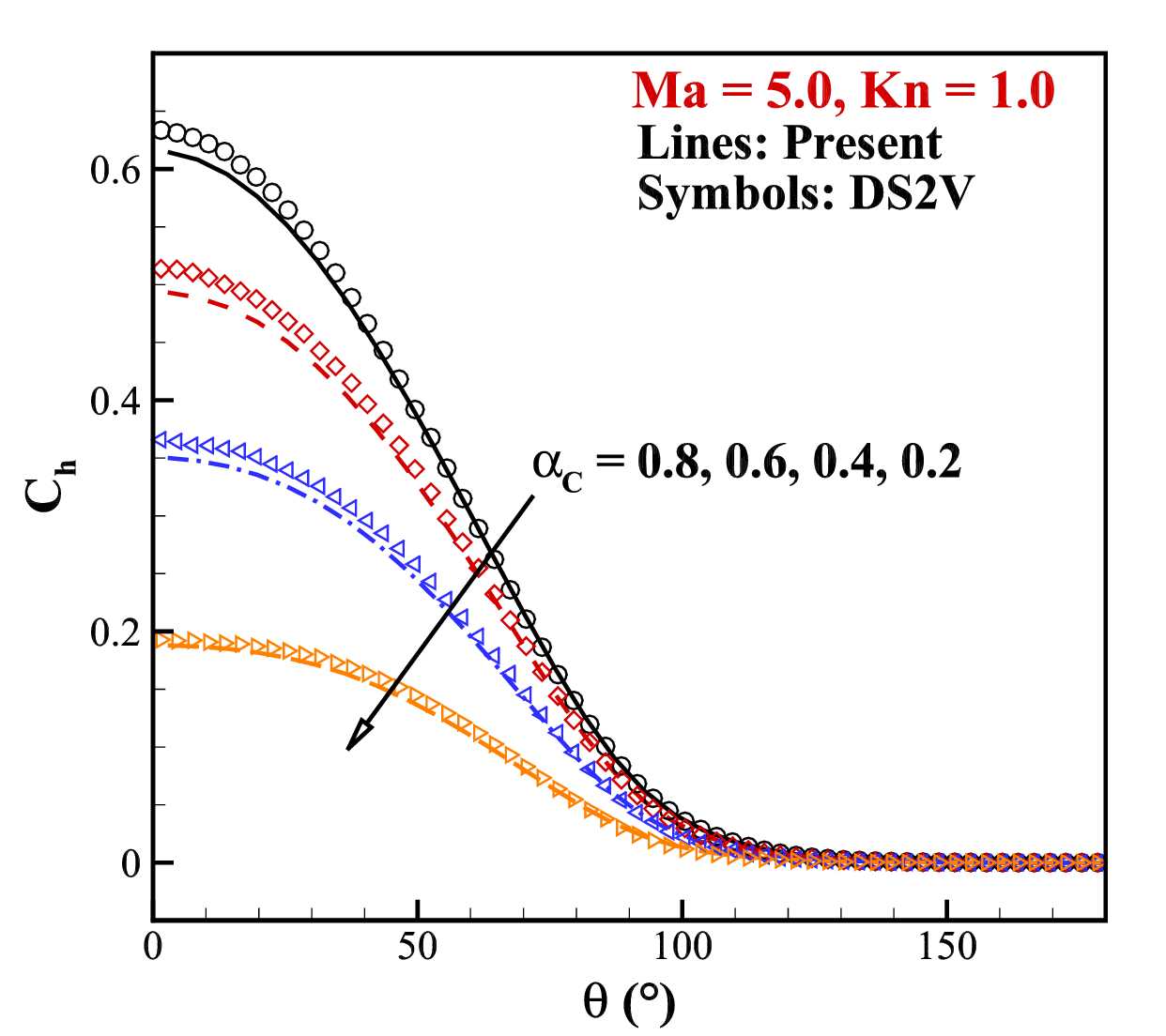}}
		\caption{\label{cylinder_CLL_5_1}{Comparison of the (a) pressure coefficient, (b) skin friction coefficient, and (c) heat transfer coefficient on the surface of cylinder with different $\alpha_{C}$ when employing the CLL boundary ($Ma = 5.0$, $Kn = 1.0$, $T_{\infty} = 273 K$, $T_{w} = 273 K$).}}
	\end{figure}

	\begin{figure}[!htp]
		\centering
		\subfigure[]{\label{cylinder_CLL_5_01_P}\includegraphics[width=0.32\textwidth]{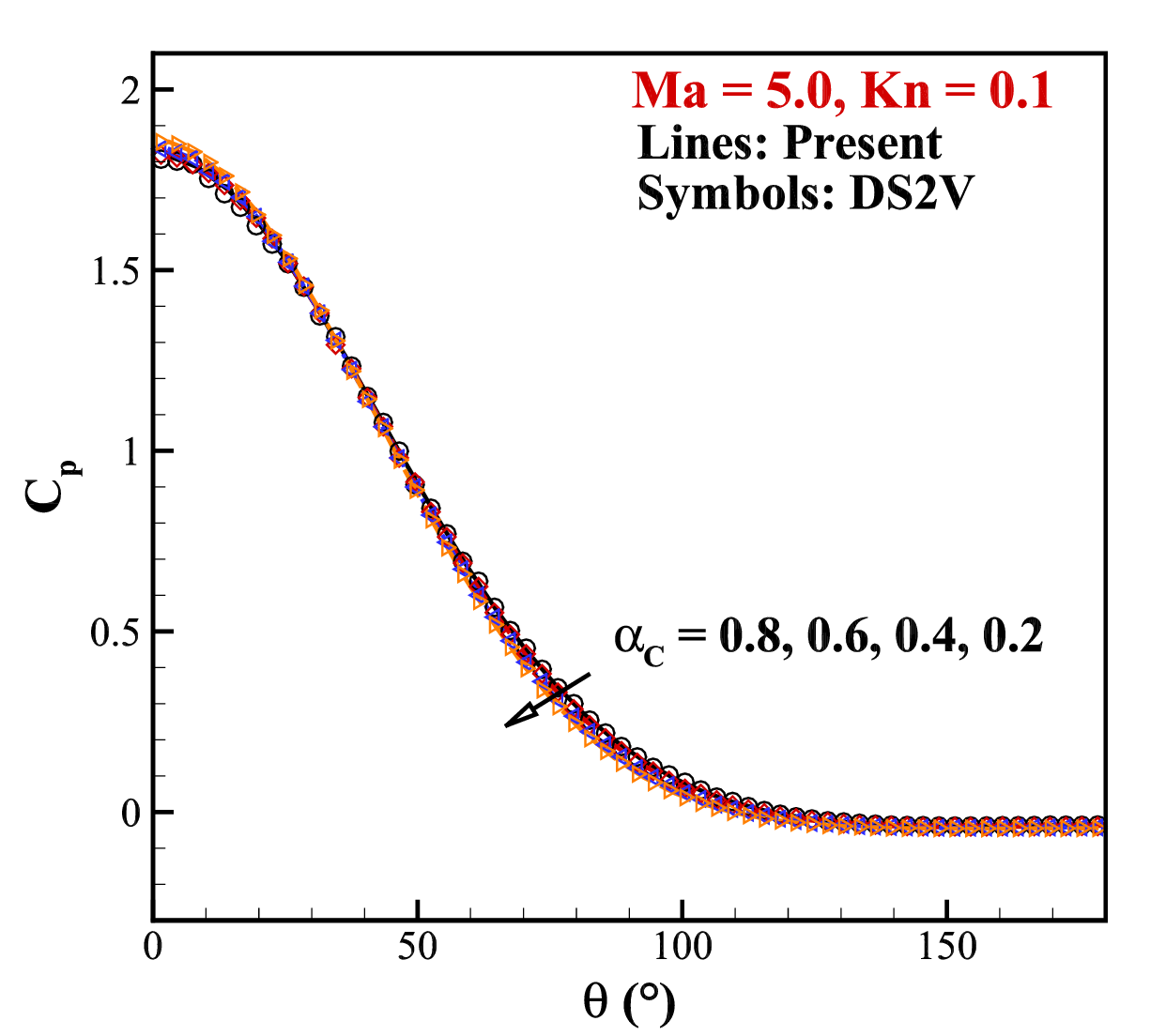}}
		\subfigure[]{\label{cylinder_CLL_5_01_S}\includegraphics[width=0.32\textwidth]{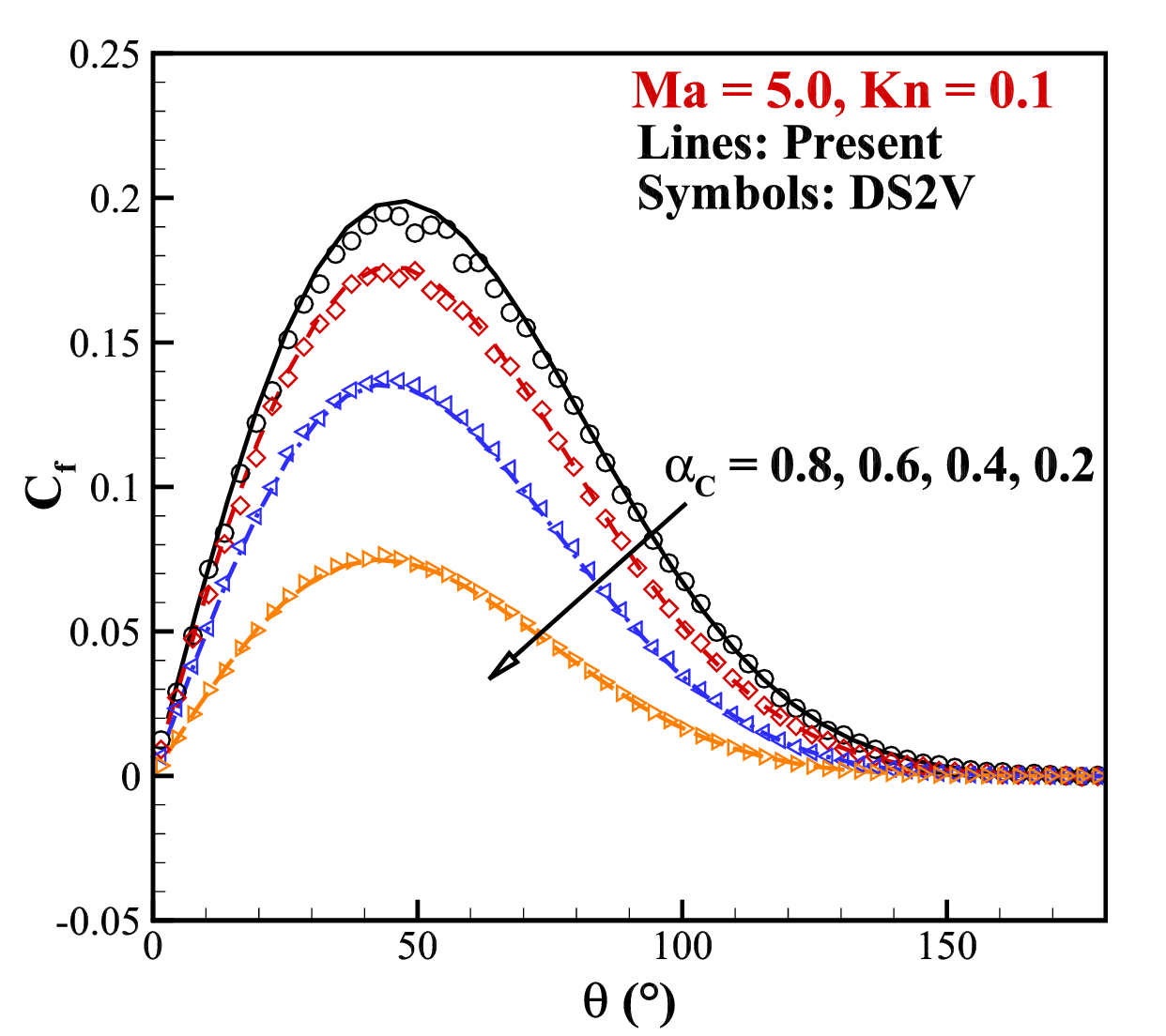}}
		\subfigure[]{\label{cylinder_CLL_5_01_H}\includegraphics[width=0.32\textwidth]{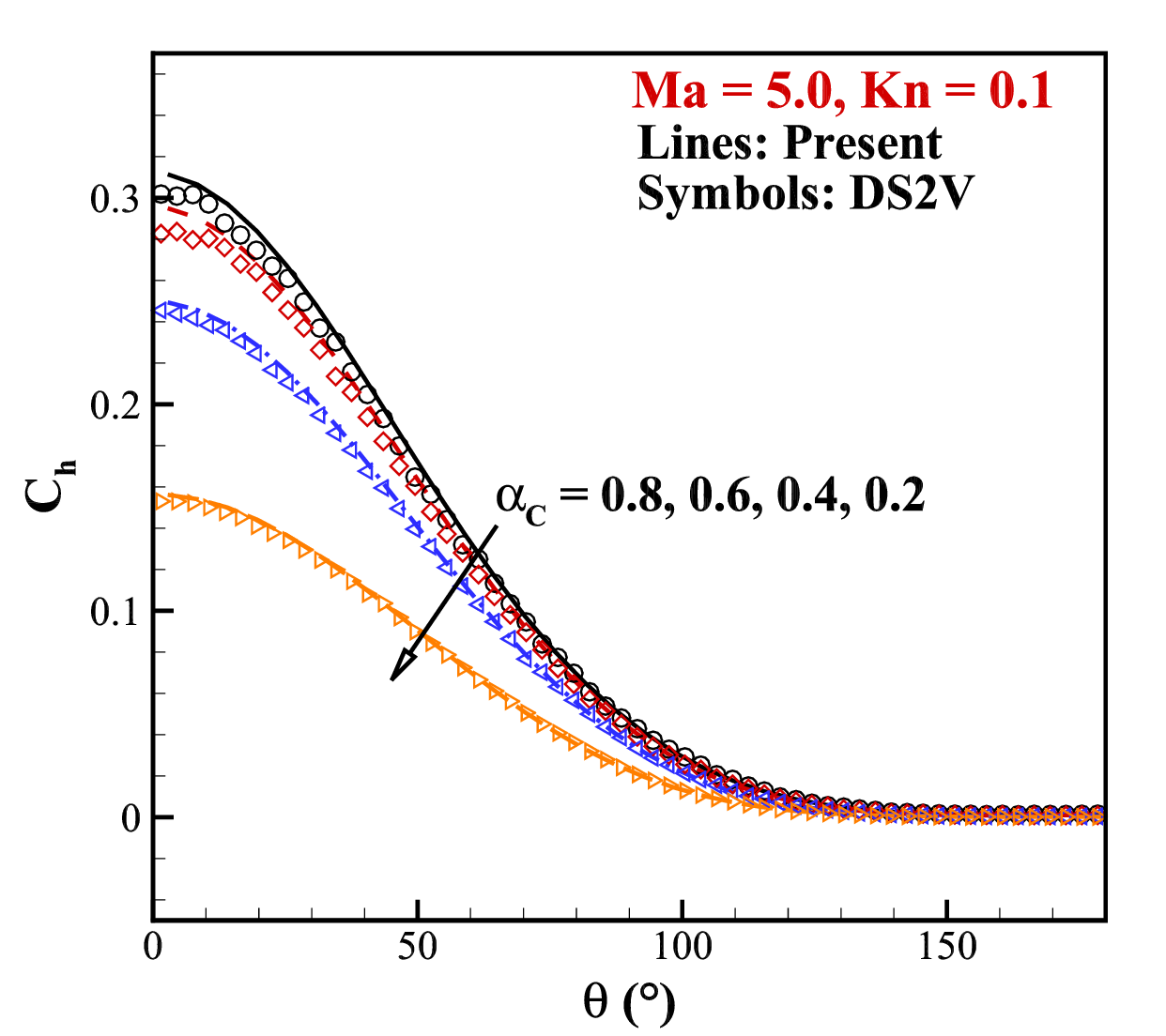}}
		\caption{\label{cylinder_CLL_5_01}{Comparison of the (a) pressure coefficient, (b) skin friction coefficient, and (c) heat transfer coefficient on the surface of cylinder with different $\alpha_{C}$ when employing the CLL boundary ($Ma = 5.0$, $Kn = 0.1$, $T_{\infty} = 273 K$, $T_{w} = 273 K$).}}
	\end{figure}

	\begin{figure}[!htp]
		\centering
		\subfigure[]{\label{cylinder_CLL_10_10_P}\includegraphics[width=0.32\textwidth]{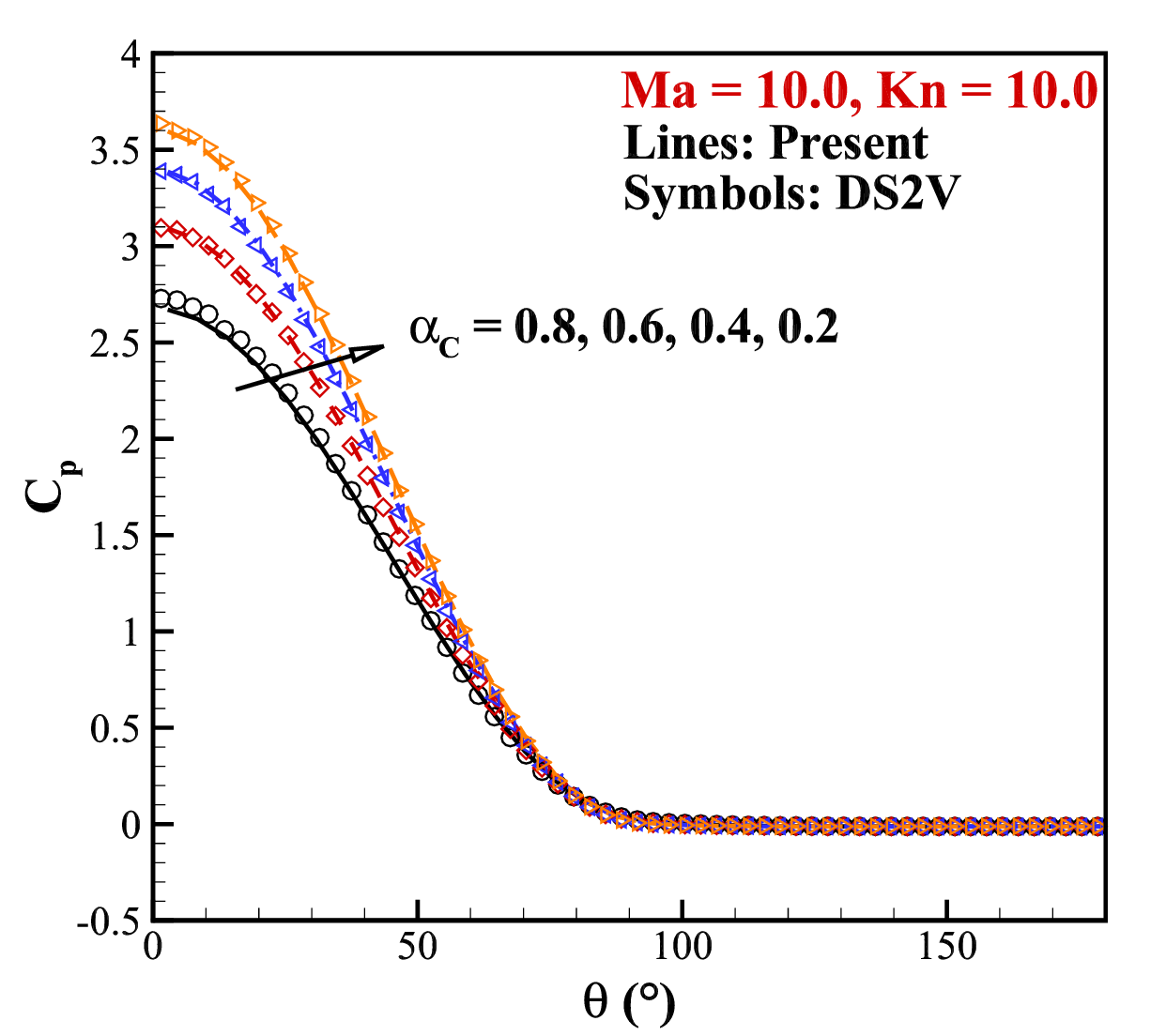}}
		\subfigure[]{\label{cylinder_CLL_10_10_S}\includegraphics[width=0.32\textwidth]{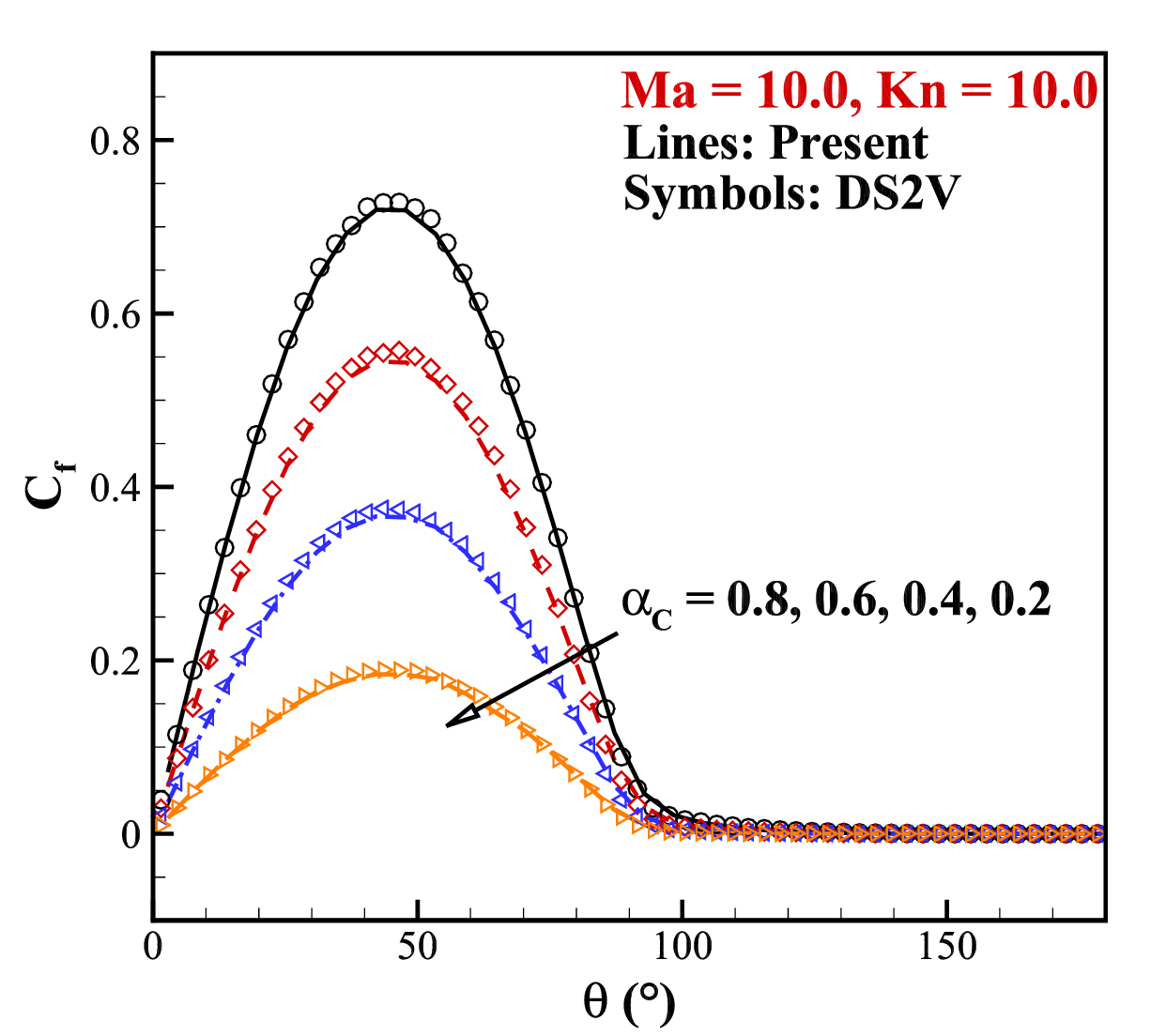}}
		\subfigure[]{\label{cylinder_CLL_10_10_H}\includegraphics[width=0.32\textwidth]{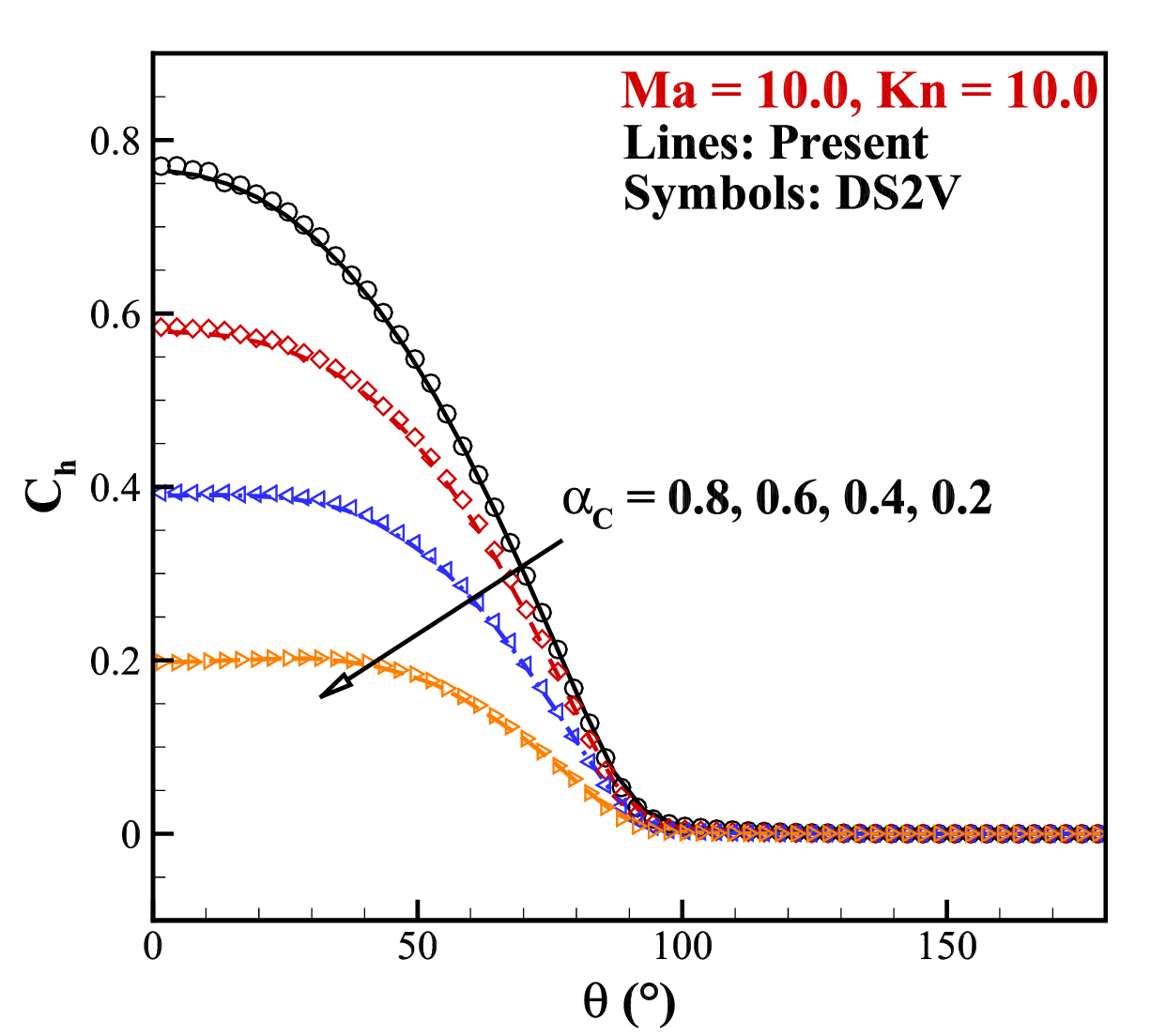}}
		\caption{\label{cylinder_CLL_10_10}{Comparison of the (a) pressure coefficient, (b) skin friction coefficient, and (c) heat transfer coefficient on the surface of cylinder with different $\alpha_{C}$ when employing the CLL boundary ($Ma = 10.0$, $Kn = 10.0$, $T_{\infty} = 273 K$, $T_{w} = 273 K$).}}
	\end{figure}

	\begin{figure}[!htp]
		\centering
		\subfigure[]{\label{cylinder_CLL_10_1_P}\includegraphics[width=0.32\textwidth]{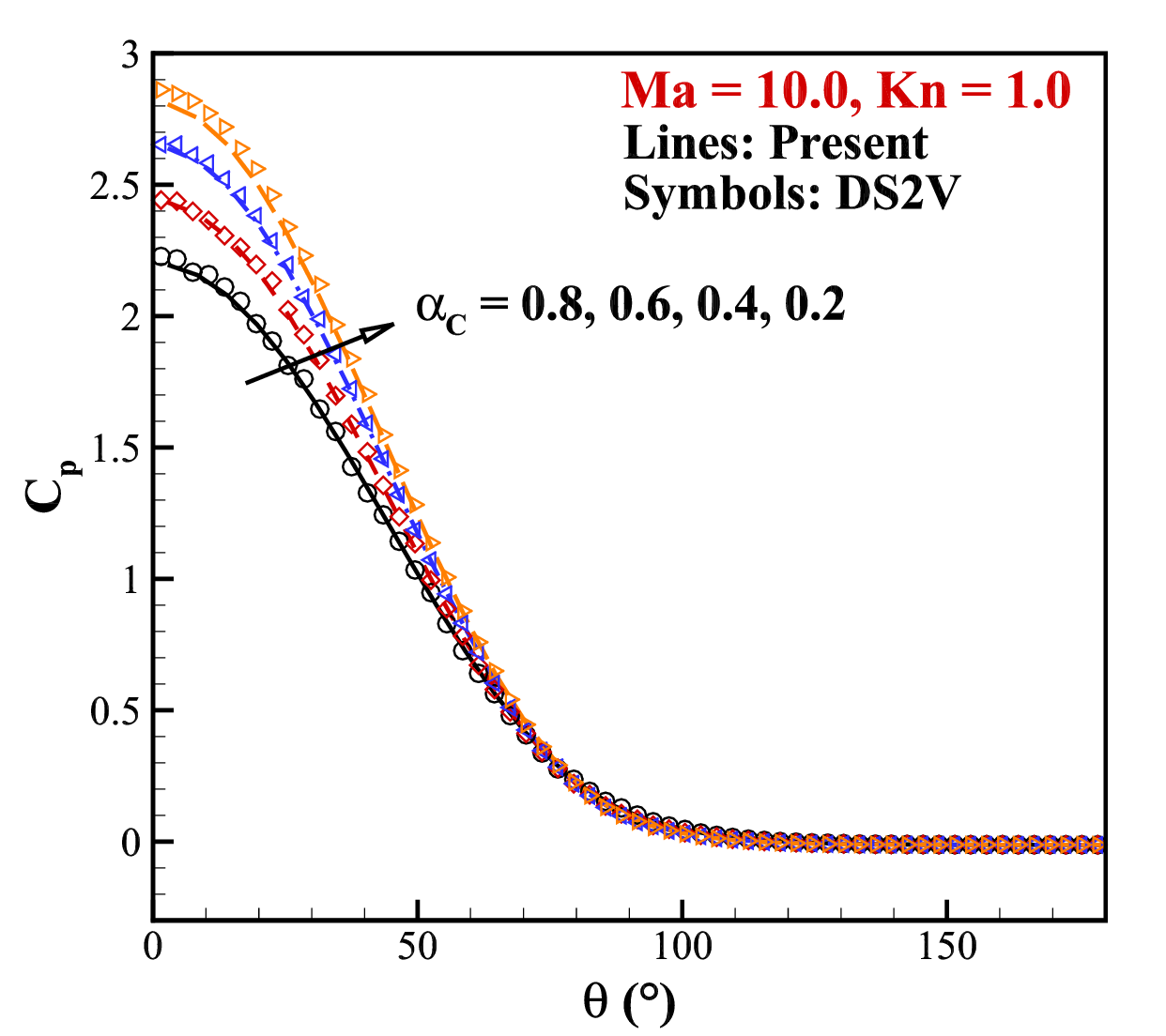}}
		\subfigure[]{\label{cylinder_CLL_10_1_S}\includegraphics[width=0.32\textwidth]{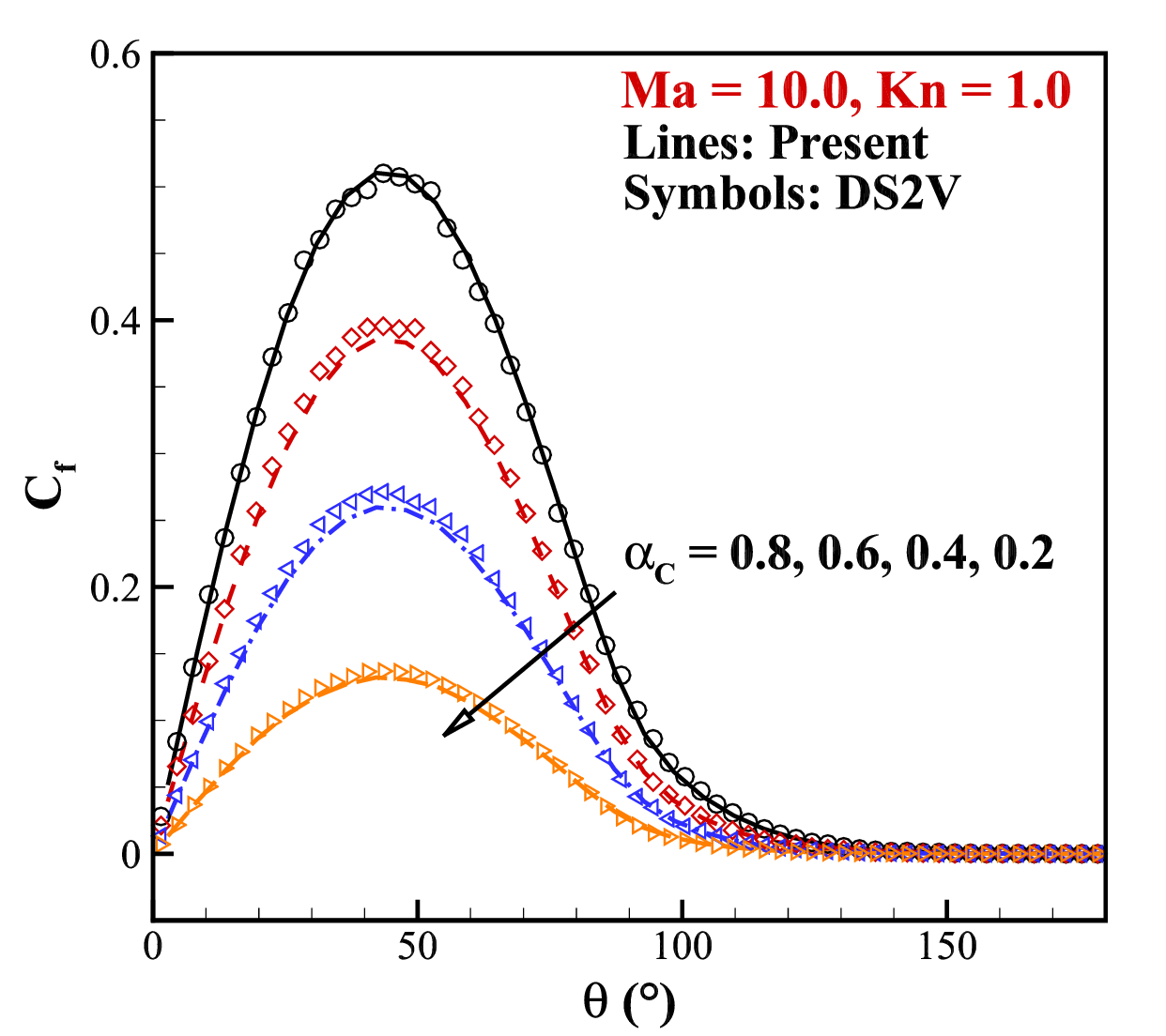}}
		\subfigure[]{\label{cylinder_CLL_10_1_H}\includegraphics[width=0.32\textwidth]{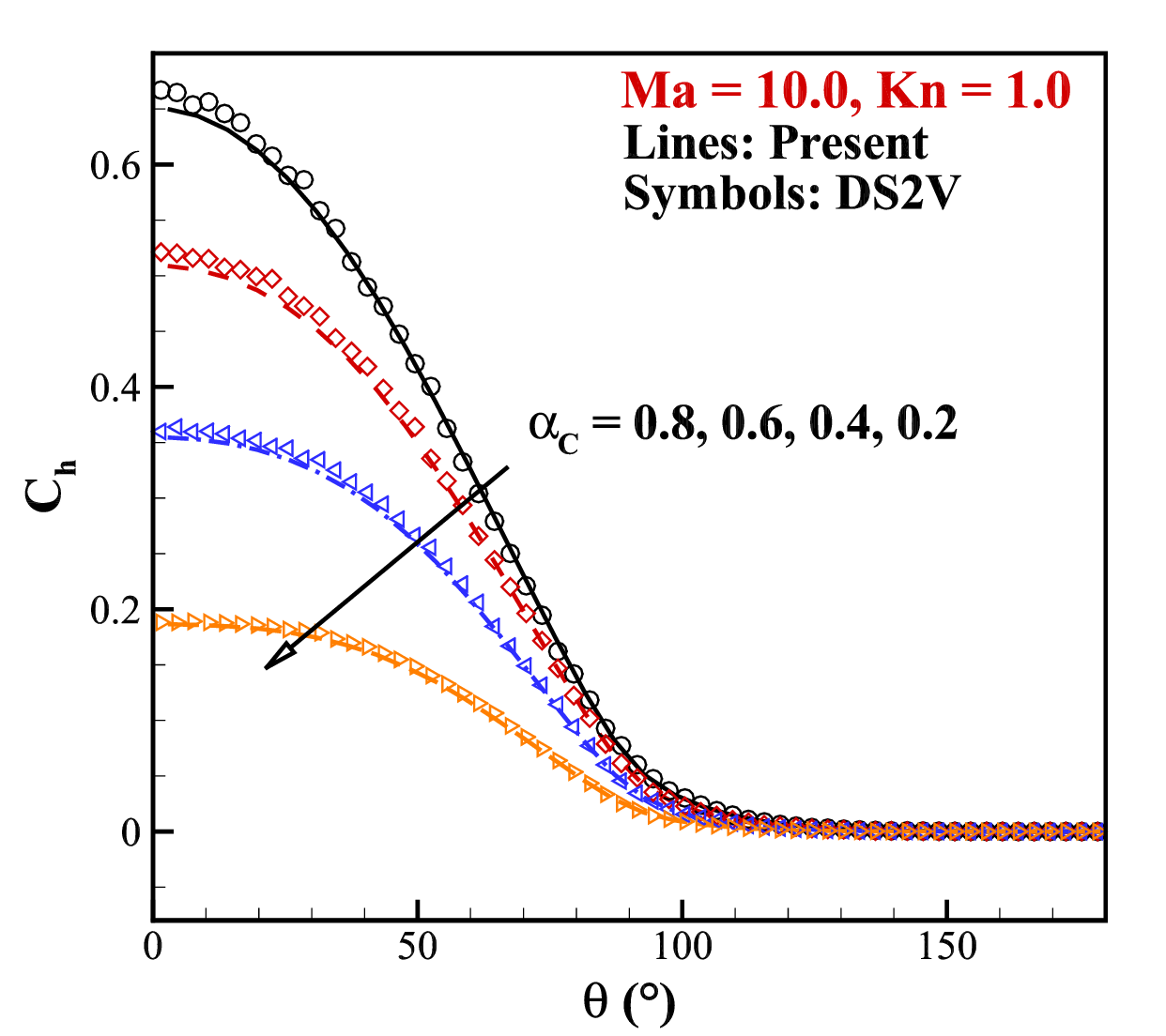}}
		\caption{\label{cylinder_CLL_10_1}{Comparison of the (a) pressure coefficient, (b) skin friction coefficient, and (c) heat transfer coefficient on the surface of cylinder with different $\alpha_{C}$ when employing the CLL boundary ($Ma = 10.0$, $Kn = 1.0$, $T_{\infty} = 273 K$, $T_{w} = 273 K$).}}
	\end{figure}

	\begin{figure}[!htp]
		\centering
		\subfigure[]{\label{cylinder_CLL_10_01_P}\includegraphics[width=0.32\textwidth]{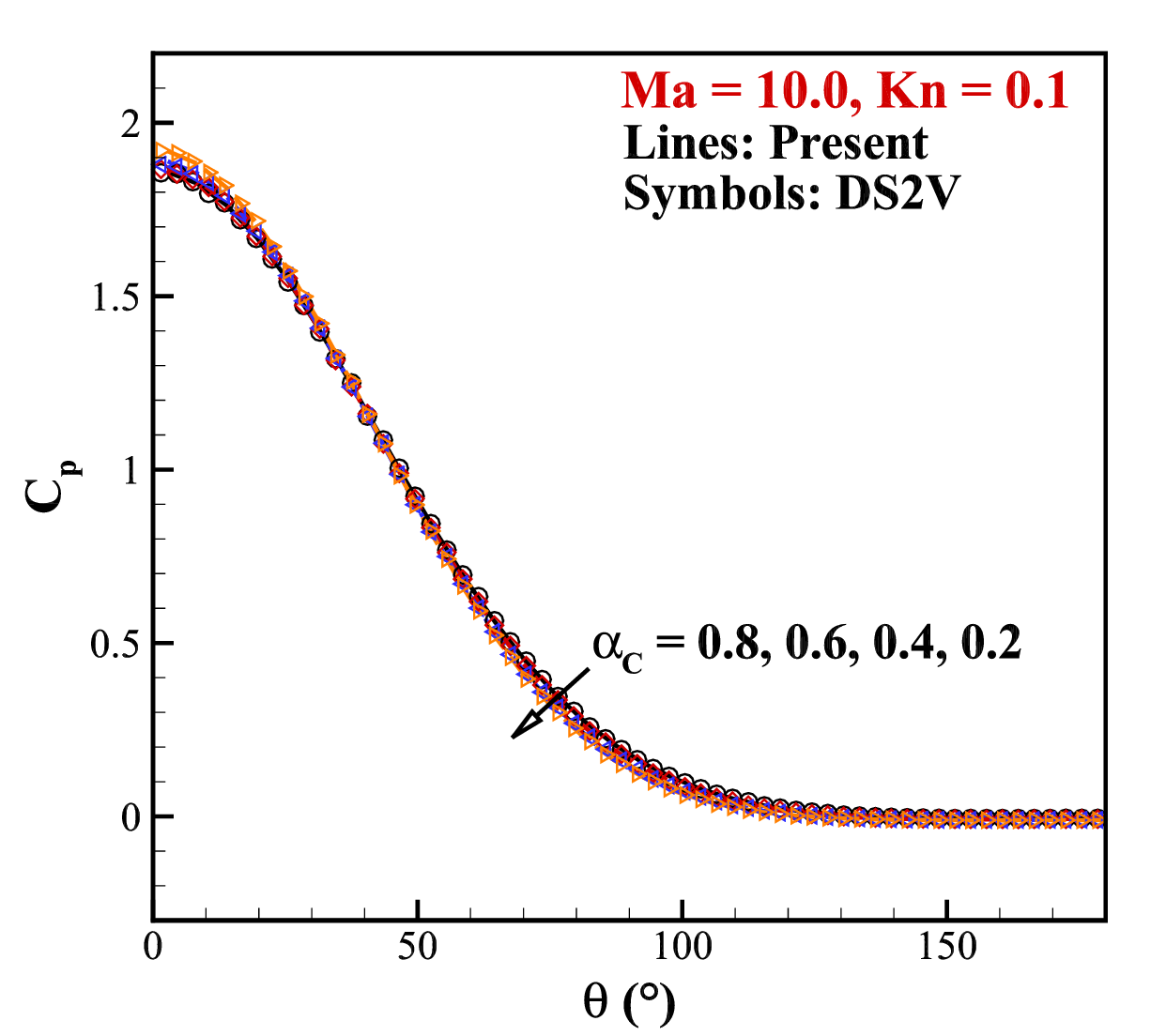}}
		\subfigure[]{\label{cylinder_CLL_10_01_S}\includegraphics[width=0.32\textwidth]{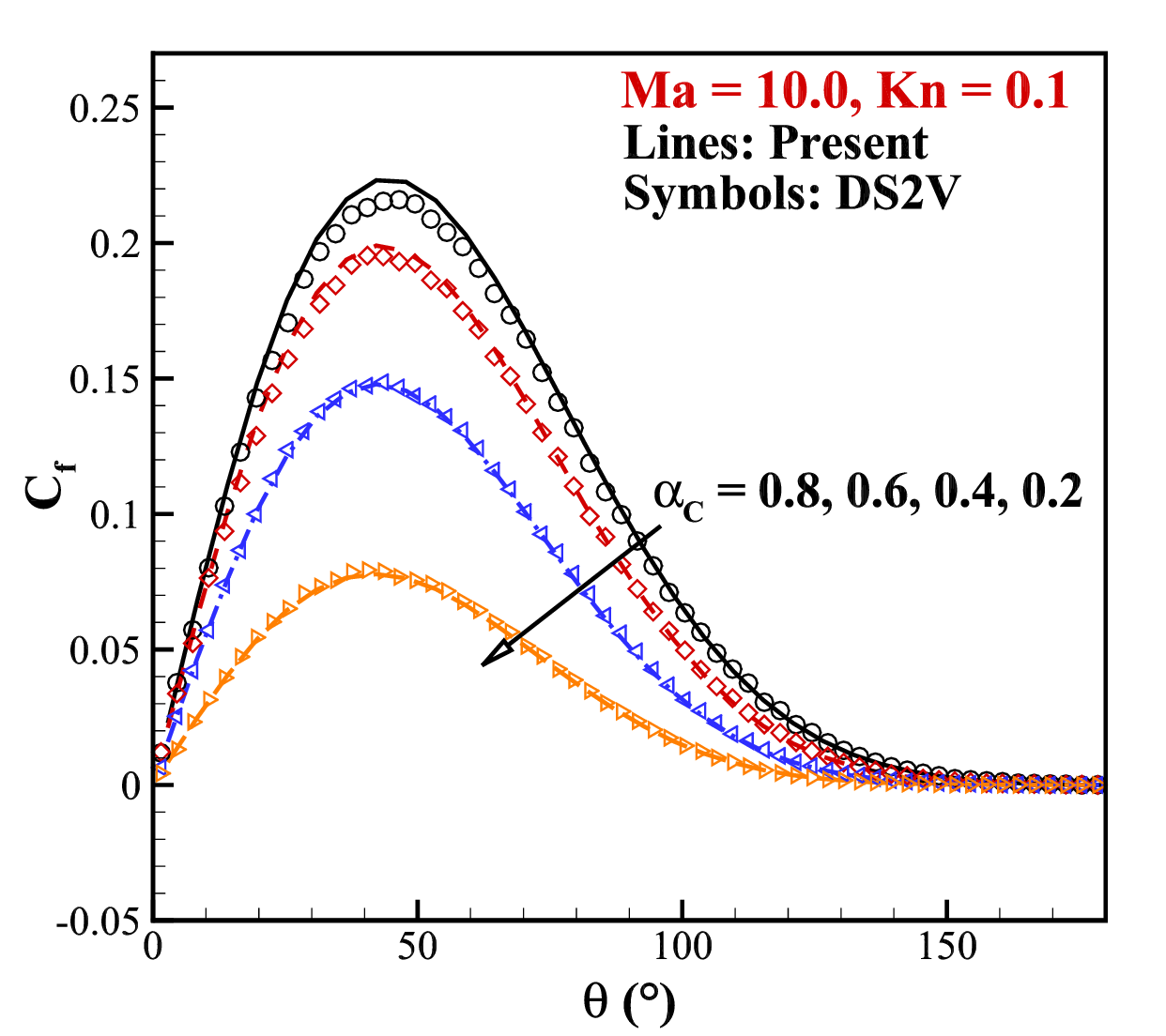}}
		\subfigure[]{\label{cylinder_CLL_10_01_H}\includegraphics[width=0.32\textwidth]{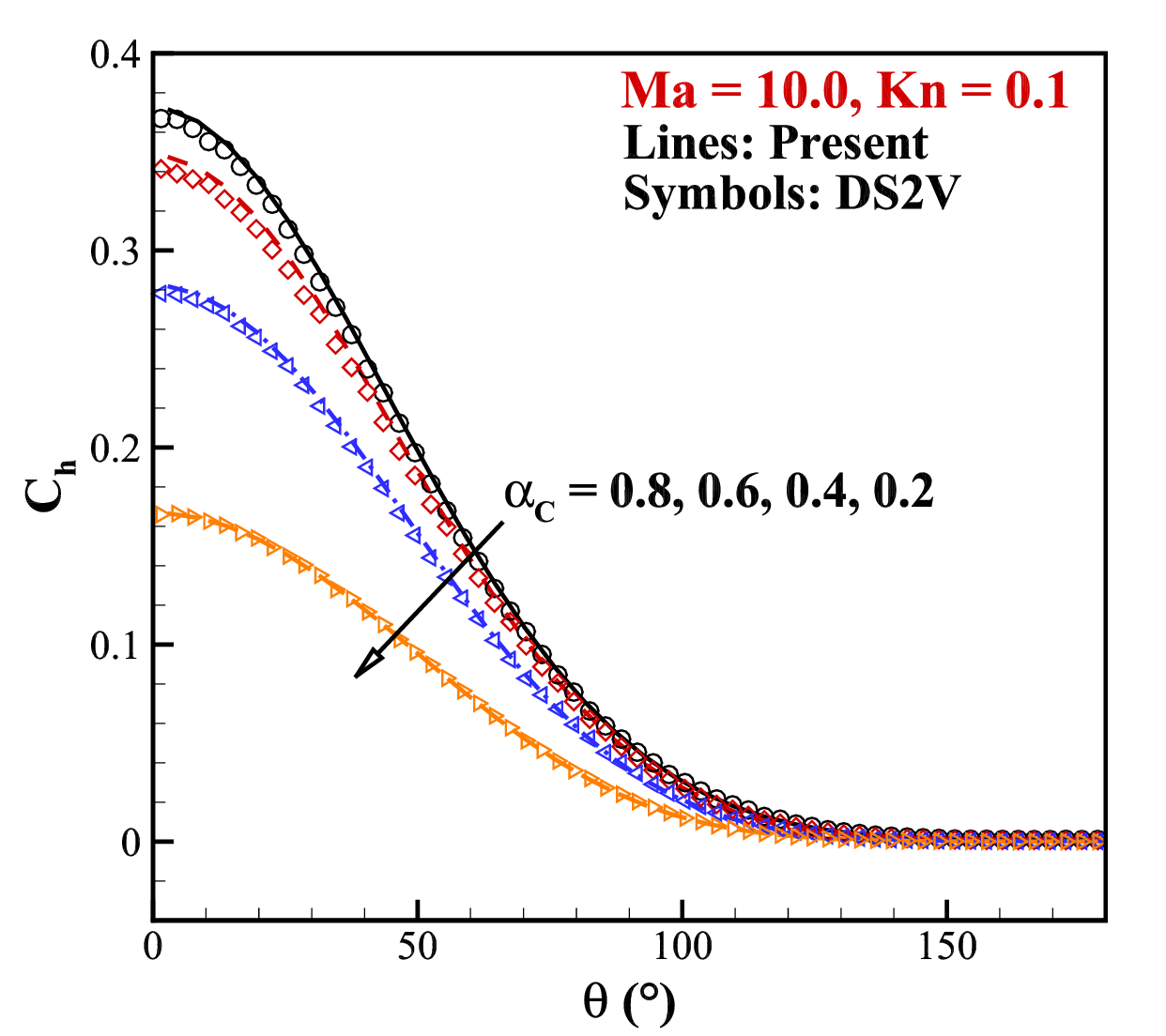}}
		\caption{\label{cylinder_CLL_10_01}{Comparison of the (a) pressure coefficient, (b) skin friction coefficient, and (c) heat transfer coefficient on the surface of cylinder with different $\alpha_{C}$ when employing the CLL boundary ($Ma = 10.0$, $Kn = 0.1$, $T_{\infty} = 273 K$, $T_{w} = 273 K$).}}
	\end{figure}

	\begin{figure}[!htp]
		\centering
		\subfigure[]{\label{cylinder_CLL_15_10_P}\includegraphics[width=0.32\textwidth]{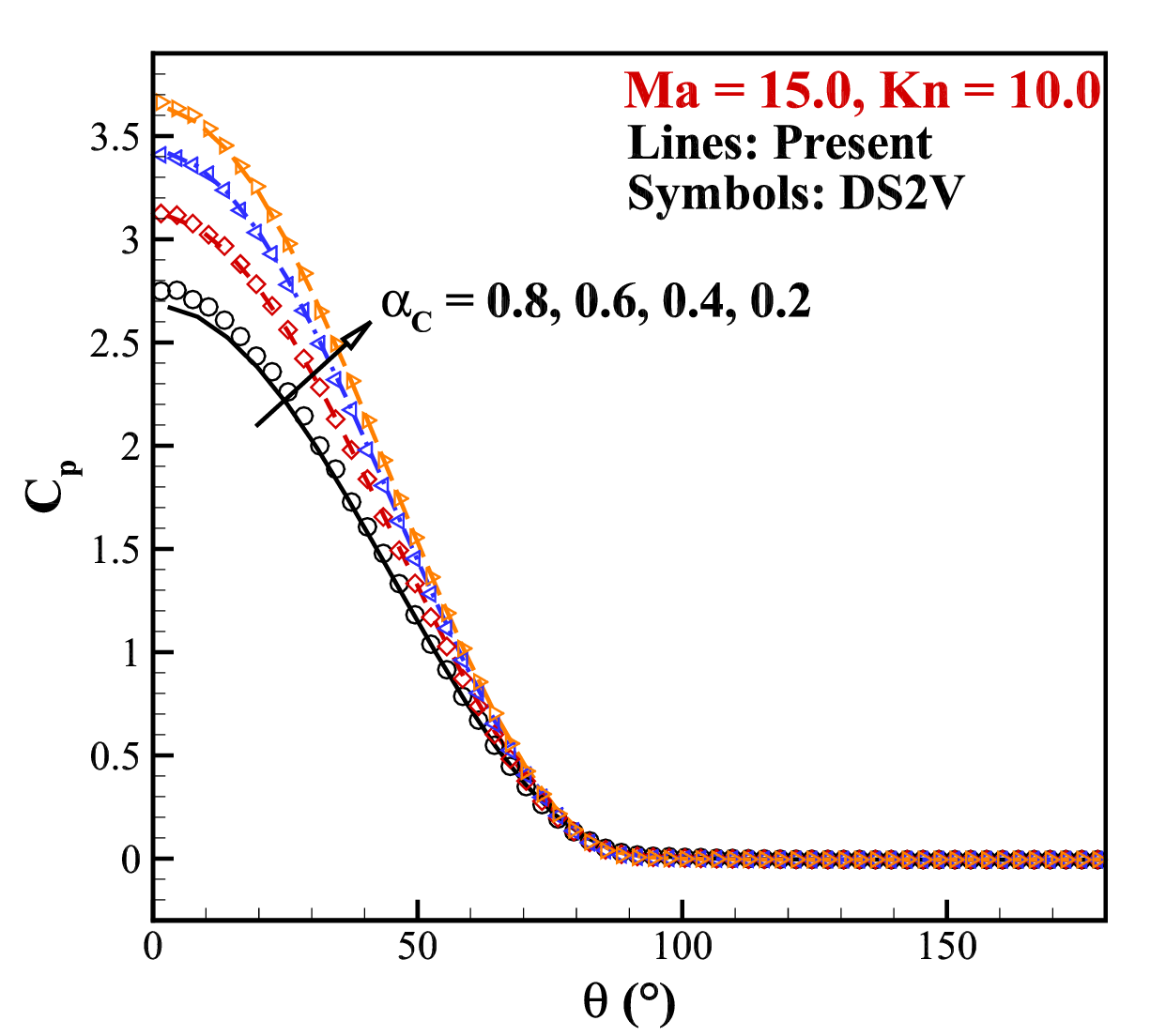}}
		\subfigure[]{\label{cylinder_CLL_15_10_S}\includegraphics[width=0.32\textwidth]{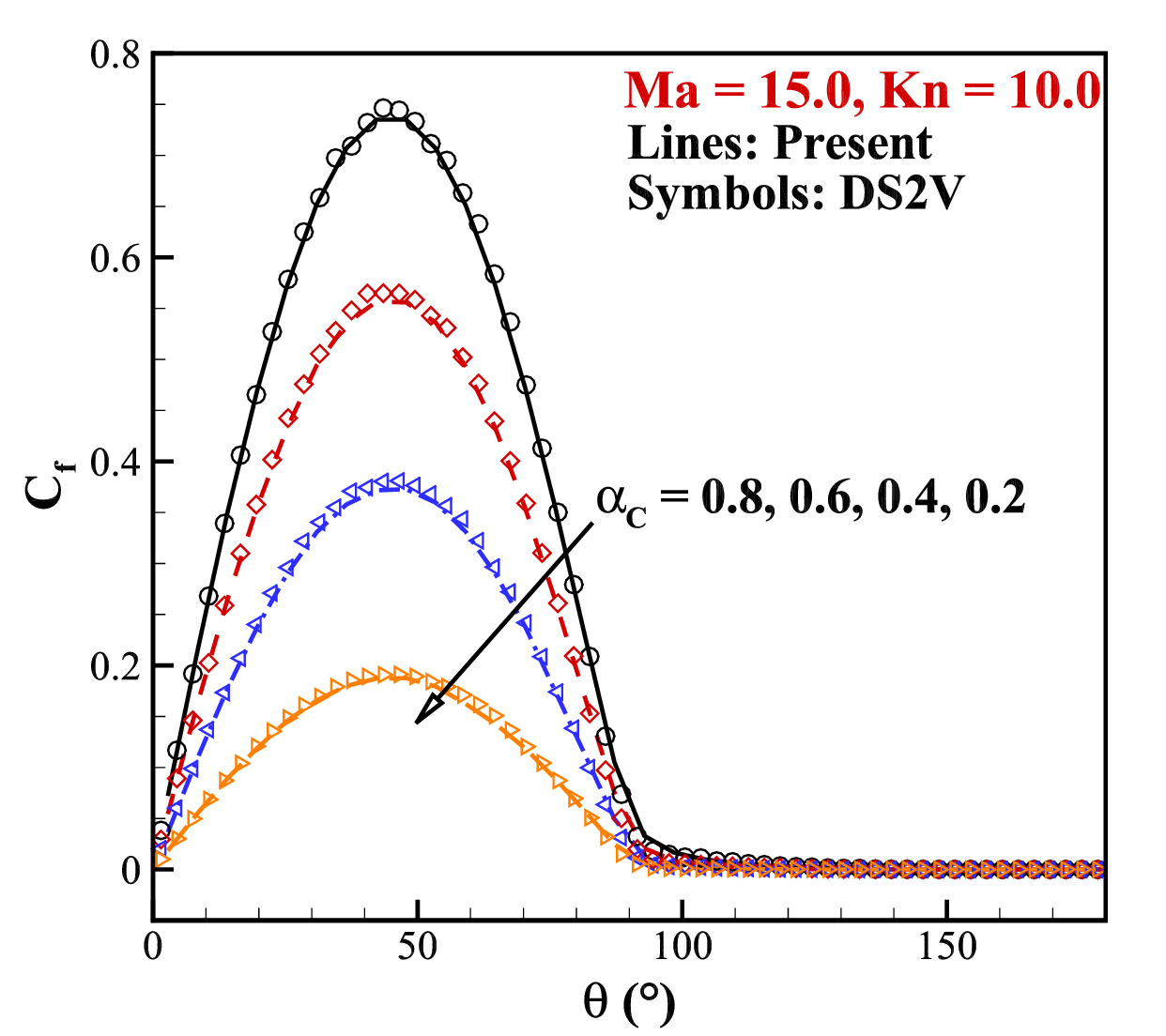}}
		\subfigure[]{\label{cylinder_CLL_15_10_H}\includegraphics[width=0.32\textwidth]{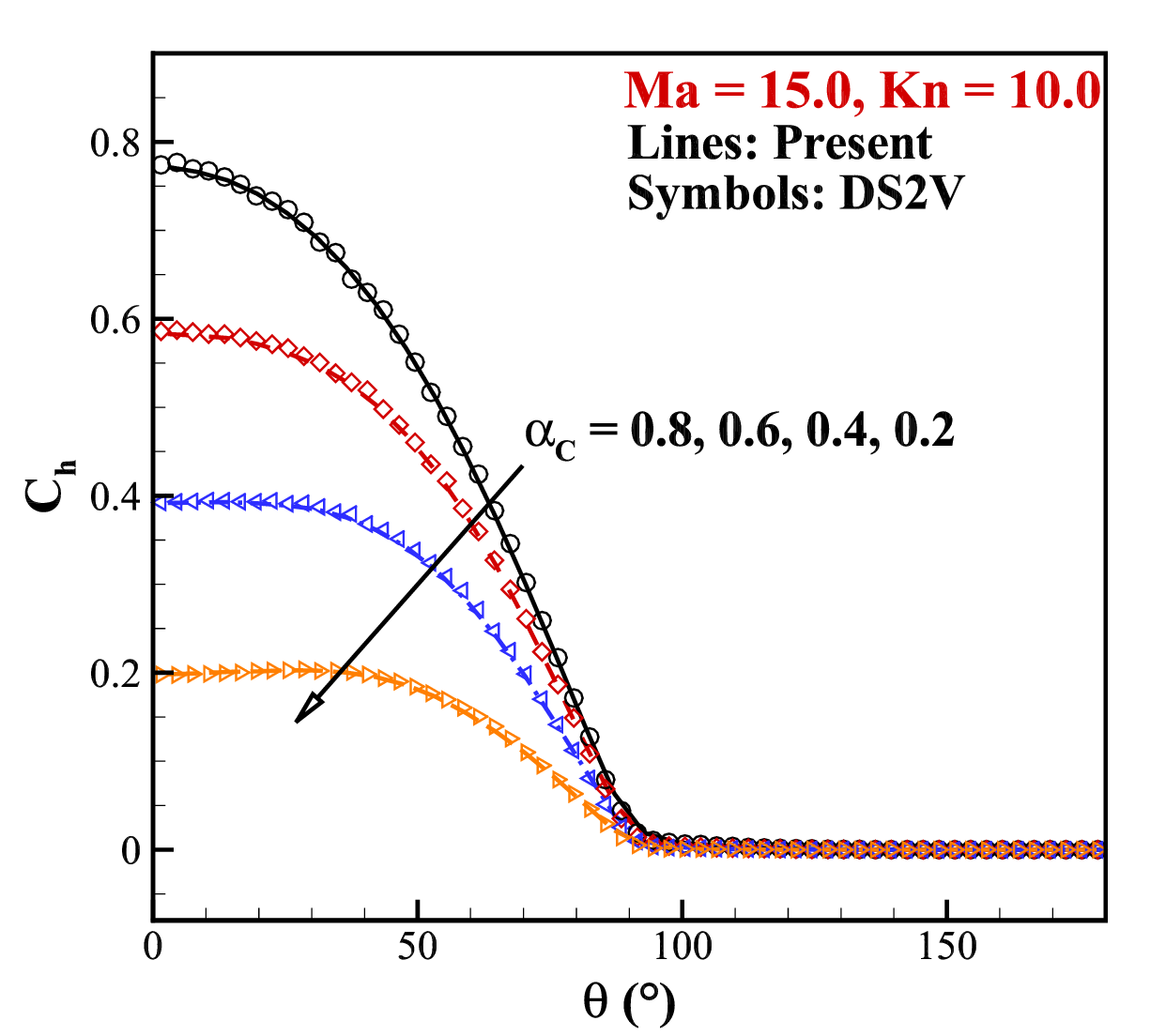}}
		\caption{\label{cylinder_CLL_15_10}{Comparison of the (a) pressure coefficient, (b) skin friction coefficient, and (c) heat transfer coefficient on the surface of cylinder with different $\alpha_{C}$ when employing the CLL boundary ($Ma = 15.0$, $Kn = 1.0$, $T_{\infty} = 273 K$, $T_{w} = 273 K$).}}
	\end{figure}
	
	\begin{figure}[!htp]
		\centering
		\subfigure[]{\label{cylinder_MAX_15_10_P_fix}\includegraphics[width=0.32\textwidth]{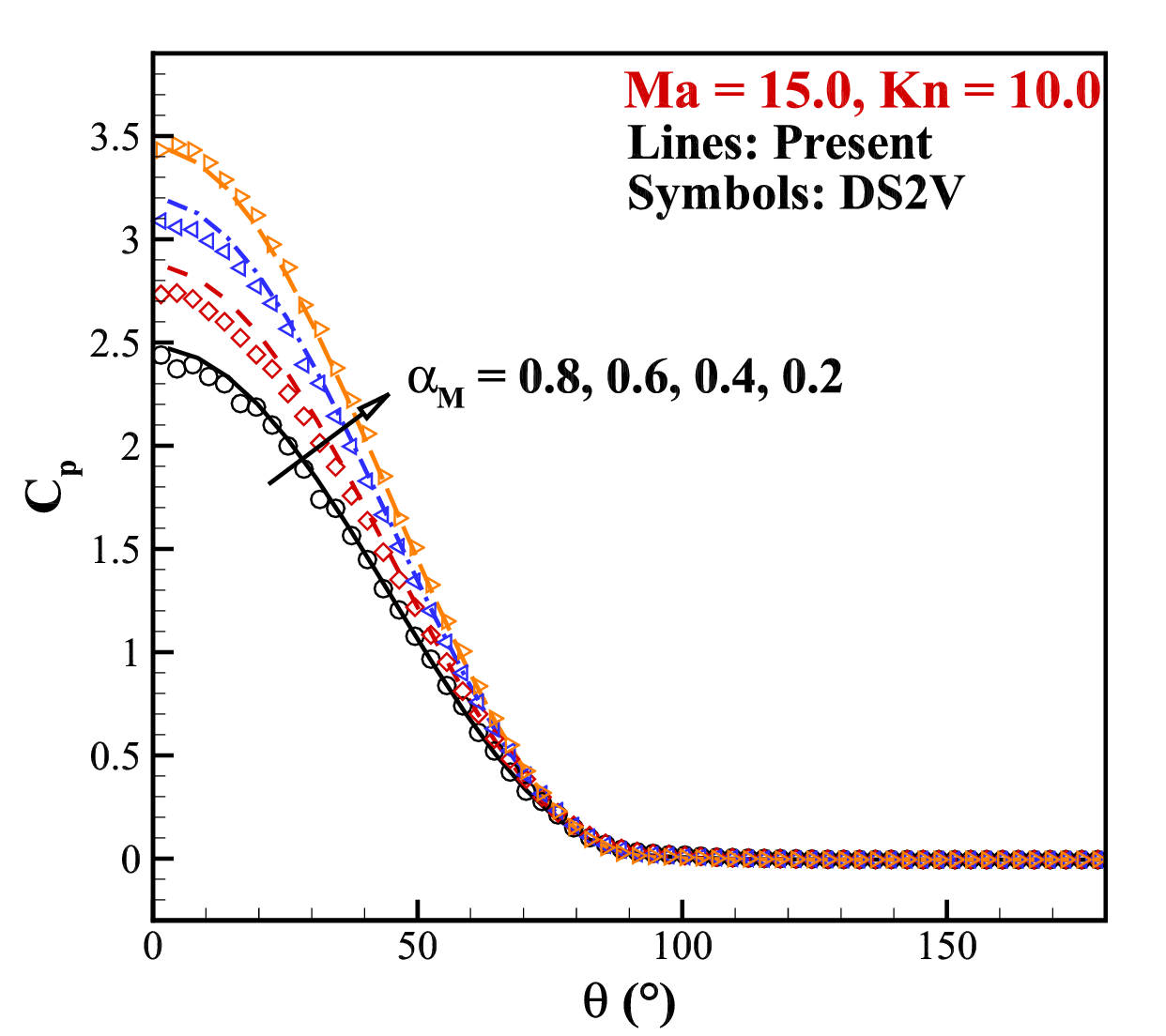}}
		\subfigure[]{\label{cylinder_MAX_15_10_S_fix}\includegraphics[width=0.32\textwidth]{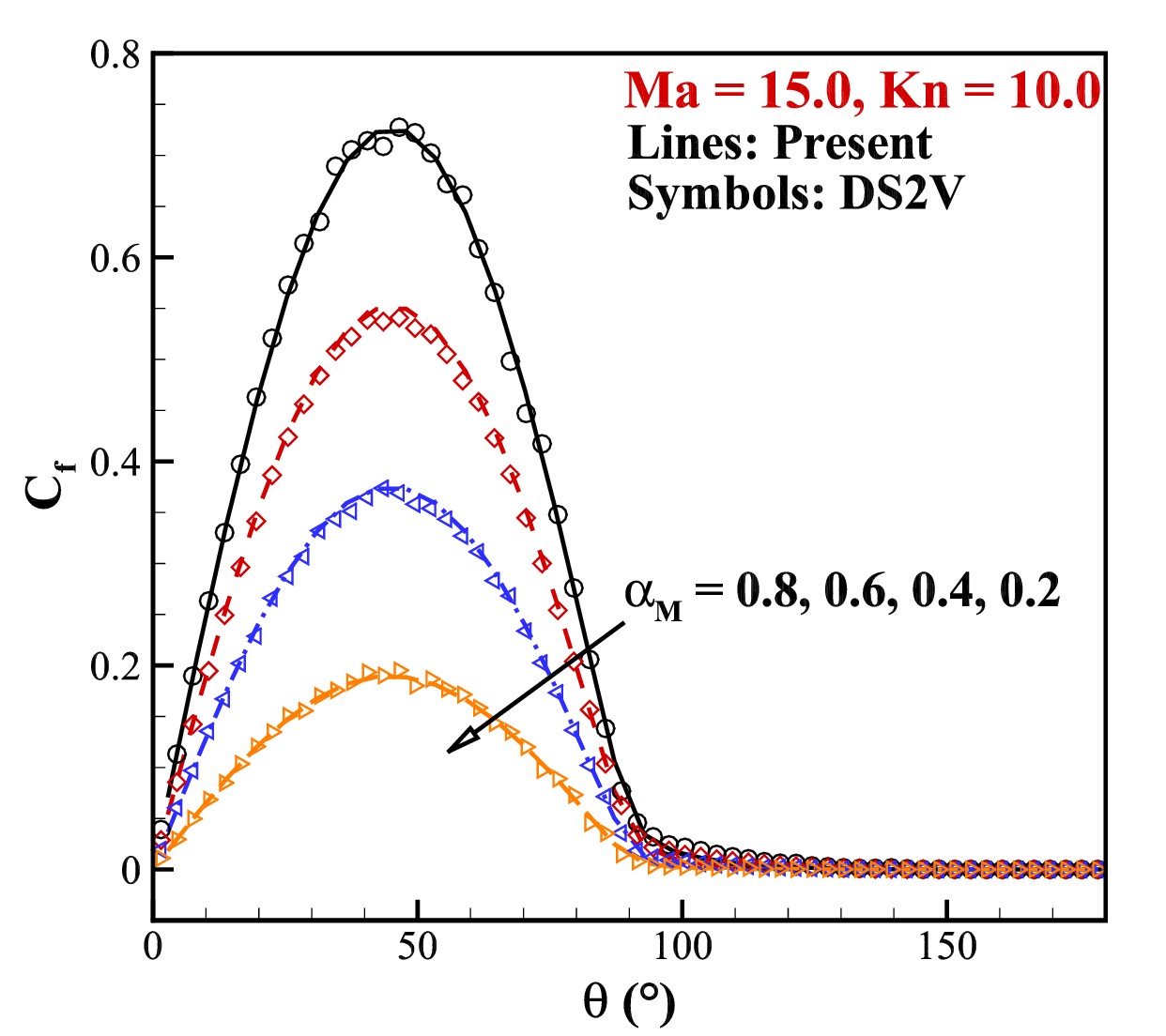}}
		\subfigure[]{\label{cylinder_MAX_15_10_H_fix}\includegraphics[width=0.32\textwidth]{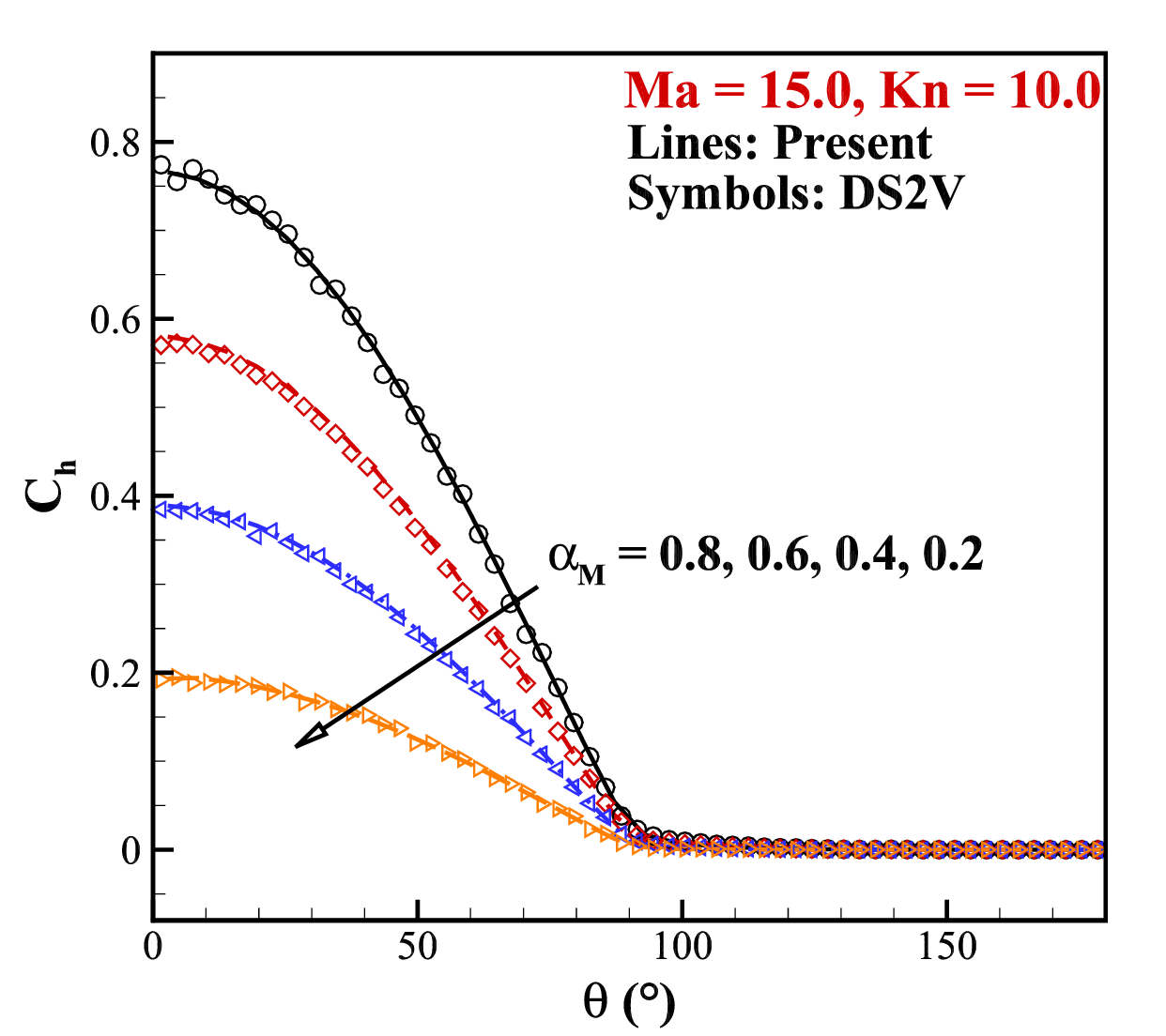}}
		\caption{\label{cylinder_MAX_15_10_fix}{Comparison of the (a) pressure coefficient, (b) skin friction coefficient, and (c) heat transfer coefficient on the surface of cylinder with different $\alpha_{M}$ when employing the Maxwell boundary (Obtained from model Eq. (\ref{CL_GSI})).}}
	\end{figure}

	\begin{figure}[!htp]
		\centering
		\subfigure[]{\label{cylinder_CLL_15_10_P_fix}\includegraphics[width=0.32\textwidth]{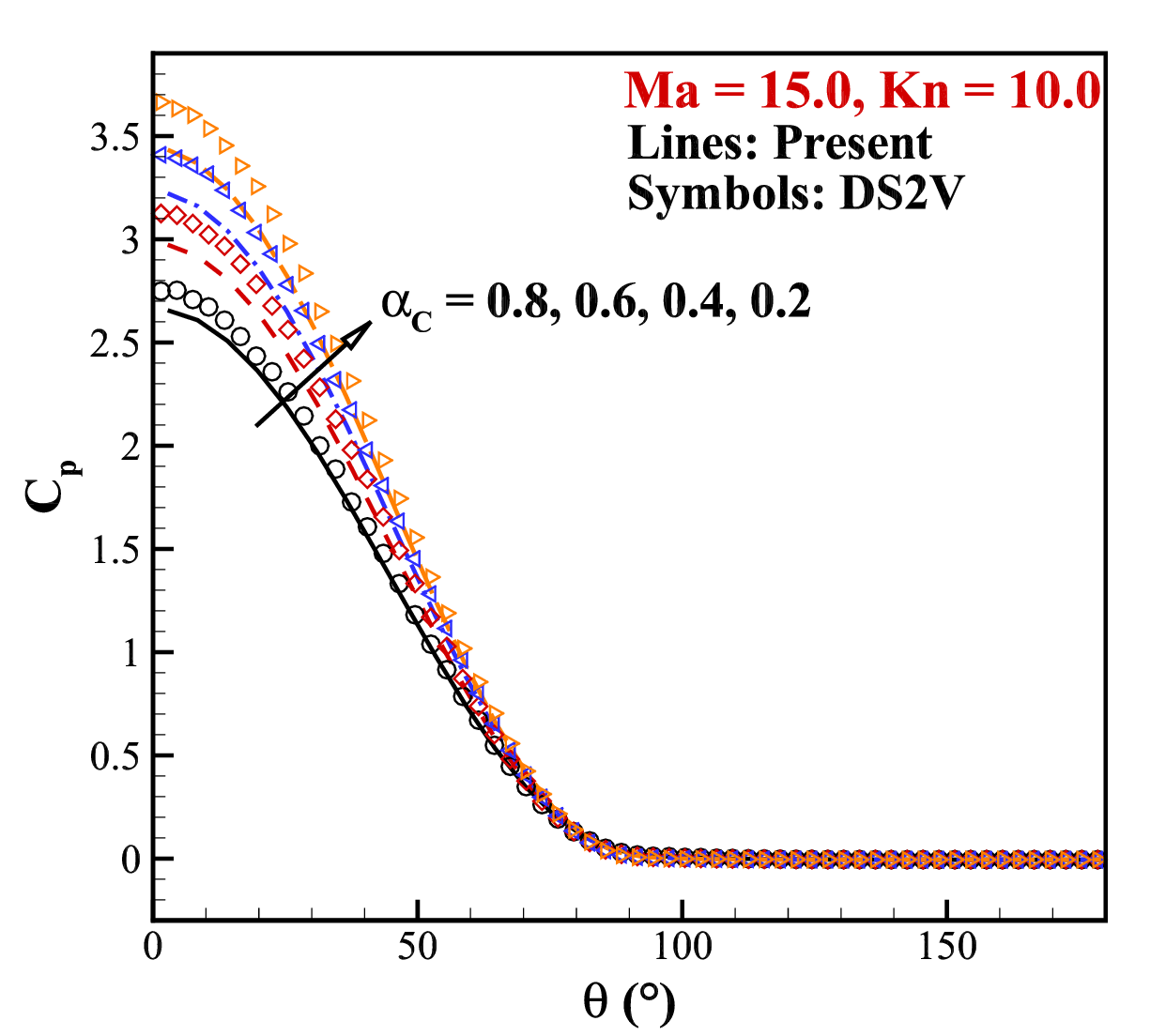}}
		\subfigure[]{\label{cylinder_CLL_15_10_S_fix}\includegraphics[width=0.32\textwidth]{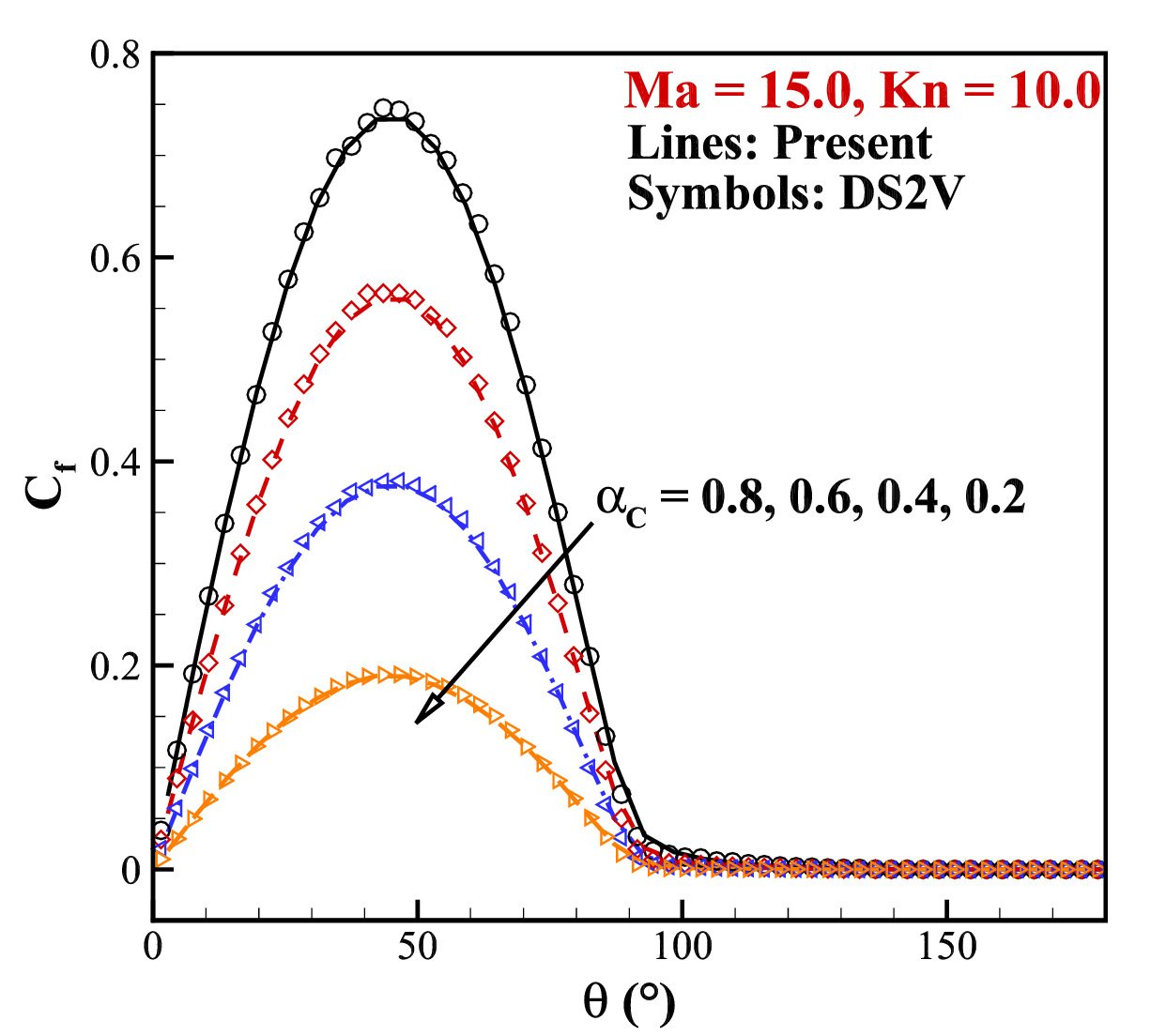}}
		\subfigure[]{\label{cylinder_CLL_15_10_H_fix}\includegraphics[width=0.32\textwidth]{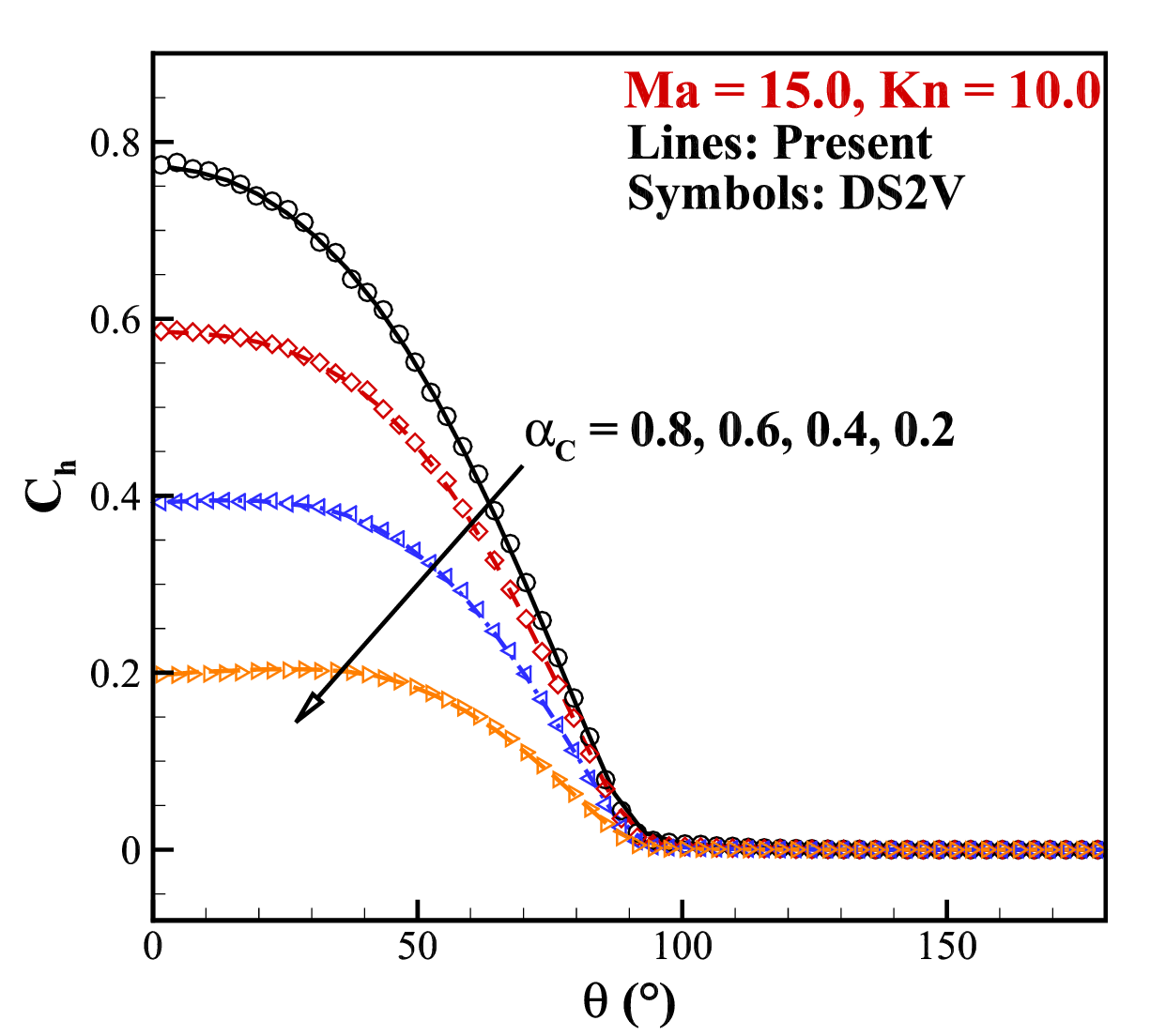}}
		\caption{\label{cylinder_CLL_15_10_fix}{Comparison of the (a) pressure coefficient, (b) skin friction coefficient, and (c) heat transfer coefficient on the surface of cylinder with different $\alpha_{C}$ when employing the CLL boundary (Obtained from model Eq. (\ref{CL_GSI})).}}
	\end{figure}
	
\end{document}